\documentclass[11pt]{article}
\usepackage{geometry}                
\usepackage{graphicx}
\usepackage{amssymb}
\usepackage{epstopdf}
\usepackage{amsthm}
\author{Seth S. Cottrell\\{\it Courant Institute of Mathematical Sciences}\\{\it New York University, 251 Mercer Street, New York, NY 10012}}

\newtheorem{thm}{Theorem}[section]
\newtheorem*{thma}{Theorem}

\title{Finding Structural Anomalies in Star Graphs Using Quantum Walks: A General Approach}
\begin{document}
\maketitle

\pagebreak

\section*{Abstract}

In previous papers about searches on star graphs several patterns have been made apparent; the speed up only occurs when graphs are ''tuned'' so that their time step operators have degenerate eigenvalues, and only certain initial states are effective.  More than that, the searches are never faster than $O\left(\sqrt{N}\right)$ time.

In this paper the problem is defined rigorously, the causes for all of these patterns are identified, sufficient and necessary conditions for quadratic-speed searches for any connected subgraph are demonstrated, the tolerance of these conditions is investigated, and it is shown that (unfortunately) we can do no better than $O\left(\sqrt{N}\right)$ time.  Along the way, a useful formalism is established that may be useful in future work involving highly symmetric graphs.

\pagebreak

\section{Introduction and Review of Quantum Walks}

A quantum walk is a quantum version of a random walk \cite{reitzner}.  In a quantum walk, a particle moves on a general structure, a graph, which is a collection of vertices and edges connecting them, and its motion is governed by amplitudes, whereas in a classical random walk it would be governed by probabilities.  There are a number of types of quantum walks.  First, the time can be either continuous \cite{farhi} or advance in discrete steps \cite{davidovich,vazirani}.  Within the discrete- time walks, there are another two types.  In the coined walk, the particle sits on the vertices and an extra degree of freedom, a quantum coin, is needed to make the dynamics of the walk unitary.  In the scattering walk, the particle sits on the edges, and no coin is necessary \cite{hillery0}.  There have been a number of experimental implementations of quantum walks, some using trapped ions \cite{ScMaScGlEnHuSc09,KaFoChStAlMeWi09} and others using photons in optical networks \cite{PeLaPoSoMoSi08}-\cite{schreiber}.

Quantum walks have proven useful in finding new quantum algorithms and expanding the applicability of ones that are known.  We are interested in searches.  In most cases there is a distinguished vertex, one whose behavior is different from the others, and the object is to find this vertex \cite{shenvi}-\cite{kendon2}.  More recently, quantum walks have been shown to be useful in finding marked structures, such as cliques \cite{marked-clique}, or structures that break the symmetry of a graph \cite{feldman,hillery1}.  In this paper a generalization of the latter problem will be pursued.

\vspace{5mm}

In the scattering model of quantum walks \cite{hillery0} the states are defined on edges, with two states on every edge; one for each of the two possible directions.  So, if two vertices connected by some edge are labeled $a$ and $b$, then $|a,b\rangle$ is the state on the edge that points from $a$ to $b$, and $|b,a\rangle$ is the state on the edge pointing from $b$ to $a$.  If the vertices are labeled $1,\cdots,m$, then a state of the system is written $|\Psi\rangle = \sum_{j,k=1}^m \alpha_{j,k}|j,k\rangle$, where $\alpha_{j,k}\in\mathbb{C}$, $0\le|\alpha_{j,k}|\le1$, and $\sum_{j,k=1}^m|\alpha_{j,k}|^2=1$.  These $\alpha$'s are probability amplitudes, and $|\alpha_{j,k}|^2$ is the probability of measuring $|\Psi\rangle$ and detecting the particle in the state $|j,k\rangle$.

Each vertex hosts a local unitary operator that maps all of the incoming states to outgoing states.  That is, if $v$ is a vertex, and $\{a_j\}$ is the set of connected vertices, then the unitary opperator defined on $v$, denoted ${\bf U}_v$, preforms a mapping ${\bf U}_v: \{|a_j,v\rangle\}\to\{|v,a_j\rangle\}$.  Notice that the notation used to describe the states indicates which vertex operator to apply.  The only operator that will act on the state $|s,t\rangle$ is ${\bf U}_t$, and the vertex operator most recently applied to $|s,t\rangle$ was ${\bf U}_s$.  The time step operator, ${\bf U}$, is defined as all of the ${\bf U}_v$ taken together; ${\bf U} = \oplus_v {\bf U}_v$, where this direct sum is taken over all vertices.  Like each ${\bf U}_v$, ${\bf U}$ is unitary, but unlike each of the vertex-specific operators, ${\bf U}$ is an endomorphism.  As such, we can talk about the eigenvalues and eigenvectors (equivalently, ''eigenstates'') of ${\bf U}$.  Because the edge state implies immediately which of the vertex operators to apply, we will in general only talk about ${\bf U}$ as a whole and abandon the vertex-specific notation.  For example, ${\bf U}|s,t\rangle = \oplus_v{\bf U}_v|s,t\rangle = {\bf U}_t|s,t\rangle$, but it is unnecessary on the right hand side to indicate which vertex operator is being used, since it is already announced by the state, ``$|s,t\rangle$''.

The eigenvalues of ${\bf U}$ all have modulus 1, and there is no ''steady state''.  In an intuituve, not-rigorous sense, because ${\bf U}$ must be unitary (it is a quantum mechanical time evolution operator) it must conserve information.  That is, for any given state there is a unique pre-image.  But a particle on a vertex at time $t$ could have been on any adjacent vertex at time $t-1$.  In the earliest attempts to define a quantum analog to discrete random walks the states were again defined on the vertices, but it was quickly found that an ancillary ``coin space'' needed to be attached to each vertex to maintain unitarity and keep track of the state's previous vertex.  The edge state formalism (e.g., $|a,b\rangle$) is an equivalent formalism that eliminates the need for ancillary spaces, while still carrying the additional information required.

\vspace{5mm}

{\bf Definition} For the purposes of this paper a ``search'' is defined to be a process used to distinguish between $N-1$ identical elements, and one marked element that a priori can be {\it any} of the $N$ total elements.  Under this definition the Shor algorithm, for example, is not a search because it is used to find numbers with certain properties that set them apart.  The set of numbers being considered are not all equally likely to be "marked".  Part of the reasoning behind this definition is that it leads naturally to graphs with very high symmetry, which substantially reduces the dimension and difficulty of the problem.

\vspace{5mm}

Define a mapping $\phi: E \to E$, where $E$ is the set of edges, to be some rearrangement of edges such that ${\bf U} = \phi^{-1}\circ{\bf U} \circ \phi$, or equivalently $\phi \circ {\bf U} = {\bf U} \circ \phi$.  $\phi$ is called a quantum graph automorphism \cite{krovi}, and it is a rearrangement of edges and vertices that leaves the effect of ${\bf U}$ invariant.  If there is a set of edges that can be mapped into each other by some quantum graph automorphism, then we say that these edges belong to the same equivalence class, and a uniform superposition of states on these edges is seen as a single edge on the "collapsed graph".

For example, consider the complete graph $K_3$ with vertices labeled $A, B, C$.  Define $A$ to be a strictly reflecting vertex (i.e., ${\bf U}|B,A\rangle=|A,B\rangle$) and $B$ and $C$ as strictly transmitting (i.e., ${\bf U}|B,C\rangle = |C,A\rangle$).  We find that the only non-trivial quantum graph automorphism is the one that exchanges $B$ and $C$.  That is, $\phi$ makes the following exchanges: $|A,B\rangle\leftrightarrow|A,C\rangle$, $|B,A\rangle\leftrightarrow|C,A\rangle$, and $|B,C\rangle\leftrightarrow|C,B\rangle$.

\begin{figure}[h!]
\centering
\includegraphics[keepaspectratio=true, scale = 0.5]{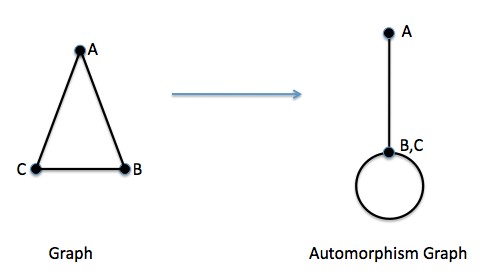}
\caption{Collapsing a graph.}
\end{figure}

Whereas the original graph consisted of six states, $G=\{|A,B\rangle, |B,A\rangle, |A,C\rangle, |C,A\rangle, |B,C\rangle, |C,B\rangle\}$, the collapsed graph consists of only three states,

$G_A=\{\frac{1}{\sqrt{2}}\left(|A,B\rangle+|A,C\rangle\right), \frac{1}{\sqrt{2}}\left(|B,A\rangle+|C,A\rangle\right), \frac{1}{\sqrt{2}}\left(|B,C\rangle+|C,B\rangle\right)\}$.  There are two things to notice here.  First, this set is closed under the action of ${\bf U}$, and second, each state in $G_A$ is mapped to itself under $\phi$.  Indeed, the states on any collapsed quantum graph are composed of exactly those states that are left invariant under the action of every $\phi$.

\vspace{5mm}

{\bf Definition} For any graph $G$ with a time step operator ${\bf U}$ we define the "collapsed graph" or "automorphism graph" as $G_A \equiv \{|\psi\rangle: \phi|\psi\rangle=|\psi\rangle, \forall \phi\textrm{ where } {\bf U} = \phi^{-1}{\bf U} \phi\}$.  That is, the collapsed graph is composed only of those states that are left invariant under the mapping of every automorphism that commutes with ${\bf U}$.

\vspace{5mm}

It's straightforward to demonstrate that $G_A$ is closed under ${\bf U}$.  If $|\psi\rangle$ is a state on $G_A$, then

$\begin{array}{ll}
|\psi\rangle \in G_A \\
\Rightarrow \phi|\psi\rangle = |\psi\rangle \\
\Rightarrow {\bf U}\phi|\psi\rangle = {\bf U}|\psi\rangle \\
\Rightarrow \phi{\bf U}|\psi\rangle = {\bf U}|\psi\rangle \\
\Rightarrow {\bf U}|\psi\rangle \in G_A
\end{array}$

By using collapsed quantum graphs we can decrease the dimension of the state space substantially.  In the example above the number of dimensions was cut in half, but the greater the symmetry the greater the decrease in dimensions.

\section{Star graphs in With an Unknown Flaw}

\begin{figure}[h!]
\centering
\includegraphics[keepaspectratio=true, scale = 0.5]{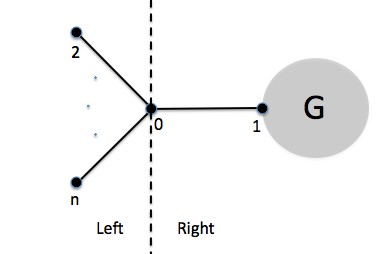}
\caption{A star graph with an unknown flaw "G".}
\end{figure}

This section describes the model that will be used throughout this paper.

In figure 2 the 0 vertex is the "hub vertex".  A hub vertex is a vertex with $N$ connections, where $N$ is allowed to vary.  It is generally assumed that the reflection and transmission coefficients at a hub are the same for each of the $N$ edges, so any incoming state, $|j,0\rangle$ will be mapped to

\begin{equation}
{\bf U}|j,0\rangle = r|0,j\rangle + t\sum_{k\ne j}|0,k\rangle
\end{equation}

Unitarity at the hub requires that $|r|^2+(N-1)|t|^2=1$ and $2Re(r^*t) + (N-2)|t|^2=0$.  For the examples in this paper we'll assume that the hub is a standard "diffusive vertex" which means that $r=-1+\frac{2}{N}$ and $t=\frac{2}{N}$, although it's easy to generalize away from this.

The subgraph, $G$, is attached to any one of the $N$ edges radiating from vertex 0 which, in analogy to a spoked wheel, is called the "hub" or "hub vertex".  Although we don't know which edge is connected to $G$ we can, without loss of generality, assume that it is attached to vertex 1.  In this way, vertex 1 is the "marked" vertex.

Vertices 0 and 1, the states between them, and everything in $G$ make up the Right side of the graph.

Vertices 2 through $N$ ``reflect with a phase of $\phi$'', which means that ${\bf U}|0,j\rangle = e^{i\phi}|j,0\rangle$.  $\phi$ is left with an unspecified value, so that it can be used to ``dial in'' certain eigenvalues of ${\bf U}$ (how this is done will be shown momentarily).

Vertices 0 and 2 through N, as well as the edges connecting them, are called the Left side of the graph.

Throughout this paper I will be referring to Right or Left eigenvalues or vectors.  This will always indicate that the thing in question is native to that side of the graph.  All eigenvectors and values 

The goal of a quantum search on a star graph is to somehow get the probability amplitude on the states $|0,1\rangle$ and $|1,0\rangle$ as high as possible, so that when a measurement is made, the result is likely to be the marked edge.  In this way the marked edge is found, and the search completed.  Typically it is assumed that measurements cannot be made on $G$ itself since, in some sense, if we had access to $G$ we wouldn't be looking for it.  So if a particle would have been measured on an edge in $G$, the measurement is assumed to be a null result.  This turns out to be a smaller problem than it might seem at first and so a better, less-specific goal is to get the state supported mostly on the Right side.

The effect of ${\bf U}$ on $|0,1\rangle$, as well as on any of the states in $G$, is different for different graphs, and will be left as a ``black box''.

\vspace{5mm}

For every value of $N\ge2$ we have another, different graph, and a new problem to solve.  However, by taking advantage of the obvious symmetries on the Left side, this graph can be collapsed substantially.
For ease of notation define the following:

\begin{eqnarray}
|in\rangle \equiv \frac{1}{\sqrt{N-1}}\sum_{j=2}^N |j,0\rangle \\[2mm]
|out\rangle \equiv \frac{1}{\sqrt{N-1}}\sum_{j=2}^N |0,j\rangle
\end{eqnarray}

This is the collapsing discussed in the previous section.  While there may be additional collapsing/simplification taking place inside of the subgraph, $G$, that will not be directly addressed in this paper.  At the hub vertex we find that ${\bf U}$ does this:

\begin{eqnarray}
{\bf U}|in\rangle = [r+(N-2)t]|out\rangle + t\sqrt{N-1}|0,1\rangle \\[2mm]
{\bf U}|1,0\rangle = r|0,1\rangle + t\sqrt{N-1}|out\rangle
\end{eqnarray}

\vspace{5mm}

Rewriting this in terms of a new perturbation variable, $\epsilon\equiv\frac{1}{N}$, we find:

\begin{eqnarray}
{\bf U}|in\rangle = (1-2\epsilon)|out\rangle + 2\sqrt{\epsilon-\epsilon^2}|0,1\rangle \\[2mm]
{\bf U}|1,0\rangle = (-1+2\epsilon)|0,1\rangle + 2\sqrt{\epsilon-\epsilon^2}|out\rangle
\end{eqnarray}

\vspace{5mm}

Note that ${\bf U}$ is a function of $\epsilon$, and define ${\bf U}_0\equiv{\bf U}|_{\epsilon=0}$.  The number of edges, $N$, connected to the hub vertex is the only variable in the graph, and studying the dependence of the eigenvalues and eigenvectors on $\epsilon = \frac{1}{N}$ will be the focus of the rest of this paper.  Unless otherwise noted, assume that every variable (eigenvalues, eigenvectors, etc.) is a function of $\epsilon$.  In general, denote $f_0\equiv f(\epsilon)|_{\epsilon=0}$.  For example, suppose $|w\rangle$ is an eigenvector of ${\bf U}$.  It may depend on $\epsilon$, because ${\bf U}$ depends on $\epsilon$.  $|w_0\rangle$ is the corresponding eigenvector of ${\bf U}_0$, defined as $|w_0\rangle\equiv|w\rangle\,|_{\epsilon=0}$, and it is constant.

When $\epsilon=0$, which is the $N\to\infty$ limit, the states of the two sides of the graph, Left $\equiv\{|in\rangle, |out\rangle\}$ and Right $\equiv\{G, |0,1\rangle, |1,0\rangle\}$, are kept separate by ${\bf U}_0$.  The eigenvalues and vectors of both the Left and Right sides are changed relatively little by $\epsilon$ (this will be proven), so once they have been found they can be used without modification.  Clearly, $\epsilon=0$ is the easiest case to work with, and dealing with very large values of $N$ (where search algorithms are useful) corresponds to very small values of $\epsilon$.  Phrased in this way, the problem lends itself naturally to a perturbative approach.

\vspace{5mm}

It's worth taking a moment to repeat that, and to point out how profound it is.  The original problem involved an infinite family of graphs indexed by, the number of edges in the star graph, $N$, each of which had a Hilbert space of dimension $2N+|G|$, where $|G|$ is the number of edge states in $G$.  But by using the symmetry of vertices 2 through $N$ to collapse the graph, the problem now only considers a {\it single} graph with a Hilbert space of dimension $4+|G|$ and a single non-constant vertex (the "hub'') that behaves like a valve between the Left and Right sides in a simple and predictable way.

\vspace{5mm}

Because the location of the 1 vertex is unknown, the initial states we have access to are of the form,

\begin{equation}
|\psi\rangle = \frac{\alpha}{\sqrt{N}}\sum_{j=1}^N |j,0\rangle + \frac{\beta}{\sqrt{N}}\sum_{j=1}^N |0,j\rangle
\end{equation}

Which can also be written,

\begin{equation}
|\psi\rangle = \alpha |in\rangle + \beta|out\rangle + O\left(\frac{1}{\sqrt{N}}\right)
\end{equation}

In other words, the initial state is (to within a small error) just a superposition of $|in\rangle$ and $|out\rangle$.

There's absolutely nothing stoping us from starting with a state like $\frac{1}{\sqrt{N}}\sum_{j=1}^N (-1)^j|0,j\rangle$.  However, this state breaks the graph's symmetry a bit, since we now have to treat the odd and even vertices separately, and the graph becomes that much more unwieldy.

A cursory look at ${\bf U}_0$ reveals that the eigenvalues from the Left side of the graph are $\lambda=\pm e^{i\frac{\phi}{2}}$, and that the corresponding eigenvectors are $\frac{1}{\sqrt{2}}\left(|out\rangle\pm e^{i\frac{\phi}{2}}|in\rangle\right)$.  It will be shown that if the Right side of the graph shares one or both of these eigenvalues, then a "rotation" (as described in section 4) may occur, and an initial state of the form $|\ell_0\rangle =  \frac{1}{\sqrt{2}}\left(|out\rangle\pm e^{i\frac{\phi}{2}}|in\rangle\right)$ will move into another eigenstate with the same eigenvalue, $|\mathfrak{r}_0\rangle$, in the Right side in $O\left(\sqrt{N}\right)$ time.

So, finding $G$ boils down to finding one of the eigenvalues and eigenvectors of the Right side, and this can be done by "dialing in" $\phi$ so that the Left side shares an eigenvalue with the Right.  This, by the way, is the great advantage of star graphs; any eigenvalue can be easily attained, and there are only two dimensions ($|in\rangle$ and $|out\rangle$).  That said, the results and techniques of this paper can be applied far more generally.

The "paired" eigenvalues of ${\bf U}$ that are involved in searches with a quadratic speed up take the form $\lambda(\epsilon) = \sum_{j=0}^\infty a_j(\pm\sqrt{\epsilon})^j$ (this will be proven).  In what follows, it will often be advantageous to write this as $\lambda = \lambda_0 e^{\pm ic\sqrt{\epsilon}}+O(\epsilon)$.  It will be shown that finding a known graph merely requires matching Left and Right eigenvalues of ${\bf U}_0$.  Finding an {\it unknown} graph means finding an eigenvalue of the Right side, which consists of finding $\lambda_0$ and dealing with $c$.

\subsection{Grover Graph Example}

For the Grover graph, $G$ is simply a vertex that reflects with a phase of $\pi$.  The Grover graph is named for the Grover algorithm, which it exactly emulates.  This example is simple enough that it can be completed by hand.  The pairing of eigenvectors, the $\sqrt{\epsilon}$ dependences and errors, as well as the $O(\sqrt{N})$ time required to execute the search are all evident here.

The basis states of the (collapsed) Grover graph are:

$\begin{array}{ll}
|\psi_1\rangle = |out\rangle \\
|\psi_2\rangle = |in\rangle \\
|\psi_3\rangle = |0,1 \rangle\\
|\psi_4\rangle = |1,0 \rangle
\end{array}$

\vspace{5mm}

The time step matrix, ${\bf U}$, is defined as before with the addition that ${\bf U}|0,1\rangle = -|1,0\rangle$.  So, using the basis states as listed above,

${\bf U}=\left(\begin{array}{cccc}
0 & 1-2\epsilon & 0 & 2\sqrt{\epsilon-\epsilon^2} \\
e^{i\phi} & 0 & 0 & 0 \\
0 & 2\sqrt{\epsilon-\epsilon^2} & 0& 2\epsilon-1 \\
0 & 0 & -1 & 0 \\
\end{array}\right)$ and
${\bf U}_0=\left(\begin{array}{cccc}
0 & 1 & 0 & 0 \\
e^{i\phi} & 0 & 0 & 0 \\
0 & 0 & 0& -1 \\
0 & 0 & -1 & 0 \\
\end{array}\right)$

\vspace{5mm}

The characteristic polynomial for ${\bf U}_0$ is $C_0(\lambda) = \lambda^4 - (e^{i\phi}+1)\lambda^2 + e^{i\phi}=(\lambda^2-e^{i\phi})(\lambda^2-1)$.  This clean factoring is symptomatic of the separation of the Right and Left sides when $\epsilon=0$.  Clearly, double roots can be found when $\phi=0$, so in what follows $\phi$ will be set to 0.  The eigenvectors on the Left are $|\ell^{(1)}\rangle = \frac{1}{\sqrt{2}} \left(|out\rangle + |in\rangle\right)$ and $|\ell^{(-1)}\rangle = \frac{1}{\sqrt{2}} \left(|out\rangle - |in\rangle\right)$, for $\lambda=\pm1$, and the eigenvectors on the Right are $|\mathfrak{r}^{(1)}\rangle = \frac{1}{\sqrt{2}} \left(|0,1\rangle - |1,0\rangle\right)$ and $|\mathfrak{r}^{(-1)}\rangle = \frac{1}{\sqrt{2}} \left(|0,1\rangle + |1,0\rangle\right)$, for $\lambda=1$ and $\lambda=-1$ respectively.

The characteristic polynomial for ${\bf U}$ (with $\phi=0$) is $C(\lambda) = \lambda^4 - 2(1-2\epsilon)\lambda^2 + 1$.  The solutions of this equation are $\lambda(\epsilon) = \pm\sqrt{1-2 \epsilon \pm 2i\sqrt{\epsilon-\epsilon^2}}$, where the $\pm$'s are independent of each other.  The four eigenvalues and their associated eigenvectors are:

$\begin{array}{ll}
\lambda^{(1)}(\epsilon)=\sqrt{1-2 \epsilon + 2i\sqrt{\epsilon-\epsilon^2}} \\
\lambda^{(2)}(\epsilon)=\sqrt{1-2 \epsilon - 2i\sqrt{\epsilon-\epsilon^2}} \\
\lambda^{(3)}(\epsilon)=-\sqrt{1-2 \epsilon + 2i\sqrt{\epsilon-\epsilon^2}} \\
\lambda^{(4)}(\epsilon)=-\sqrt{1-2 \epsilon - 2i\sqrt{\epsilon-\epsilon^2}} \\
\end{array}$

$|V^{(1)}(\epsilon)\rangle=\frac{1}{2}\left(\begin{array}{c}
\sqrt{1-2\epsilon+2i\sqrt{\epsilon-\epsilon^2}} \\
1 \\
-i\sqrt{1-2\epsilon+2i\sqrt{\epsilon-\epsilon^2}} \\
i
\end{array}\right)$
$|V^{(2)}(\epsilon)\rangle=\frac{1}{2}\left(\begin{array}{c}
\sqrt{1-2\epsilon-2i\sqrt{\epsilon-\epsilon^2}} \\
1 \\
i\sqrt{1-2\epsilon-2i\sqrt{\epsilon-\epsilon^2}} \\
-i
\end{array}\right)$

$|V^{(3)}(\epsilon)\rangle=\frac{1}{2}\left(\begin{array}{c}
\sqrt{1-2\epsilon+2i\sqrt{\epsilon-\epsilon^2}} \\
-1 \\
-i\sqrt{1-2\epsilon+2i\sqrt{\epsilon-\epsilon^2}} \\
-i
\end{array}\right)$
$|V^{(4)}(\epsilon)\rangle=\frac{1}{2}\left(\begin{array}{c}
\sqrt{1-2\epsilon-2i\sqrt{\epsilon-\epsilon^2}} \\
-1 \\
i\sqrt{1-2\epsilon-2i\sqrt{\epsilon-\epsilon^2}} \\
i
\end{array}\right)$

Notice that looping $\epsilon$ around 0 permutes $\{\lambda^{(1)}(\epsilon), \lambda^{(2)}(\epsilon)\}$ and $\{\lambda^{(3)}(\epsilon), \lambda^{(4)}(\epsilon)\}$.  This property is important later.  Written more simply,

$\begin{array}{ll}
\lambda^{(1)}(\epsilon) = e^{i\sqrt{\epsilon}} + O(\epsilon)\\
\lambda^{(2)}(\epsilon) = e^{-i\sqrt{\epsilon}} + O(\epsilon)\\
\lambda^{(3)}(\epsilon) = -e^{i\sqrt{\epsilon}} + O(\epsilon)\\
\lambda^{(4)}(\epsilon) = -e^{-i\sqrt{\epsilon}} + O(\epsilon)
\end{array}$

$|V^{(1)}(\epsilon)\rangle=\frac{1}{2}\left(\begin{array}{c}
1 \\
1 \\
-i \\
i
\end{array}\right)+O(\sqrt{\epsilon}) = \frac{1}{\sqrt{2}}\left(|\ell^{(1)}\rangle-i|\mathfrak{r}^{(1)}\rangle\right)+O(\sqrt{\epsilon})$

$|V^{(2)}(\epsilon)\rangle=\frac{1}{2}\left(\begin{array}{c}
1 \\
1 \\
i \\
-i
\end{array}\right)+O(\sqrt{\epsilon})=\frac{1}{\sqrt{2}}\left(|\ell^{(1)}\rangle+i|\mathfrak{r}^{(1)}\rangle\right)+O(\sqrt{\epsilon})$

$|V^{(3)}(\epsilon)\rangle=\frac{1}{2}\left(\begin{array}{c}
1 \\
-1 \\
-i \\
-i
\end{array}\right)+O(\sqrt{\epsilon})=\frac{1}{\sqrt{2}}\left(|\ell^{(-1)}\rangle-i|\mathfrak{r}^{(-1)}\rangle\right)+O(\sqrt{\epsilon})$

$|V^{(4)}(\epsilon)\rangle=\frac{1}{2}\left(\begin{array}{c}
1 \\
-1 \\
i \\
i
\end{array}\right)+O(\sqrt{\epsilon})=\frac{1}{\sqrt{2}}\left(|\ell^{(-1)}\rangle+i|\mathfrak{r}^{(-1)}\rangle\right)+O(\sqrt{\epsilon})$

Notice that the two eigenspaces of ${\bf U}_0$ are both two dimensional, and are spanned differently.  For example, the 1-eigenspace is spanned by $\{|\ell^{(1)}\rangle, |\mathfrak{r}^{(1)}\rangle\}$ as well as by $\{|V^{(1)}(0)\rangle, |V^{(2)}(0)\rangle\}$.  Moreover, the space spanned by $\{|V^{(1)}(\epsilon)\rangle, |V^{(2)}(\epsilon)\rangle\}$ for $0<\epsilon\ll1$ is nearly the same.  Using the eigenvectors of ${\bf U}_0$, instead of the "true" eigenvectors of ${\bf U}$ does introduce errors, but they are small (this will be proven).

The only reasonable initial states we have access too, under the assumption that the target vertex, 1, is unknown, are even superpositions of the form $\frac{1}{\sqrt{N}}\sum_{j=1}^N |0,j\rangle$ and $\frac{1}{\sqrt{N}}\sum_{j=1}^N |j,0\rangle$.  These states are located almost completely on the Left side.

This means that the initial states will always start on the Left, up to a tiny error.  Setting the initial state to be (approximately) in the 1-eigenspace: $|\psi\rangle=\frac{1}{\sqrt{2N}}\sum_{j=1}^N\left(|0,j\rangle+|j,0\rangle\right) = \frac{1}{\sqrt{2}}\left(|in\rangle+|out\rangle\right)+O(\sqrt{\epsilon}) = |\ell^{(1)}\rangle +O(\sqrt{\epsilon})$

\vspace{5mm}

Applying ${\bf U}$ repeatedly yields:

$\begin{array}{ll}
{\bf U}^m|\psi\rangle \\
= {\bf U}^m\left(|\ell^{(1)}\rangle +O(\sqrt{\epsilon})\right) \\
= {\bf U}^m|\ell^{(1)}\rangle +O(\sqrt{\epsilon}) \\
= \frac{1}{\sqrt{2}}{\bf U}^m|V^{(1)}\rangle + \frac{1}{\sqrt{2}}{\bf U}^m|V^{(2)}\rangle + O(\sqrt{\epsilon}) \\
= \frac{1}{\sqrt{2}}e^{im\sqrt{\epsilon}}|V^{(1)}\rangle + \frac{1}{\sqrt{2}}e^{-im\sqrt{\epsilon}}|V^{(2)}\rangle + O(\sqrt{\epsilon}) \\
= \frac{1}{2\sqrt{2}}\left(\begin{array}{c}
e^{im\sqrt{\epsilon}} + e^{-im\sqrt{\epsilon}}\\
e^{im\sqrt{\epsilon}} + e^{-im\sqrt{\epsilon}}\\
-ie^{im\sqrt{\epsilon}} + ie^{-im\sqrt{\epsilon}}\\
ie^{im\sqrt{\epsilon}} - ie^{-im\sqrt{\epsilon}}
\end{array}\right) + O(\sqrt{\epsilon}) \\
= \frac{1}{\sqrt{2}}\left(\begin{array}{c}
\cos{(m\sqrt{\epsilon})} \\
\cos{(m\sqrt{\epsilon})} \\
-i\sin{(m\sqrt{\epsilon})}  \\
i\sin{(m\sqrt{\epsilon})}
\end{array}\right) + O(\sqrt{\epsilon}) \\
\end{array}$

So, if $m = \frac{\pi}{2\sqrt{\epsilon}} = \frac{\pi}{2}\sqrt{N}$, then to within an error of $O\left(\frac{1}{\sqrt{N}}\right)$ the system will be on the Right side, in either the state $|0,1\rangle$ or $|1,0\rangle$.  A measurement at this time will complete the search.

\vspace{5mm}

It's worth pointing out that if we hadn't set $\phi=0$ then the "true" eigenvectors $|V^{(k)}(\epsilon)\rangle$ would have been almost entirely constrained to one side or the other.  This is because $|V^{(k)}(0)\rangle$ span the same eigenspaces as the Right and Left eigenvectors of ${\bf U}_0$.  If the Right and Left sides share no common eigenvalues, then we have four distinct eigenvalues and four distinct eigenspaces.  So, rather than being a rearrangment of the eigenvectors of ${\bf U}_0$, as was the case above, the $|V^{(k)}(\epsilon)\rangle$ are merely tiny perturbations of each of them individually.

Grover originally described his algorithm with a diffusion operator and an oracle operator.  The diffusion operator is here replaced by the hub vertex, and the oracle with the differently-phased reflections from the remaining vertices.  The result is exactly the same, and we can see that Grover's proof of optimality is merely a statement about the nature of the hub.  That is, there's no way to modify the hub alone to get a better speed up.

\section{Algebraic Functions and the Behavior of Zeros}
In this section we'll consider the relationship between the zeros of a polynomial and the coefficients of that polynomial.  Later these results will be applied to characteristic polynomials.  For more background on the math used in this section see \cite{kato} and \cite{ahlfors}.

\vspace{5mm}

{\bf Definition} A "globally analytic function" is an analytic function that is defined as every possible analytic continuation of an analytic function from a particular point in a domain.  Globally analytic functions can be many-valued.

For example, the globally analytic function $f(z)=\sqrt{z}$ has two branches (when $z\ne0$): $f^{(0)}(re^{i\theta})=\sqrt{r}e^{i\frac{\theta}{2}}$ and $f^{(1)}(re^{i\theta})=\sqrt{r}e^{i\frac{\theta}{2}+\pi}$ with the branch cut (arbitrarily) taken at $\theta =0$.

Just to emphasize the point that globally analytic functions can be many-valued, if $f(z)= \sqrt[4]{z}$, then $f(16)=2, 2i, -2, -2i$, that is; $f$ is "4-valued".  For the different branches, $f^{(0)}(16)=2$, $f^{(1)}(16)=2i$, $f^{(2)}(16)=-2$, and $f^{(3)}(16)=-2i$.

\vspace{5mm}

\begin{thm}
If $f(z)$ is globally analytic in an annulus around $0$, and is $m$-valued, then $f(z)$ can be expressed as a Puiseux series (a Laurent series with certain rational powers) of the form $f(z)=\sum_{n=-\infty}^\infty A_n z^\frac{n}{m}$.  Moreover, the m different branches of $f$, $f^{(k)}$, can be separated by an arbitrary branch cut through the annulus and expressed as $f^{(k)}(z) = \sum_{n=-\infty}^\infty A_n \omega^{\alpha n} z^\frac{n}{m}$, where $\omega$ is a primitive $m$th root of unity, $\omega = e^{i\frac{2\pi}{m}}$.
\end{thm}

{\it Proof} The proof of this theorem is included in the appendix.

\vspace{5mm}

Define $P(z,\epsilon)$ to be a polynomial in $z$ and $\epsilon$, written $P(z,\epsilon) = \sum_{j=0}^d a_j(\epsilon)z^j$, where each $a_j(\epsilon)$ is a polynomial.

Assume that $a_d(0)\ne 0$ and that $P(z,0)$ has no zeros with a multiplicity higher than 1.  That is, $P(z,0)$ has $d$ independent roots: $\lambda^{(1)}, \cdots, \lambda^{(d)}$.

\vspace{5mm}

\begin{thm} 
There exists an open disk $D$, containing 0, and $d$ analytic functions, $f^{(1)}, \cdots, f^{(d)}$, such that:

$\begin{array}{ll}
(i) & P\left(f^{(k)}(\epsilon), \epsilon\right)=0, \epsilon\in D\\
(ii) & f^{(k)}(0) = \lambda^{(k)} \\
(iii) & P(\lambda, \epsilon)=0, \epsilon\in D \Rightarrow \lambda=f^{(k)}(\epsilon), for\; some\; k
\end{array}$
\end{thm}

Note that $f_k$ are indexed functions, and not necessarily branches of the same globally analytic function.  It is true that when $P(z,\epsilon)$ is not simultaneously reducible in both $z$ and $\epsilon$ the zeros are all branches of a single globally analytic function of $\epsilon$, however it isn't necessary to know that here.

{\it Proof} In appendix.

\vspace{5mm}

Issues can crop up with the theorem above.  Specifically, if $P(z,0)$ has a repeated zero.  When this is the case, we can concern ourselves with the annulus $0<|z-\lambda^{(k)}|<\delta$, and $0<\epsilon<\sigma$, where $\delta$ and $\sigma$ are as defined in the theorem above.

Before dealing with higher order roots we need a few more tools.

\vspace{5mm}

\begin{thm} If $P(z,\epsilon)$ is an irreducible polynomial in $z$ and $\epsilon$, then all of the double roots of $P$ are isolated in the $\epsilon$-plane.  That is, if for some value $\epsilon_0$, $P(z,\epsilon_0)$ has a double root, then there exists $\delta>0$ such that when $0<|\epsilon-\epsilon_0|<\delta$, $P(z,\epsilon)$ does not have a double root in $z$.
\end{thm}

{\it Proof} In appendix.

\vspace{5mm}

\begin{thm} 
In the neighborhood of a zero of $P(z,\epsilon)$ of multiplicity $s>1$, the zeros take the form $f_\alpha(\epsilon) = \sum_{n=-\infty}^\infty A_n\omega^{n\alpha}\epsilon^\frac{n}{H}$, where $H<s$.  Specifically, the zeros are branches of one or more $H_i$-valued global analytic functions, with the given Puiseux series expansion, such that $\sum H_i=s$.
\end{thm}

{\it Proof} In appendix

\vspace{5mm}

Notice that the only new case that this last theorem applies to are repeated roots.  We've continued to assume that $a_d(\epsilon)\ne 0$.

\subsection{A Note on the Analyticity of the Characteristic Polynomial}

In the section above it was assumed that the polynomial $P(z,\epsilon)$ is a polynomial with respect to both $z$ and $\epsilon$.  Yet a quick look at ${\bf U}$ reveals that this may not necessarily be the case for the characteristic polynomial, since ${\bf U}$ contains entries of $O(\sqrt{\epsilon})$.

However, it can be shown that for the situation being considered in this write-up $C(z,\epsilon) = |{\bf U}(\epsilon) - z {\bf I}|$ is always a polynomial with respect to $\epsilon$.  In fact, it is affine with respect to $\epsilon$.

\begin{thm}
Assume that ${\bf U}$ is a time step matrix as described so far.  That is, there is a Left and Right side and these are connected only through a hub vertex with $N$ edges where the reflection and transmission coefficients are $r=-1+\frac{2}{N}$ and $t=\frac{2}{N}$.  Then $C(z, \epsilon) = \left|{\bf U} - z {\bf I}\right|$ is an affine  polynomial of $\epsilon=\frac{1}{N}$, and can be written $C(z,\epsilon) = C_0(z) + \epsilon f(z)$.
\end{thm}

{\it Proof} The proof of this is in the appendix, in section 9.2.

\section{Pairing}

In the last section it was seen that the zeros, $\lambda^{(k)}$, of a polynomial with coefficients that are polynomials of $\epsilon$ take the form

\begin{equation}
\lambda^{(k)}(\epsilon) = \sum_{j=-\infty}^\infty A_j\omega^{j\alpha}\epsilon^\frac{j}{H}
\end{equation}

and these eigenvalues are grouped together into sets of size $H$ which permute when $\epsilon$ loops around zero.  For any given eigenvalue of ${\bf U}_0$ there can be more than one of these sets, and they may have different sizes.  When the polynomial in question is the characteristic polynomial of the transition matrix of a quantum walk, $C(z,\epsilon) = a_0(\epsilon)+a_1(\epsilon)z+\cdots+a_{d-1}(\epsilon)z^{d-1}+z^d$, then some new restrictions are brought into play.

The perturbation is assumed to be set up in such a way that for some path in the $\epsilon$-plane starting at $\epsilon=0$ ${\bf U}$ is unitary and along this path $|\lambda^{(k)}(\epsilon)|=1$.  In a star graph ${\bf U}$ is unitary for all positive integer values of $N$, so since $\epsilon=\frac{1}{N}$ this path is along the positive real axis in the $\epsilon$-plane.

\begin{thm}
The eigenvalues, $\lambda$, of the matrix for a quantum walk, ${\bf U}$, with a characteristic polynomial that is a polynomial in both $\lambda$ and $\epsilon$, can only take the form of $\lambda(\epsilon) = \sum_{j=0}^\infty A_j\epsilon^j$ or $\lambda^{(k)}(\epsilon) = \sum_{j=0}^\infty (-1)^{jk}A_j\left(\sqrt{\epsilon}\right)^j$.
\end{thm}

{\it Proof} Since ${\bf U}_0$ is unitary, $\lim_{\epsilon\to0}\lambda(\epsilon)$ exists and is finite ($|\lambda^{(k)}(0)|=1$, for all $k$).  Therefore the Puiseux series for $\lambda^{(k)}(\epsilon)$ has no negative terms.

For a particular eigenvalue of ${\bf U}_0$, $\lambda_0$, with multiplicity at least $H$, the various branches satisfy $\lambda^{(k)}(\epsilon)-\lambda_0 = \sum_{j=1}^\infty A_j\omega^{j\alpha}\epsilon^\frac{j}{H}$.  We can explore how the $H$ branches, $\{\lambda^{(1)}(\epsilon), \cdots,\lambda^{(H)}(\epsilon)\}$, separate from $\lambda_0$ by looking at the angle (phase difference) between two branches, $\lambda^{(k)}(\epsilon) - \lambda_0$ and $\lambda^{(j)}(\epsilon) - \lambda_0$.  If $\Delta$ is the angle between two complex numbers, $x$ and $y$, then $|x||y|\cos{(\Delta)} = Re(x\bar{y})$.  So,

$\begin{array}{ll}
|\lambda^{(k)}(\epsilon) - \lambda_0||\lambda^{(j)}(\epsilon) - \lambda_0| \cos{(\Delta)} = Re\left( (\lambda^{(k)}(\epsilon) - \lambda_0)\overline{(\lambda^{(j)}(\epsilon) - \lambda_0)} \right) \\
\Rightarrow |A_1\omega^k\epsilon^\frac{1}{H}+O(\epsilon^\frac{2}{H})||A_1\omega^j\epsilon^\frac{1}{H}+O(\epsilon^\frac{2}{H})| \cos{(\Delta)} = Re\left( (A_1\omega^k\epsilon^\frac{1}{H}+O(\epsilon^\frac{2}{H}))\overline{(A_1\omega^j\epsilon^\frac{1}{H}+O(\epsilon^\frac{2}{H}))} \right) \\
\Rightarrow \left(|A_1|^2|\omega|^{k+j} |\epsilon|^\frac{2}{H}+O(\epsilon^\frac{3}{H})\right)\cos{(\Delta)} = Re\left( |A_1|^2 \omega^k \bar{\omega}^j|\epsilon|^\frac{2}{H}+O(\epsilon^\frac{3}{H})) \right) \\
\Rightarrow \left(|A_1|^2 + O(\epsilon^\frac{1}{H})\right)\cos{(\Delta)} = Re\left( |A_1|^2 \omega^{k-j} + O(\epsilon^\frac{1}{H})) \right) \\
\Rightarrow \cos{(\Delta)} = \frac{Re\left( |A_1|^2 \omega^{k-j} + O(\epsilon^\frac{1}{H})) \right)}{|A_1|^2 + O(\epsilon^\frac{1}{H})} \\
\Rightarrow \cos{(\Delta)} = Re\left(\omega^{k-j}\right)+ O(\epsilon^\frac{1}{H}) \\
\Rightarrow \cos{(\Delta)} = \cos{\left(\frac{2\pi}{H}(k-j)\right)}+ O(\epsilon^\frac{1}{H}) \\
\Rightarrow \Delta = \arccos{\left(\cos{\left(\frac{2\pi}{H}(k-j)\right)}+ O(\epsilon^\frac{1}{H})\right)} \\
\Rightarrow \Delta = \frac{2\pi}{H}(k-j)+ O(\epsilon^\frac{1}{H}) \\
\end{array}$

So, the angle between any two branches is at least $\frac{2\pi}{H}$.  Indeed, the $H$ directions that the different zeros take as they move away from $\lambda_0$ are initially evenly spaced.  The only effect of $\epsilon$ taking different paths from zero is to rotate every $(\lambda^{(k)}(\epsilon) - \lambda_0)$.  If there is a path the maintains the unitarity of ${\bf U}(\epsilon)$, then for values of $\epsilon$ along that path $|\lambda^{(k)}(\epsilon)|=1$.  For small values of $\epsilon$ the unit circle is essentially a line.  Specifically, if $\lambda^{(k)}(\epsilon)$ are restricted to the unit circle and $\lambda^{(k)}(\epsilon)-\lambda_0$ are evenly spaced, then either the angle between $\lambda^{(1)}(\epsilon)-\lambda_0$ and $\lambda^{(2)}(\epsilon)-\lambda_0$ must be $\pi$, or $\lambda(\epsilon)$ is single-valued ($H=1$).  But that rules out values of $H$ above 2.

$\square$

\vspace{5mm}

{\bf Definition} Two eigenvalues are said to be "paired" when a small loop around 0 in the $\epsilon$-plane causes them to switch ($H=2$).  As seen above, paired eigenvalues vary on the order of $O(\sqrt{\epsilon})$, and non-paired eigenvalues vary by $O(\epsilon)$.  In addition, the associated eigenvectors and eigenprojections, ${\bf P}$, are also said to be paired.

\vspace{5mm}

\begin{thm}
$||{\bf P}^{(k)}(\epsilon) - {\bf P}^{(k)}(0)||=O(\sqrt{\epsilon})$ and $|V^{(k)}(\epsilon)\rangle=|V^{(k)}(0)\rangle+O(\sqrt{\epsilon})$.
\end{thm}

{\it Proof} The proof of this is included in the appendix.

\vspace{5mm}

In that proof it was shown that ${\bf P}^{(k)}(\epsilon)$ can be expressed as a power series in $\sqrt{\epsilon}$.  This means that since we can define $|V^{(k)}\rangle$ using these projections, not only do we find that the eigenvectors can likewise be written as power series in $\sqrt{\epsilon}$, but we automatically gain normalization: $|\langle V^{(k)} |V^{(k)}\rangle|=1, \forall \epsilon$.  What follows barely deserves to be a theorem, but it needs to be emphasized.

\begin{thm}
If the vector $|W\rangle$ is some combination of eigenvectors of ${\bf U}$, $|W\rangle = \sum_{j}a_j|V^{(j)}\rangle$ where the $a_j$ are independent of $\epsilon$, then $|W\rangle = |W_0\rangle + O(\sqrt{\epsilon})$, where $|W_0\rangle \equiv \lim_{\epsilon\to0} |W\rangle$.

\end{thm}

{\it Proof} Since all eigenprojections can be written as power series in $\sqrt{\epsilon}$, eigenvectors with distinct eigenvalues can as well and it follows that a linear combination of such eigenvectors can also be written as a power series in $\sqrt{\epsilon}$.  $|W\rangle = \sum_{j=0}^\infty (\sqrt{\epsilon})^j|W_j\rangle = |W_0\rangle + O(\sqrt{\epsilon})$.

$\square$

\vspace{5mm}

An extremely useful consequence of these theorems is that eigenvectors of ${\bf U}_0$, which are easy to find, can be used as approximations of the exact eigenvectors of ${\bf U}$, which are difficult to find, are variable, and needlessly complex.

\vspace{5mm}

\begin{thm}
Define ${\bf U}|V^{(j)}(\epsilon)\rangle = \lambda^{(j)}(\epsilon)|V^{(j)}(\epsilon)\rangle$ for all $j$, $\mathcal{S} = span\{|V^{(1)}(0)\rangle, \cdots,|V^{(k)}(0)\rangle\}$, and ${\bf P}_{\mathcal{S}}$ as the projection operator onto $\mathcal{S}$.

If $|u\rangle \in \mathcal{S}$, then $\forall m$

$i) \quad{\bf P}_{\mathcal{S}^\perp}{\bf U}_0^m|u\rangle = 0$

$ii) \quad{\bf P}_{\mathcal{S}^\perp}{\bf U}^m|u\rangle = O(\sqrt{\epsilon})$

That is, if $|u\rangle\in\mathcal{S}$, then ${\bf U}_0^m|u\rangle$ is also in $\mathcal{S}$, and ${\bf U}^m|u\rangle$ is almost entirely in $\mathcal{S}$.
\end{thm}

{\it Proof} The proof of this theorem is trivial, but takes up a lot of space.  It can be found in the appendix.

\vspace{5mm}

The essential idea behind this theorem is that if an initial state starts in an eigenstate, or collection of eigenstates, of ${\bf U}_0$, then it will approximately stay there under arbitrarily many applications of ${\bf U}$.  This means that, when setting up a search, only one eigenspace/eigenvalue of ${\bf U}_0$ needs to be considered.  This keeps the situation much simpler.

\subsection{Necessary and Sufficient Conditions for Pairing}

In this section it will be shown that the eigenvectors of ${\bf U}_0$ can be disjointly divided up into "constant eigenvectors" and "active eigenvectors".

\begin{thm}[the three-case theorem]
If $\lambda_0$ is a root of $C_0(z)$ with multiplicity $s$, then only one of the following cases applies to the "$\lambda_0$ family" of roots of $C(z,\epsilon)$, $\{\lambda^{(1)}(\epsilon), \cdots, \lambda^{(s)}(\epsilon)\}$, where $\lambda^{(k)}(0)=\lambda_0$, $\forall k$.

i) $\lambda^{(k)}(\epsilon)=\lambda_0$, $\forall k$.  That is, all of the roots are constant.

ii) $\lambda^{(k)}(\epsilon)=\lambda_0$, for all but one value of $k$.  This root takes the form $\lambda^{(k)}(\epsilon)=\lambda_0e^{ib\epsilon}+O(\epsilon^2)$.

iii) $\lambda^{(k)}(\epsilon)=\lambda_0$, for all but two values of $k$.  These two are paired and take the form $\lambda_\pm = \lambda_0e^{\pm ic\sqrt{\epsilon}}+O(\epsilon)$.

\end{thm}

{\it Proof} From theorem 3.5, the characteristic equation can be written $C(z,\epsilon) = C_0(z) + \epsilon f(z)$.  Making the substitution $z=\lambda_0+\delta$ this becomes $0 = C(\lambda_0+\delta,\epsilon) = \sum_{j=s} a_j\delta^j + \epsilon \sum_{j=t}b_j\delta^j$, where $t=deg(\lambda_0,f)$.  If $s\ge t$, then

$\begin{array}{ll}
0 = \sum_{j=s} a_j\delta^j + \epsilon \sum_{j=t}b_j\delta^j \\
\Rightarrow \epsilon \sum_{j=t}b_j\delta^j = -\sum_{j=s} a_j\delta^j \\
\Rightarrow \epsilon \left(b_t\delta^t + O(\delta^{t+1})\right) = -a_s\delta^s + O(\delta^{s+1}) \\
\Rightarrow \epsilon = -\frac{a_s}{b_t}\delta^{s-t} + O(\delta^{s-t+1}) \\
\Rightarrow \delta = O\left(\epsilon^{\frac{1}{s-t}}\right)
\end{array}$

However, by theorem 4.1 we know that unitarity implies $s-t\le2$.  Therefore, if $\lambda_0$ is a zero of $C_0(z)$ with multiplicity $s$, then it must also be a zero of $f(z)$ with a multiplicity of at least $s-2$.  This leads to three cases:

i) If $deg(\lambda_0,f)\ge s$, then $0 = C(z,\epsilon) = (z-\lambda_0)^s(g(z)+\epsilon h(z))$, where $g(\lambda_0)\ne 0$.  Here the entire $\lambda_0$ family is identically equal to $\lambda_0$, and has no $\epsilon$ dependence.

ii) If $deg(\lambda_0,f) = s-1$, then $0 = C(z,\epsilon) = (z-\lambda_0)^{s-1}((z-\lambda_0)g(z)+\epsilon h(z))$, where $g(\lambda_0), h(\lambda_0)\ne 0$.  In this case $s-1$ of the members of the $\lambda_0$ family are constant, and the last root is a solution of $0=(z-\lambda_0)g(z)+\epsilon h(z) = (g(\lambda_0)\delta + O(\delta^2))+\epsilon (h(\lambda_0) + O(\delta))$.  But this implies that $\delta=O(\epsilon)$.

iii) If $deg(\lambda_0,f) = s-2$, then $0 = C(z,\epsilon) = (z-\lambda_0)^{s-2}((z-\lambda_0)^2g(z)+\epsilon h(z))$, where $g(\lambda_0), h(\lambda_0)\ne 0$.  Therefore, all but two members of the $\lambda_0$ family are constant, and the remaining two are of the form $\lambda_0+O(\sqrt{\epsilon})$.  But by theorem 3.4, and the definition of pairing, an eigenvalue of this form can only show up with another eigenvalue to which it is paired.  It follows that, since there are no other possible eigenvalues to pair with, the two non-constant members of the $\lambda_0$ family must be paired with each other.

$\square$

\vspace{5mm}

Since there is only one paired set of eigenvectors for a given $\lambda_0$, we can label them $|V^{+}\rangle$ and $|V^{-}\rangle$, where ${\bf U} |V^{\pm}\rangle = \lambda_0 e^{\pm ic\sqrt{\epsilon}+O(\epsilon)}|V^{\pm}\rangle = \lambda_0 e^{\pm ic\sqrt{\epsilon}}|V^{\pm}\rangle + O(\epsilon)$.

In the next few theorems the following properties will be important.

\begin{thm} Assume that $|V\rangle \equiv |V_0\rangle + \sqrt{\epsilon}|V_1\rangle+O(\epsilon)$, $|W\rangle \equiv |W_0\rangle + \sqrt{\epsilon}|W_1\rangle+O(\epsilon)$, $\langle V|V\rangle = \langle W|W\rangle = 1$ $\forall \epsilon$, and $\langle V|W\rangle=0$ $\forall \epsilon$, then

i) $\langle V_0|V_0\rangle = 1$

ii) $\langle V_0|V_1\rangle+\langle V_1|V_0\rangle=0$

iii) $\langle V_0|W_1\rangle + \langle V_1|W_0\rangle = 0$

\end{thm}

{\it Proof} These three results are immediate, upon inspection.

$\square$

\vspace{5mm}

\begin{thm}
Paired eigenvectors "straddle the hub".  That is, if $|V\rangle$ is a paired eigenvector, then $|V_0\rangle$ has both a Left and Right side component.
\end{thm}

{\it Proof} Paired eigenvectors have eigenvalues that always take the form $\lambda=\lambda_0e^{ic\sqrt{\epsilon}}+O(\epsilon)$.

$\begin{array}{ll}
\langle V|{\bf U}|V\rangle = \lambda\langle V|V\rangle = \lambda \\
\Rightarrow \left(\langle V_0| + \sqrt{\epsilon}\langle V_1|\right)\left({\bf U}_0+\sqrt{\epsilon}{\bf U}_1\right)\left(|V_0\rangle + \sqrt{\epsilon}|V_1\rangle\right) + O(\epsilon) = \lambda_0 + ic\lambda_0\sqrt{\epsilon} + O(\epsilon) \\
\Rightarrow \langle V_0|{\bf U}_0|V_0\rangle + \sqrt{\epsilon}\left(\langle V_1|{\bf U}_0|V_0\rangle + \langle V_0|{\bf U}_0|V_1\rangle + \langle V_0|{\bf U}_1|V_0\rangle\right) = \lambda_0 + ic\lambda_0\sqrt{\epsilon} \\
\Rightarrow \lambda_0 + \sqrt{\epsilon}\left(\lambda_0\langle V_1|V_0\rangle + \lambda_0\langle V_0|V_1\rangle + \langle V_0|{\bf U}_1|V_0\rangle\right) = \lambda_0 + ic\lambda_0\sqrt{\epsilon} \\
\Rightarrow \sqrt{\epsilon}\left(\lambda_0\left[\langle V_1|V_0\rangle + \langle V_0|V_1\rangle\right] + \langle V_0|{\bf U}_1|V_0\rangle\right) = ic\lambda_0\sqrt{\epsilon} \\
\Rightarrow\lambda_0\left[0\right] + \langle V_0|{\bf U}_1|V_0\rangle = ic\lambda_0 \\
\Rightarrow \langle V_0|{\bf U}_1|V_0\rangle = ic\lambda_0
\end{array}$

Since ${\bf U}_1$, the $O\left(\sqrt{\epsilon}\right)$ terms of ${\bf U}$, is only involved in the transimission between the Left and Right sides, in order for $\langle V_0|{\bf U}_1|V_0\rangle$ to be non-zero, $|V_0\rangle$ must show up on both sides.  That is; if $|V_0\rangle$ was entirely supported on the Right side, then ${\bf U}_1|V_0\rangle$ would be entirely supported on the Left side.

$\square$

\vspace{5mm}

\begin{thm} Assume that the Right side $\lambda_0$-eigenspace of ${\bf U}_0$ is $D$ dimensional.

1) If the $\lambda_0$-eigenspace of ${\bf U}_0$ is bound in $G$, then the $\lambda_0$-eigenspace of ${\bf U}$ is $D$ dimensional and all of the associated eigenvectors are constant.  This is case i) of the Three Case Theorem.

2) If the $\lambda_0$-eigenspace of ${\bf U}_0$ is in contact with the hub vertex, then the Right sided $\lambda_0$-eigenspace of ${\bf U}$ is $D$-1 dimensional and the $D$-1 associated eigenvectors are constant and bound in $G$.  This leaves one eigenvector which is non-constant in $\epsilon$, and is in contact with the hub vertex.  This is either case ii) or case iii) of the Three Case Theorem.

\end{thm}

{\it Proof} The first result is trivial.  If an eigenvector is bound in $G$, then varying $\epsilon$ (which only affects reflection and transmission across the hub vertex) can't have any impact on it.  So, for eigenvectors bound in $G$, ${\bf U}|V\rangle = {\bf U}_0|V\rangle = \lambda_0|V\rangle$.  The contra-positive is likewise clear; if an eigenvector is non-constant, then it must be in contact with the hub vertex.

The second result is far more hard won, and is included in the appendix.

\vspace{5mm}

In the proof of the above theorem (in the appendix), the Left side was given a particular form; it consisted of the states $|in\rangle$ and $|out\rangle$, with ${\bf U}|out\rangle = e^{i\phi}|in\rangle$.  It may seem strange that a particular form for the Left side can be used to say such general things about the Right side.  But keep in mind that what's been shown is that a certain number of the Right side $\lambda_0$ eigenvectors are bound in $G$.  Once we know that an eigenvector on the Right is definitely not in contact with the hub vertex, then it doesn't matter what the Left side is doing.

This last theorem can be stated far more succinctly as,

\begin{thm}Excluding a special case, a Right side eigenvector is constant and has a constant eigenvalue if and only if it is not in contact with the hub vertex.
\end{thm}

That special case is $\lambda_0^2+e^{i\phi}=0$.  When this happens the eigenvectors on each side are "balanced", and there is no net flow of probability across the hub.  In the Grover graph example, this happens when $e^{i\phi} = -1$, which means that all of the edges become identical to the marked edge.  As a result, the quantum graph can be collapsed to just two states: $|A\rangle = \frac{1}{\sqrt{N}}\sum_{j=1}^N |0,j\rangle$ and $|B\rangle = \frac{1}{\sqrt{N}}\sum_{j=1}^N |j,0\rangle$.  Without two sides, there is no net flow.  In general, even though there may not be a further collapsing of the graph, when $\lambda_0^2+e^{i\phi}=0$ there is no net flow of probability across the hub.

\vspace{5mm}

{\bf Definition} This unique non-constant member of the $\lambda_0$ family of eigenvectors, $|w\rangle$, depends on the Left side.  But while $|w\rangle$ may depend on $\phi$ and $\epsilon$, $|w_0\rangle\equiv\lim_{\epsilon\to0}|w\rangle$ does not.  It is unique so long as $\lambda_0^2\ne e^{i\phi}$ (so long as there is no pairing).  $|w_0\rangle$ is precisely the $\lambda_0$ eigenvector of ${\bf U}_0$ lost when moving from the $\epsilon=0$ case to the $\epsilon\ne0$ case.  This $|w_0\rangle$ is a uniquely defined Right side $\lambda_0$ eigenvector of ${\bf U}_0$, and it is called the "Right side active $\lambda_0$ eigenvector".  An essentially identical proof demonstrates the existence of "Left side active $\lambda_0$ eigenvectors".

\vspace{5mm}

{\bf Definition} All of the other $\lambda_0$ eigenvectors are bound in $G$, and so are called "bound $\lambda_0$ eigenvectors".  Bound eigenvectors are independent of $\epsilon$, or any structure on the far side of the hub.  If the $\lambda_0$-eigenspace of ${\bf U}_0$ is $D$-dimensional overall, then the space of bound $\lambda_0$ eigenvectors of ${\bf U}$ will be $D$, $D$-1, or $D$-2 dimensional depending on whether the Left and Right sides have active eigenvectors.

\vspace{5mm}

{\bf Definition} The "active subspace" is the span of all of the active eigenvectors of ${\bf U}_0$, for all eigenvalues.  Since it is the compliment of the span of all of the bound eigenvectors, and each non-constant eigenvector is perpendicular to every bound eigenvector, every non-constant eigenvector is contained in the the active subspace.

\vspace{5mm}

\begin{thm}[The fundamental pairing theorem]
The $\lambda_0$-eigenspace is in contact with both the Left and Right sides of the hub vertex if and only if there exists paired vectors $|V^\pm\rangle$ with eigenvalues of the form $\lambda_0e^{\pm ic\sqrt{\epsilon}}+O(\epsilon)$.
\end{thm}

{\it Proof} The proof of this is somewhat involved and so is included in the appendix.  Among the useful results of the Fundamental Pairing Theorem is the fact that ${\bf U}$ can be expressed in the following form:

\begin{equation}
{\bf U}\left(\begin{array}{cc} |\ell_0\rangle \\ |\mathfrak{r}_0\rangle \end{array}\right) = \lambda_0\left(\begin{array}{cc} cos(c\sqrt{\epsilon}) & i\,sin(c\sqrt{\epsilon}) \\ i\,sin(c\sqrt{\epsilon}) & cos(c\sqrt{\epsilon}) \end{array}\right)\left(\begin{array}{cc} |\ell_0\rangle \\ |\mathfrak{r}_0\rangle \end{array}\right)+O(\epsilon)
\end{equation}

And, with correctly chosen complex phases, the paired eigenvectors take the form:

\begin{equation}
|V^\pm\rangle = \frac{|\ell_0\rangle\pm|\mathfrak{r}_0\rangle}{\sqrt{2}} + O\left(\sqrt{\epsilon}\right)
\end{equation}

So, on each side of the hub vertex we have a unique active eigenvector and paired eigenvectors are just combinations of the active eigenvectors from both sides, up to an error of $O(\sqrt{\epsilon})$.  This makes the situation pretty simple, and we can ignore the bound states entirely.

\subsection{The Efficiency of Searches Using Paired Eigenvectors}

In order to execute an search in $O\left(\sqrt{N}\right)$ it is necessary to used paired eigenspaces, since this is where changes of $O(\sqrt{\epsilon})$ can take place.  The question addressed in this section is, given an initial state on one side of the hub vertex, how much of it can be transferred to the other?  That is, what is the lower bound on the probability of a successful measurement under ideal conditions?  Happily, the answer is $1-|O(\sqrt{\epsilon})|$.

In the course of proving the fundamental pairing theorem an important additional fact was also proven, and is worth noting separately.

\begin{thm}
Paired eigenvectors are always evenly divided across the hub vertex.  That is, if ${\bf P}$ is a projection onto either the Left or Right side and $|V^\pm\rangle$ is a paired eigenvector, then $|\langle V_0^\pm|{\bf P}|V_0^\pm\rangle| = \frac{1}{2}$.
\end{thm}

{\it Proof} As established in the fundamental pairing theorem, $|V^\pm_0\rangle = \frac{1}{\sqrt{2}}\left(|\ell_0\rangle\pm|\mathfrak{r}_0\rangle\right)$.  For the Left side projector,

$\begin{array}{ll}
|\langle V_0^\pm|{\bf P}_L|V_0^\pm\rangle| \\
= \frac{1}{2}|\left(\langle\ell_0|\pm\langle \mathfrak{r}_0|\right){\bf P}_L\left(|\ell_0\rangle\pm|\mathfrak{r}_0\rangle\right)| \\
= \frac{1}{2}|\left(\langle\ell_0|\pm\langle \mathfrak{r}_0|\right)|\ell_0\rangle| \\
= \frac{1}{2}|\langle\ell_0|\ell_0\rangle + 0| \\
=\frac{1}{2}
\end{array}$

The same holds for the Right side projector.

$\square$

\vspace{5mm}

\begin{thm}
If the paired eigenvalues are of the form $\lambda^\pm = \lambda_0e^{\pm ic\sqrt{\epsilon}}+O(\epsilon)$, then $c = \left|\langle\ell_0|{\bf U}_1|\mathfrak{r}_0\rangle\right| = 2 \left|\langle out| \ell_0\rangle |\,| \langle 1,0|\mathfrak{r}_0\rangle\right|$.
\end{thm}

{\it Proof} In the last subsection it was established that

${\bf U}\left(\begin{array}{cc} |\ell_0\rangle \\ |\mathfrak{r}_0\rangle \end{array}\right) = \lambda_0\left(\begin{array}{cc} cos(c\sqrt{\epsilon}) & i\,sin(c\sqrt{\epsilon}) \\ i\,sin(c\sqrt{\epsilon}) & cos(c\sqrt{\epsilon}) \end{array}\right)\left(\begin{array}{cc} |\ell_0\rangle \\ |\mathfrak{r}_0\rangle \end{array}\right)+O(\epsilon)$.

\vspace{5mm}

From which it follows that,

$\begin{array}{ll}
\langle\ell_0|{\bf U}|\mathfrak{r}_0\rangle \\[2mm]
= \langle\ell_0|{\bf U}_0|\mathfrak{r}_0\rangle + \langle\ell_0|{\bf U}_1|\mathfrak{r}_0\rangle\sqrt{\epsilon} + O(\epsilon)\\[2mm]
= \lambda_0\langle\ell_0|\mathfrak{r}_0\rangle + \langle\ell_0|{\bf U}_1|\mathfrak{r}_0\rangle\sqrt{\epsilon} + O(\epsilon)\\[2mm]
= 0 + \langle\ell_0|{\bf U}_1|\mathfrak{r}_0\rangle\sqrt{\epsilon} + O(\epsilon)\\[2mm]
= \langle\ell_0|{\bf U}_1|\mathfrak{r}_0\rangle\sqrt{\epsilon} + O(\epsilon) \\[2mm]
\Rightarrow i\,\lambda_0 sin(c\sqrt{\epsilon})+O(\epsilon) = \langle\ell_0|{\bf U}_1|\mathfrak{r}_0\rangle\sqrt{\epsilon} + O(\epsilon) \\[2mm]
\Rightarrow i\,\lambda_0 c\sqrt{\epsilon}+O(\epsilon) = \langle\ell_0|{\bf U}_1|\mathfrak{r}_0\rangle\sqrt{\epsilon} + O(\epsilon) \\ [2mm]
\Rightarrow c = -i\,\lambda_0^* \langle\ell_0|{\bf U}_1|\mathfrak{r}_0\rangle
\end{array}$

Because $1 = \left|\lambda^\pm\right|$ we can infer that $c$ is real, and we can therefore use the somewhat simpler expression $c = \left|\langle\ell_0|{\bf U}_1|\mathfrak{r}_0\rangle\right|$.  Finally,

$\begin{array}{ll}
c = \left|\langle\ell_0|{\bf U}_1|\mathfrak{r}_0\rangle\right| \\[2mm]
= \left|\langle\ell_0|out\rangle \langle out|{\bf U}_1 |1,0\rangle \langle 1,0|\mathfrak{r}_0\rangle\right| \\[2mm]
= 2 \left|\langle\ell_0|out\rangle |\,| \langle 1,0|\mathfrak{r}_0\rangle\right| \\[2mm]
= 2 \left|\langle out| \ell_0\rangle |\,| \langle 1,0|\mathfrak{r}_0\rangle\right| \\[2mm]
\end{array}$

$\square$

\vspace{5mm}

\begin{thm}
${\bf U}^m$, where $m=\left\lfloor\frac{\pi\sqrt{N}}{2|\langle\mathfrak{r}_0|{\bf U}_1|\ell_0\rangle|}\right\rfloor$, exchanges the Left and Right $\lambda_0$ active eigenvectors, $|\ell_0\rangle$ and $|\mathfrak{r}_0\rangle$, almost completely (to within $O(\sqrt{\epsilon})$).
\end{thm}

{\it Proof} To show that ${\bf U}$ interchanges $|\ell_0\rangle$ and $|\mathfrak{r}_0\rangle$ it suffices to show that $|\langle \mathfrak{r}_0|{\bf U}^m|\ell_0\rangle|=1+O(\sqrt{\epsilon})$.  Note that $m = \left\lfloor\frac{\pi}{2c\sqrt{\epsilon}}\right\rfloor = \frac{\pi}{2c\sqrt{\epsilon}}+O(1)$, since the floor function rounds down by at most 1.

$\begin{array}{ll}
|\langle \mathfrak{r}_0|{\bf U}^m|\ell_0\rangle| \\[2mm]
= \frac{1}{2}\left|\left(\langle V^+_0|-\langle V^-_0|\right){\bf U}^m\left(|V^+_0\rangle + |V^-_0\rangle\right)\right| \\[1mm]

= \frac{1}{2}\left|\left(\langle V^+|-\langle V^-|\right){\bf U}^m\left(|V^+\rangle + |V^-\rangle\right)\right| + O(\sqrt{\epsilon})\\[1mm]
= \frac{1}{2}\left|\left(\langle V^+|-\langle V^-|\right)\left((\lambda_0e^{ic\sqrt{\epsilon}+O(\epsilon)})^m|V^+\rangle + (\lambda_0e^{-ic\sqrt{\epsilon}+O(\epsilon)})^m|V^-\rangle\right)\right| + O(\sqrt{\epsilon})\\[2mm]
= \frac{1}{2}|\lambda_0^m| \left |\left(\langle V^+|-\langle V^-|\right)\left(e^{ic\sqrt{\epsilon}+O(\epsilon)})^m|V^+\rangle + (e^{-ic\sqrt{\epsilon}+O(\epsilon)})^m|V^-\rangle\right)\right| + O(\sqrt{\epsilon})\\[2mm]
= \frac{1}{2}\left|\left(e^{ic\sqrt{\epsilon}+O(\epsilon)}\right)^{\frac{\pi}{2c\sqrt{\epsilon}}+O(1)} - \left(e^{-ic\sqrt{\epsilon}+O(\epsilon)}\right)^{\frac{\pi}{2c\sqrt{\epsilon}}+O(1)}\right| + O(\sqrt{\epsilon})\\[3mm]
= \frac{1}{2}\left|e^{i\frac{\pi}{2}+O(\sqrt{\epsilon})} - e^{-i\frac{\pi}{2}+O(\sqrt{\epsilon})}\right| + O(\sqrt{\epsilon})\\[2mm]
= \frac{1}{2}\left|e^{i\frac{\pi}{2}} - e^{-i\frac{\pi}{2}} + O(\sqrt{\epsilon})\right| + O(\sqrt{\epsilon})\\[2mm]
= \left|sin\left(\frac{\pi}{2}\right)\right| + O(\sqrt{\epsilon})\\[2mm]
=1 + O(\sqrt{\epsilon})
\end{array}$

$\square$

\vspace{5mm}

So, when $m = \left\lfloor\frac{\pi\sqrt{N}}{2|\langle\mathfrak{r}_0|{\bf U}_1|\ell_0\rangle|}\right\rfloor$, $|\langle\mathfrak{r}_0|{\bf U}^m|\ell_0\rangle|=1+O(\sqrt{\epsilon})$.  This means that searches with $|\ell_0\rangle$ as the initial state are almost guaranteed to succeed, and all of the work that goes into setting up the search: choosing an initial state, knowing what to look for afterward, and figuring out how many times to iterate ${\bf U}$, has now been pushed back to finding the active eigenvectors $|\ell_0\rangle$ and $|\mathfrak{r}_0\rangle$.

Even this is relatively straightforward.  $|\mathfrak{r}_0\rangle$ can be found be looking for the one eigenvector of ${\bf P}_R{\bf U}_0$ with eigenvalue $\lambda_0$ that is in contact with the hub vertex.  This can be done in a few ways.  For example, by removing each bound $\lambda_0$ eigenvector, or by comparing the $\lambda_0$-eigenspaces of ${\bf U}_0$ and a quantum graph with the hub vertex's behavior replaced with something simple, like ${\bf U}|1,0\rangle = 0$.  Trivially, if the Right side $\lambda_0$-eigenspace is one dimensional and in contact with the hub, then the $\lambda_0$ eigenvector {\it is} the active eigenvector.

The value of $N$ or $\epsilon$, and the entire Left side, are unimportant to determining the active eigenvectors.

\subsection{Best Choice Selection and the Best Time Bound}

The bigger $c$ is the faster a search will proceed.  So, when the eigenvalue can be selected, the best choice is the eigenvalue with the largest value of $c$.

The specificity of the star graph described in section 2 allows us to be more precise about the value of $c$.  Regardless of the value of $\phi$, $|\langle out|\ell_0\rangle | = \frac{1}{\sqrt{2}}$.  Therefore, $c = 2|\langle out|\ell_0\rangle |\,| \langle 1,0|\mathfrak{r}_0\rangle| = \sqrt{2}|\langle 1,0|\mathfrak{r}_0\rangle|$.

If $G$ is unknown, then $|\mathfrak{r}_0\rangle$ and $c$ cannot be known either.  The most we can say is that, for a star graph, $0\le c\le\sqrt{2}$.  Since $m=\left\lfloor\frac{\pi}{2c}\sqrt{N}\right\rfloor$, not being able to bound $c$ from below means not being able to bound $m$ from above.

Define $d$ to be the number of Right side active eigenvectors, denoted $|\mathfrak{r}_0^{(j)}\rangle$ for $j=1, \cdots, d$.  Each of these eigenvectors has a value of $c$, $c^{(j)}$.  While there is no lower bound for all of these, we can construct a lower bound for at least some of them.

\begin{thm}
For Left side eigenvectors of the form $|\ell_0\rangle = \frac{1}{\sqrt{2}}\left(|out\rangle+ e^{i\frac{\phi}{2}}|in\rangle\right)$, if $d$ is the number of Right side active eigenvectors, then $\sum_{j=1}^d \left(c^{(j)}\right)^2 = 2$ and there exists at least one $j$ such that $c^{(j)}\ge\sqrt{\frac{2}{d}}$.
\end{thm}

{\it Proof} $|1,0\rangle$ is adjacent to the hub vertex and therefore $|1,0\rangle$ is entirely contained in the active subspace.  It follows that $1 = \sum_{j=1}^d \left|\langle 1,0|\mathfrak{r}_0^{(j)}\rangle\right|^2 = \frac{1}{2}\sum_{j=1}^d \left(c^{(j)}\right)^2$.  Since $E\left[\left(c^{(j)}\right)^2\right] = \frac{2}{d}$, we know that there is at least one $j$ such that $\left(c^{(j)}\right)^2 \ge\frac{2}{d}$, which implies that $c^{(j)} \ge \sqrt{\frac{2}{d}}$.

$\square$

\vspace{5mm}

{\bf Definition} There is at least one eigenvalue such that $m\le\frac{\pi}{2\sqrt{2}}\sqrt{dN}$.  This is the "best time bound".  If the eigenvalue with the highest value of $c$ is used, then the search will require this many time steps {\it at most}.

\begin{equation}
best\,\,time\,\, bound\,=\frac{\pi\sqrt{dN}}{2\sqrt{2}}\le\frac{\pi\sqrt{(|G|+2)N}}{2\sqrt{2}}
\end{equation}

Where $|G|$ is the number of states in $G$.  Examples can be constructed where there is an arbitrarily small value of $c^{(j)} = \sqrt{2}\left|\langle1,0|\mathfrak{r}_0\rangle\right|$ for some values of $j$, but not for all.  So long as there is a bound on the size of $G$, there is at least one eigenvalue where $c^{(j)}$ can be bound from below.  Equivalently, there is always at least one eigenvalue such that when the initial state is prepared in the corresponding Left eigenstate, the number of time steps required for a search is bound from above.

\pagebreak

\section{Tolerances}

For any of a number of reasons the Left and Right eigenvalues may not exactly match.  However, the pairing doesn't break immediately, but instead fades away quickly as the disagreement between the eigenvalues increases.  The eigenvalues don't have to be exactly equal in order for a $O(\sqrt{N})$ search, but they do have to be {\it nearly} equal by an amount dictated by $\epsilon$.

This section will make heavy use of the math introduced in section 3.

\subsection{Grover Graph example revisited}

This is exactly the same situation previously seen in section 2.1, but now we are allowing $e^{i\phi}$ to take values other than 1.  The advantage of the Grover graph is that it is simple enough that it is exactly solvable, even with this generalization.

\vspace{5mm}

The characteristic polynomial is $C(z) = z^4 - (1-2\epsilon)(1+e^{i\phi})z^2 + e^{i\phi}$.

The four eigenvalues, expressed as one global function, is

$\lambda =\pm\sqrt{\frac{1}{2}(1-2\epsilon)(e^{i\phi}+1) \pm \sqrt{\left(\frac{e^{i\phi}-1}{2}\right)^2-(e^{i\phi}+1)^2(\epsilon-\epsilon^2)}}$.

Where the $\pm$'s are independent.  In the original example, $\phi = 0$, so $\left(\frac{e^{i\phi}-1}{2}\right)^2=0$ and looping $\epsilon$ around zero permutes pairs of eigenvalues.  When $\phi=0$, then to lowest order $\lambda = \pm1\pm i\sqrt{\epsilon}+O\left(\epsilon\right) = \pm e^{\pm i\sqrt{\epsilon}}+O\left(\epsilon\right)$.

When $\phi\ne0$ the characteristic polynomial no longer has a double zero when $\epsilon=0$.  We can still expand around $\epsilon=0$, but we find that the  power series is in $\epsilon$, not $\sqrt{\epsilon}$.

\begin{equation}
\lambda = \left\{\begin{array}{ll}
\pm e^{i\frac{\phi}{2}} \left( 1 + i \, cot\left(\frac{\phi}{2}\right) \epsilon \right) + O\left(\epsilon^2\right) \\
\pm \left( 1 - i \, cot\left(\frac{\phi}{2}\right)\epsilon \right) + O\left(\epsilon^2\right)
\end{array}\right.
= \left\{\begin{array}{ll}
\pm e^{i\frac{\phi}{2}} e^{i\,cot\left(\frac{\phi}{2}\right)\epsilon} + O\left(\epsilon^2\right) \\
\pm e^{-i\,cot\left(\frac{\phi}{2}\right)\epsilon} + O\left(\epsilon^2\right)
\end{array}\right.
\end{equation}

Clearly the pairing has been lost.  However, the double zero required for pairing hasn't vanished, merely moved.  This new location is $\epsilon_0$.  To find it we set the discriminant equal to zero and solve for $\epsilon$.  In this particular case the discriminant is exactly what's found under the inner radical, $\mathcal{D} = (e^{i\phi}-1)^2 - 4(e^{i\phi}+1)^2\epsilon + 4(e^{i\phi}+1)^2\epsilon^2$.  We find immediately that $\epsilon_0 = \frac{1}{2}\pm\frac{1}{e^{i\phi}+1}e^{i\frac{\phi}{2}} = \frac{1}{2} \pm \frac{1}{2cos\left(\frac{\phi}{2}\right)}$.  We're interested in the zero that corresponds to $\epsilon_0|_{\phi=0} = 0$, and find that $\epsilon_0 = \frac{1}{2}-\frac{1}{2cos\left(\frac{\phi}{2}\right)} =-\frac{1}{16}\phi^2+O(\phi^4)$.  Expanding $\lambda$ as a power series in $\sqrt{\epsilon - \epsilon_0} = \sqrt{\epsilon - \frac{1}{2}+\frac{1}{2cos\left(\frac{\phi}{2}\right)}}$, which we find can be done, the eigenvalues can now be written,

\begin{equation}
\lambda = \pm e^{i\frac{\phi}{4}} \pm ie^{i\frac{\phi}{4}}\sqrt{cos\left(\frac{\phi}{2}\right)}\sqrt{\epsilon-\epsilon_0}+O(\epsilon-\epsilon_0) = \pm e^{i\frac{\phi}{4}} e^{\pm i\sqrt{\epsilon-\epsilon_0}}+O(\phi^2\sqrt{\epsilon-\epsilon_0}, \epsilon-\epsilon_0)
\end{equation}

Notice that if $|\epsilon| \gg |\epsilon_0|$, then the quadratic behavior is recovered.  This will be discussed in substantially more detail below.

\vspace{5mm}

There's something subtle that happens with the eigenvectors here as well.  When $\phi=0$, all four of the eigenvectors of ${\bf U}$ are evenly divided between the Left and Right sides, to within $O\left(\sqrt{\epsilon}\right)$.  That is, ${\bf U}$ has four eigenvectors and when $\phi=0$ the limit of all of them, as $\epsilon\to0$, are evenly divided between the two sides since all four are paired.

When $\phi\ne0$, ${\bf U}_0$ has four distinct eigenvalues: $\{1, e^{i\frac{\phi}{2}}, -1, -e^{i\frac{\phi}{2}} \}$.  Therefore, the four eigenvectors of ${\bf U}$, in the limit as $\epsilon\to0$, converge to two couples of vectors.  Two on the Left for $\lambda_0 = e^{i\frac{\phi}{2}}, -e^{i\frac{\phi}{2}}$, and two on the Right for $\lambda_0 = 1, -1$.

Clearly there is a dramatic difference between these limits.  This is a "twisting" of the eigenvectors of ${\bf U}$ that occurs around a double zero.  A careful reading of theorem 4.2 reveals that it says that eigenvectors are continuous when they have distinct eigenvalues.  When an eigenvalue is degenerate this "continuity of eigenvectors" becomes a "continuity of eigenspaces".  There are two degenerate eigenspaces when $\epsilon = \frac{1}{2}-\frac{1}{2cos\left(\frac{\phi}{2}\right)}$ (when $\phi=0$ these are the $\lambda=\pm1$ eigenspaces of the original graph).  Picking either, the degenerate space turns into two paired eigenvectors as $\epsilon$ moves away from this point, and it turns into two non-paired and single sided eigenvectors as $\phi$ moves away from this point.

From either direction the two eigenvectors of ${\bf U}$ converge to the same degenerate eigen{\it space}.  The fact that the vectors they converge to are different is unimportant.

\subsection{Altering the Graph}

We know that eigenvalues can pair, and that their values vary on the order of $\sqrt{\epsilon}=\frac{1}{\sqrt{N}}$, when ${\bf U}_0$ has an identical eigenvalue on both the Left and Right sides.  However, we can hope that there should be some leeway, and that there will be pairing for {\it nearly} equal eigenvalues.  The concern here is that changing the graph will turn a double root of $C(z,0)$ into a closely-spaced pair of \underline{distinct} roots of $C(z,0)$.  Since pairing and quadratic speed searches depend on the existence of double roots, these may be lost along with the double root.

Define a new parameter, $\xi$, which describes a change in the graph, such as a change in the reflection and transmission coefficients of some of the vertices.  Further, define $\xi$ such that $C(z,\epsilon,\xi)$ is a polynomial in all of its arguments, and so that $C(z,0,0)$ has a double root in $z$.  $\xi = e^{i\phi}-1\approx i\phi$, from section 5.1, is the motivating example.

\vspace{5mm}

\begin{thm}
If ${\bf U}$ has entries that are analytic functions of a given variable, $\xi$, then the eigenvectors of ${\bf U}$, and by extension the eigenvalues, vary by $O(\xi)$.
\end{thm}

{\it Proof} Using the same argument seen in the proof of theorem 4.2 (which is in the appendix) we can show that the resolvent, ${\bf R}(\zeta,\epsilon,\xi) = \left({\bf U} - \zeta {\bf I}\right)$, can be expressed as a power series in $\zeta$, $\sqrt{\epsilon}$, and $\xi$.

Since ${\bf R}(\zeta,\epsilon,\xi)$ is a power series in $\xi$, it follows that the eigenprojections, ${\bf P}^{(k)}(\epsilon,\xi) = -\frac{1}{2\pi i}\oint_{\lambda^{(k)}} {\bf R}(\zeta,\epsilon,\xi)d\zeta$, and the eigenvectors, $|V^{(k)}\rangle$, are also expressible as power series in $\xi$.

Because each eigenvector can be written as $|V\rangle = |V_0\rangle + \xi |V_1\rangle + \xi^2 |V_2\rangle + \cdots$, and because by definition ${\bf U} = {\bf U}_0 + \xi{\bf U}_1 + \xi^2{\bf U}_2 + \cdots$ it follows that the eigenvalues cannot have fractional exponents in $\xi$.

$\begin{array}{ll}
{\bf U}|V\rangle = \lambda|V\rangle \\
\Rightarrow ({\bf U}_0+O(\xi))(|V_0\rangle + O(\xi)) = (\lambda_0+O(\xi^s))(|V_0\rangle + O(\xi)) \\
\Rightarrow {\bf U}_0|V_0\rangle + O(\xi) = \lambda_0|V_0\rangle + O(\xi^s) + O(\xi) \\
\Rightarrow O(\xi) = O(\xi^s) + O(\xi) \\
\Rightarrow s\ge1
\end{array}$

$\square$

\vspace{5mm}

Notice that this theorem does not rule out ordinary pairing of $O(\sqrt{\epsilon})$, since ${\bf U}$ has entries of the form $2\sqrt{\epsilon-\epsilon^2}=2\sqrt{\epsilon} + O\left(\epsilon^{\frac{3}{2}}\right)$.

$\xi$ has been defined so that $C(z,0,0)$ has a double root.  By applying the arguments of section 3 to isolated roots of $C(z,\epsilon,\xi)$ we find that those roots, $\lambda^{(k)}(\epsilon, \xi)$, are analytic with respect to both $\epsilon$ and $\xi$.

\begin{thm}
Changing $\xi$ causes the location of double-zeros to drift.  That is, a new value of $\epsilon$ that depends on $\xi$, labeled $\epsilon_0$, may be found such that $C(z,\epsilon_0(\xi),\xi)$ has a double root in $z$, and $\epsilon_0(\xi)$ is a continuous function of $\xi$.
\end{thm}

{\it Proof} Consider the discriminant of $C$, defined as $\mathcal{D} = \prod_{j>k}(\lambda^{(j)}-\lambda^{(k)})^2$, where $\{\lambda^{(k)}\}$ are the roots of $C$ in the variable $z$.  The well-known and relevant properties of $\mathcal{D}(\epsilon, \xi)$ are 1) $\mathcal{D}(\epsilon, \xi)=0$ if and only if $C(z,\epsilon,\xi)$ has a double root in $z$, and 2) if $C(z,\epsilon,\xi)$ is a polynomial in its arguments, then $\mathcal{D}(\epsilon, \xi)$ is also a polynomial in its arguments.

Because $C(z,0,0)$ has a double root, we know that $\mathcal{D}(0,0)=0$.  Using the same trick that was used to describe the behavior of $\lambda$ with respect to $\epsilon$ (theorem 3.2), we can describe the new location of the double root in $\epsilon$ space with respect to $\xi$ as $\epsilon_0(\xi) = \frac{1}{2\pi i}\oint t\frac{\mathcal{D}_t(t, \xi)}{\mathcal{D}(t, \xi)}dt = \sum_j b_j(\sqrt[s]{\xi})^j$ where $s$ is less than or equal to the multiplicity of the root in $\mathcal{D}(\epsilon,\xi)$ in the variable $\epsilon$.  The important thing to notice here is that $\epsilon_0(\xi)$ is continuous with respect to $\xi$.

$\square$

\vspace{5mm}

Note that since we're assuming that $C(z,0,0)$ has a double root, $\epsilon_0(0)=0$.

As shown in thm. 3.4, the only time that the Puiseux series of $\lambda^{(k)}(\epsilon)$ can have non-integer powers is when the expansion is taken about the location in the $\epsilon$-plane of a double root of $C(z,\epsilon)$.  With the introduction of $\xi$, as shown in the last theorem, the location of the double root is a function, $\epsilon_0(\xi)$.  In order for the Puiseux series of $\lambda^{(k)}(\epsilon)$ to have half-integer powers in $\epsilon$ it must be expanded about $\epsilon_0$ and so it takes the form $\lambda^{(k)}(\epsilon,\xi) = \sum_{n=0}^\infty a_n(\xi)\left(\sqrt{\epsilon-\epsilon_0(\xi)}\right)^n$.

There are no new issues with expanding around $\epsilon=\epsilon_0$ as opposed to $\epsilon=0$, and the above series can certainly be constructed.

\subsection{Nearly-Paired Eigenvalues}

Assume that the graph is altered in the manner described in the last subsection.  Define the Left and Right eigenvalues when $\epsilon=0$ as $\lambda_\ell$ and $\lambda_r$ respectively.  These are both analytic functions of $\xi$, and thus the phase difference between them can also be described as an analytic function of $\xi$.  Define this phase difference as $\delta$.

\begin{equation}
e^{i\delta} = \lambda_\ell \lambda_r^*
\end{equation}

So long as $\lambda_\ell \lambda_r^*$ is invertible as a function of $\xi$, we can express $\xi$ as a power series in $\delta$ and $\delta \propto \xi + O(\xi^2)$.  This is a fairly reasonable assumption.  In the Grover Graph example provided at the beginning of this section $\delta = \frac{\phi}{2}$.

\vspace{5mm}

So, we can reasonably assume that ${\bf U}$, the eigenvectors, and the eigenvalues are all analytic functions of $\delta = \lambda_\ell \lambda_r^*$.  In addition the point in the $\epsilon$-plane about which the eigenvalues permute, $\epsilon_0$, is a continuous function of $\delta$.

We can now write $\lambda$ as a power series in $\delta$ and $\sqrt{\epsilon-\epsilon_0}$. 

\begin{equation}
\lambda^{\pm} = \sum_{j=0}^\infty (\pm1)^{j}a_j(\delta)\left(\sqrt{\epsilon-\epsilon_0(\delta)}\right)^j
\end{equation}

Since $e^x$ is an analytic function, we can express this in a somewhat more convenient form:

\begin{equation}
\lambda^{\pm} = e^{i\sum_{j=0}^\infty (\pm1)^{j}b_j(\delta)\left(\sqrt{\epsilon-\epsilon_0(\delta)}\right)^j}
\end{equation}

Already we can make a few observations.  Since these eigenvalues must have modulus 1 over a range of values of $\epsilon$ and $\delta$, it follows that $b_j(\delta)$ is real for all $j$.  This is important for the following proof.

\begin{thm}
If the Left and Right eigevalues have a phase difference of $\delta$, then the location of the double root, $\epsilon_0$, is given by $\epsilon_0 = -\left(\frac{\delta}{2c}\right)^2 + O\left(\delta^4\right)$.
\end{thm}

{\it Proof} Since $\lambda$ is a power series is $\sqrt{\epsilon}$ and $\delta$, it follows that $\lambda(0,\delta) = e^{ib_0(\delta)+ ib_1(\delta)\sqrt{-\epsilon_0(\delta)} + ib_2(\delta)\left(-\epsilon_0(\delta)\right) + \cdots}$ is a power series in $\delta$.  Therefore, either $\epsilon_0(\delta)$ is a power series in $\delta^2$, or all of the odd terms ($a_1, a_3, \ldots$) are zero.  However this can't be the case, since $b_1(0) = c$ (this is the same c used throughout the rest of this paper).  So, $\epsilon_0$ is a power series in $\delta^2$.

From the definition of $\delta$, $e^{i\delta} = \lambda_{\ell}\lambda_r^*$, we find that:

$\begin{array}{ll}
e^{i\delta} = \lambda_{\ell}\lambda_r^* \\[2mm]
= \left[e^{i\sum_{j=0}^\infty b_j(\delta)\left(\sqrt{-\epsilon_0}\right)^j}\right]\left[e^{i\sum_{j=0}^\infty (-1)^{j}b_j(\delta)\left(\sqrt{-\epsilon_0}\right)^j}\right]^* \\[2mm]
= \left[e^{i\sum_{j=0}^\infty b_j(\delta)\left(\sqrt{-\epsilon_0}\right)^j}\right]\left[e^{-i\sum_{j=0}^\infty (-1)^{j}b_j(\delta)\left(\sqrt{-\epsilon_0}\right)^j}\right] \\[2mm]
= e^{i\sum_{j=0}^\infty \left(1-(-1)^j\right)b_j(\delta)\left(\sqrt{-\epsilon_0}\right)^j} \\[2mm]
\Rightarrow \delta = \sum_{j=0}^\infty \left(1-(-1)^j\right)b_j(\delta)\left(\sqrt{-\epsilon_0}\right)^j \\[2mm]
= 2b_1(\delta)\sqrt{-\epsilon_0} + 2b_3(\delta)\left(\sqrt{-\epsilon_0}\right)^3 + \cdots \\[2mm]
= 2b_1(\delta)\sqrt{-\epsilon_0} + O\left(\delta^3\right) \\[2mm]
= 2\left(b_1(0) + O(\delta)\right)\sqrt{-\epsilon_0} + O\left(\delta^3\right) \\[2mm]
= 2c\sqrt{-\epsilon_0} + O\left(\delta^2\right) \\[2mm]
\Rightarrow 2c\sqrt{-\epsilon_0} = \delta + O\left(\delta^2\right) \\[2mm]
\Rightarrow \epsilon_0 = -\left(\frac{\delta}{2c}\right)^2 + O\left(\delta^3\right)
\end{array}$

Finally, since $\epsilon_0$ is a power series in $\delta^2$, $\epsilon_0 = -\left(\frac{\delta}{2c}\right)^2 + O\left(\delta^4\right)$.

$\square$

\vspace{5mm}

Notice also that

$\begin{array}{ll}
\lambda_{\ell}\lambda_r \\[2mm]
= \left(e^{i\sum_{j=0}^\infty b_j(\delta)\left(\sqrt{0-\epsilon_0(\delta)}\right)^j}\right) \left(e^{i\sum_{j=0}^\infty (-1)^{j}b_j(\delta)\left(\sqrt{0-\epsilon_0(\delta)}\right)^j}\right) \\[2mm]
= e^{i2\sum_{j=0}^\infty b_{2j}(\delta)\left(-\epsilon_0(\delta)\right)^j} \\[2mm]
= e^{i2b_{0}(\delta)} e^{i2\sum_{j=1}^\infty b_{2j}(\delta)\left(-\epsilon_0(\delta)\right)^j} \\[2mm]
= e^{i2b_{0}(\delta)} e^{O\left(\delta^2\right)} \\[2mm]
= e^{i2b_{0}(\delta)} + O\left(\delta^2\right) \\[2mm]
\Rightarrow e^{ib_{0}(\delta)} = \sqrt{\lambda_{\ell}\lambda_r} + O\left(\delta^2\right)
\end{array}$

\vspace{5mm}

We can now write,

$\begin{array}{ll}
\lambda^{\pm} = e^{i\sum_{j=0}^\infty (\pm1)^{j}b_j(\delta)\left(\sqrt{\epsilon-\epsilon_0(\delta)}\right)^j} \\[2mm]
= e^{ib_{0}(\delta)}  e^{\pm i b_1(\delta) \sqrt{\epsilon-\epsilon_0(\delta)} + i\sum_{j=2}^\infty (-1)^{jk}b_j(\delta)\left(\sqrt{\epsilon-\epsilon_0(\delta)}\right)^j} \\[2mm]
= \left(\sqrt{\lambda_{\ell}\lambda_r} + O\left(\delta^2\right)\right) e^{\pm i\left( c + O(\delta) \right) \sqrt{\epsilon -\epsilon_0(\delta)} + O\left(\epsilon - \epsilon_0\right)} \\[2mm]
= \sqrt{\lambda_{\ell}\lambda_r} e^{\pm ic \sqrt{\epsilon -\epsilon_0(\delta)} + O\left(\delta\sqrt{\epsilon -\epsilon_0(\delta)}, \epsilon - \epsilon_0\right)} + O\left(\delta^2\right) \\[2mm]
= \sqrt{\lambda_{\ell}\lambda_r} e^{\pm ic \sqrt{\epsilon - \epsilon_0}} + O\left(\delta^2, \delta\sqrt{\epsilon -\epsilon_0}, \epsilon - \epsilon_0\right) \\[2mm]
\end{array}$

\vspace{5mm}

So, the double zeros of the characteristic polynomial and the pairing of the eigenvectors and eigenvalues aren't lost.  The new double zero is halfway between $\lambda_\ell$ and $\lambda_r$.  The pairing still exists, however it's in terms of $\sqrt{\epsilon + \left(\frac{\delta}{2c}\right)^2 + O\left(\delta^4\right)}$, not $\sqrt{\epsilon}$.

\subsection{Nearly-Paired Eigenvectors}

Eigenvalues and eigen{\it spaces} vary by $O\left(\delta,\sqrt{\epsilon}\right)$.  This is an important distinction to make.  The fundamental pairing theorem is essentially the statement that when $\delta=0$, then $|V_0^{\pm}\rangle \equiv \lim_{\epsilon\to0} |V^{\pm}\rangle = \frac{1}{\sqrt{2}}\left(|\ell_0\rangle \pm |\mathfrak{r}_0\rangle\right)$.  However, when $\delta\ne0$ we find that $|V_0^{\pm}\rangle$ must each converge independently to $|\ell_0\rangle$ and $|\mathfrak{r}_0\rangle$.  $|V_0^{\pm}\rangle$ is defined as a limit of eigenvectors of ${\bf U}$, and must itself be an eigenvector of ${\bf U}_0$.  But if $\lambda_\ell \ne \lambda_r$, then no combination of the Left and Right active eigenvectors can be eigenvectors of ${\bf U}_0$.

Therefore, $\lim_{\delta,\epsilon\to0}|V^{\pm}(\delta,\epsilon)\rangle$ doesn't exist.  Clearly, the "angle" between $|V^{\pm}(\delta,\epsilon)\rangle$ and the active eigenvectors, $|\ell_0\rangle$ and $|\mathfrak{r}_0\rangle$, is somehow dependent on $\delta$ and possibly some relationship between $\delta$ and $\epsilon$.

\vspace{5mm}

With some foresight, define:

\begin{eqnarray}
|V^+(\delta,\epsilon)\rangle = cos\left(\omega\right)|\ell_0\rangle + sin\left(\omega\right)|\mathfrak{r}_0\rangle + O(\delta, \sqrt{\epsilon})\\
|V^-(\delta,\epsilon)\rangle = -sin\left(\omega\right)|\ell_0\rangle + cos\left(\omega\right)|\mathfrak{r}_0\rangle + O(\delta, \sqrt{\epsilon})
\end{eqnarray}

That is, we can {\it define} $\omega$ using $cos(\omega) \equiv \langle \ell_0 | V^+(\delta,\epsilon)\rangle + O(\delta, \sqrt{\epsilon})$.

\vspace{5mm}

\begin{thm}
The angle between the paired eigenvectors and the active eigenvectors, $\omega$, is to lowest order a function of $\frac{\delta^2}{4c^2 \epsilon}$.
\end{thm}

{\it Proof} The proof of this is included in the appendix.

\vspace{5mm}

In the first version of this proof, when dealing with exactly-matched eigenvalues, there was no issue with taking the limit $\epsilon \to 0$ to find the value of $\omega$.  However, in this case we have terms involving $\frac{\delta^2}{\epsilon}$ making that difficult.

\subsection{Tuning}

We now make the declaration that $\frac{\delta^2}{4c^2\epsilon} \equiv t = O(1)$, and that $O\left(\delta,\epsilon\right)$ is still small.  In this way we can take the limit as both $\epsilon$ and $\delta$ go to zero, but fix a relationship between them.  Then,

$\begin{array}{ll}
sin^2(2\omega) = \left(1 + t + O\left( t\delta, \delta, \epsilon\right)\right)^{-1}  \\[2mm]
\Rightarrow sin^2(2\omega) = \left(1 + t + O\left(\delta, \epsilon\right)\right)^{-1} \\[2mm]
\Rightarrow sin^2(2\omega) = \frac{1}{1 + t } \\[2mm]
\Rightarrow t\,sin^2(2\omega) = 1 - sin^2(2\omega) \\[2mm]
\Rightarrow t = \frac{cos^2(2\omega)}{sin^2(2\omega)} \\[2mm]
\Rightarrow t = cot^2(2\omega) \\[2mm]
\end{array}$

\vspace{5mm}

When $t\approx 0$, $N\ll\left(\frac{2c}{\delta}\right)^2$  and $\omega \approx \frac{\pi}{4}$.  In this case the graph behaves like a normal, paired system.  That is, the mis-matching of the eigenvalues isn't large enough to affect the algorithm.  The paired eigenvectors are each equal combinations of both the Right and Left active eigenvectors (to within $O(\sqrt{\epsilon},\delta)$), and quadratic speed searches can be executed using the vectors $|\ell_0\rangle$ and $|\mathfrak{r}_0\rangle$.

For $t \approx 0$, in the $\{|\ell\rangle,|\mathfrak{r}\rangle\}$ basis,

${\bf U}= \lambda_0 
\left(\begin{array}{cc}
cos(c\sqrt{(1+t)\epsilon}) & -i\,sin\left(c\sqrt{(1+t)\epsilon}\right) \\
-i\,sin\left(c\sqrt{(1+t)\epsilon} \right) & cos(c\sqrt{(1+t)\epsilon}) \\
\end{array}\right) + O(\epsilon)$

\vspace{5mm}

When $t\gg1$, $N\gg\frac{\delta^2}{4c^2}$ and $\omega\approx0$.  The system does not behave like a paired system, but instead behaves as though there are no matched eigenvalues at all.  So, for large values of $t$ the initial state stays where it is.  The Left and Right eigenstates are decoupled, and are no longer useful for a search.

For $t \gg 1$, in the $\{|\ell\rangle,|\mathfrak{r}\rangle\}$ basis,

${\bf U}= \lambda_0 
\left(\begin{array}{cc}
e^{ic\sqrt{(1+t)\epsilon}} & 0 \\
0 & e^{-ic\sqrt{(1+t)\epsilon}} \\
\end{array}\right) + O(t\epsilon)$

\vspace{5mm}

"$t$" describes how "well tuned" a pair of active eignvectors are for given values of $\epsilon$ and $\delta$.  It also provides an easy way to move back and forth between $\epsilon$ and $\epsilon-\epsilon_0$, since $\epsilon-\epsilon_0 = \epsilon+\left(\frac{\delta}{2c}\right)^2 + O\left(\delta^4\right) = \epsilon(1+t) + O\left(\delta^4\right)$.  Graphs with a large value of $t$ are "poorly tuned", and graphs with a small value of $t$ are "well tuned".

\vspace{5mm}

We now wish to calculate $P(m) = |\langle\mathfrak{r}_0|{\bf U}^m|\ell_0\rangle|^2$, which is the probability of a successful search, using $|\ell_0\rangle$ as an initial state and $|\mathfrak{r}_0\rangle$ as a target state, after $m$ time steps.  How this depends on $t$ will be considered in the following theorem.

\vspace{5mm}

\begin{thm}There is a better than 50\% chance of a successful search of the $N$ edges of the hub vertex using the states $|\ell_0\rangle$ and $|\mathfrak{r}_0\rangle$ after $m=\left\lfloor \frac{\pi}{2c}\sqrt{N} \right\rfloor$ iterations of the time step operator, whenever

\begin{equation}
\delta < c\sqrt{\frac{2}{N}}
\end{equation}

where $\delta$ is the difference in phase between the Left and Right eigenvalues, and $c=\left|\langle\mathfrak{r_0|{\bf U}}|\ell_0\rangle\right|$.
\end{thm}

{\it Proof} The proof of this theorem will be included in the appendix.

\vspace{5mm}

So, when the error between the Left and Right eigenvalues is less than $c \sqrt{\frac{2}{N}}$, then we can ignore that error, and the algorithm will work normally more than half of the time.  We can do slightly better.  A carfeul reading of the proof reveals that the lower bound on the probability is closer to $\approx58.7\%$.  The usual ''$O(\sqrt{\epsilon}, \delta)$'' error does appear here, however that extra $8\%$ can be used to ignore these terms for sufficiently large values on $N$ ($N \ge O(100)$) and sufficiently small values of $\delta$ ($\delta \le O(0.1)$).

\pagebreak

\section{Summary of Results}

Take as given a star graph as described in section 2, with an attached subgraph $G$ and a time step operator ${\bf U}$.  Define $\epsilon=\frac{1}{N}$, where $N$ is the number of edges attached to the hub vertex.

-There are two kinds of eigenvectors of ${\bf U}_0$: bound eigenvectors, and actvie eigenvectors.  For a given eigenvalue, $\lambda_0$, of ${\bf U}_0$ there is at most one active eigenvector, $|\mathfrak{r}_0\rangle$.  The ``Right side $\lambda_0$ active eigenvector'' is unique, and will exist if and only if the $\lambda_0$ eigenspace of ${\bf U}_0$ is in contact with the Right side of the hub.

Bound eigenvectors are completely isolated inside of $G$ and are independent of $\epsilon$.

\vspace{2mm}

-The same statements applies to the Left side (although frequently throughout this paper we have assumed it to have a fixed, simple structure), and the Left side active $\lambda_0$ eigenvector is denoted $|\ell_0\rangle$.

\vspace{2mm}

-There is a pairing if and only if both the Left and Right sides have an active eigenvector with the same eigenvalue.  A pairing means that there are eigenvalues of ${\bf U}$ of the form $\lambda^\pm = \lambda_0e^{\pm ic\sqrt{\epsilon}}+O(\epsilon)$.  To first order, these eigenvalues vary by $O(\sqrt{\epsilon})$, instead of by $O(\epsilon)$ as non-paired eigenvectors do.  Paired eigenvalues always appear in pairs.

\vspace{2mm}

-The associated eigenvectors of these two eigenvalues are $|V^\pm\rangle$ which have the property that $|V^\pm_0\rangle = \frac{1}{\sqrt{2}}\left(|\ell_0\rangle\pm|\mathfrak{r}_0\rangle\right)$, for properly chosen phases.

\vspace{2mm}

-There can be no eigenvectors or eigenvalues that vary faster than $O\left(\sqrt{\epsilon}\right)$.

\vspace{2mm}

-$|\ell_0\rangle$ and $|\mathfrak{r}_0\rangle$ are rotated almost entirely into each other in $O\left(\sqrt{N}\right)$ time.  That is, there exists $m$ such that $|\langle\mathfrak{r}_0|{\bf U}^m|\ell_0\rangle|=1+O(\sqrt{\epsilon})$.

\vspace{2mm}

-That value of $m$ is $m = \left\lfloor\frac{\pi\sqrt{N}}{2c}\right\rfloor$, where $c$ is defined by the paired eigenvalues, $\lambda_0e^{\pm ic \sqrt{\epsilon}}$.  Writing the ${\bf U}$ as ${\bf U} = {\bf U}_0 + \sqrt{\epsilon}{\bf U}_1+\cdots$, the value of $c$ is $c=|\langle\mathfrak{r}_0|{\bf U}_1|\ell_0\rangle|$.  For a Left side consisting of only the states $|out\rangle$ and $|in\rangle$, we find that $c = \sqrt{2}|\langle1,0|\mathfrak{r}_0\rangle|$.

\vspace{2mm}

-We can quickly find a $\lambda_0$, $|\ell_0\rangle$, and $|\mathfrak{r}_0\rangle$ because all of them are determined entirely by ${\bf U}_0$, which tends to be easy to work with.  This is because, in addition to being block-diagonal (the blocks corresponding to the Left and Right sides), ${\bf U}_0$ is often sparse, and unlike ${\bf U}$, it has no dependence on $\epsilon$.   With $|\ell_0\rangle$ and $|\mathfrak{r}_0\rangle$ in hand, the values of $c$ and $m$ follow immediately.

\vspace{2mm}

-$\lambda_0$ doesn't need to be an exact double eigenvalue in order for a quadratic speed up to occur, however the allowable difference in the eigenvalues, $\delta$, is smaller for larger values of $N$.  That is; larger searches need more exact control.

When $\delta<c\sqrt{\frac{2}{N}}$, the probability of a successful search after $m = \left\lfloor\frac{\pi\sqrt{N}}{2c}\right\rfloor$ iterations of ${\bf U}$ is greater than half.

\subsection{Errors}

In practice, there are errors are produced by:

-Approximations of all of the important terms (i.e., $\lambda = \lambda_0e^{ic\sqrt{\epsilon}} +O(\epsilon)$).

-The initial state is not exactly equal to $|\ell_0\rangle$, since it has a component on the Right side.

-$|\ell_0\rangle$ and $|\mathfrak{r}_0\rangle$ are eigenvectors of ${\bf U}_0$, not ${\bf U}$, but are approximated by linear combinations of $|V^\pm\rangle$, the actual eigenvectors of ${\bf U}$ with eigenvalues $\lambda_\pm$.

-Rounding error due to the fact that $m$ must be an integer.

-$\lambda_0$ may not be an exact double root, but instead a pair such that $|\lambda_0-\lambda_0^\prime| = \delta\lambda \ll\sqrt{\epsilon}$

\vspace{5mm}

All of these produce errors of $O(\sqrt{\epsilon})$ or less in the final state.  This is unlikely to be a problem on any individual run of the algorithm, and can be easily dealt with by repetition.

\pagebreak

\section{Generalizations and Future Work}

{\bf Generalized Values of $r$ and $t$}

In section 2, $r$ and $t$ were introduced, along with the unitarity conditions:

\begin{eqnarray}
|r|^2+(N-1)|t|^2=1\\
2Re(tr^*)+(N-2)|t|^2=0
\end{eqnarray}

Throughout this paper the standard solution, $r=-1+2\epsilon$, $t=2\epsilon$, have been used, where $\epsilon=\frac{1}{N}$ as usual.  These are found by assuming $arg(r)=\pi$ and $arg(t)=0$.  However, by allowing arbitrary phase angles we find a family of solutions:

\begin{eqnarray}
r = \frac{1-2 \epsilon}{\sqrt{1 - 4\sin^2{(x-y)} (\epsilon - \epsilon^2)}}e^{ix}\\
t = \frac{-2\cos{(x-y)} \epsilon}{\sqrt{1 - 4\sin^2{(x-y)} (\epsilon - \epsilon^2)}}e^{iy}
\end{eqnarray}

where $cos(x-y)<1$.  This condition is necessary for the second unitary condition to have solutions.  In the collapsed graph we find $R_R=r$ and $R_L=r+(N-2)t$ to be the reflection coefficients for the hub vertex from the Right and Left sides respectively, and $T=t\sqrt{N-1}$ as the transmission coefficient.  By unitarity, $|R_R|^2+|T|^2=|R_L|^2+|T|^2=1$ and $\overline{R_R}T+R_L\overline{T}=0$.  We can use this last relation to quickly find $R_L = -\overline{R_R}\left(\frac{T}{\overline{T}}\right)=-\overline{R_R}e^{i2y}$.

We can now use this to re-visit theorems 3.5 and 4.1.  In the proof of theorem 3.5 we found that $T^2-R_RR_L=1$ for the standard $r$ and $t$, but this generalized solution may put that clean result at risk.  Luckily, $T^2-R_RR_L = |T|^2e^{i2y} - (R_R)(-\overline{R_R}e^{i2y}) = \left(|T|^2+|R_R|^2\right)e^{i2y}=e^{i2y}$.

In the proof of 3.5 it was shown that $C(z,\epsilon) = p_1(z)+p_2(z)R_R(\epsilon) + p_3(z)R_L(\epsilon)$, where $p_1, p_2, p_3$ are polynomials.  So, while we still have a clean result, $C(z, \epsilon)$ is no longer a polynomial of $\epsilon$.  But if a substitution is made, $r=\frac{(1-2\epsilon)e^{ix}}{\sqrt{1 - 4\sin^2{(x-y)} (\epsilon - \epsilon^2)}} = e^{ix} - 2cos^2(x-y)e^{ix}\epsilon +O(\epsilon^2) \equiv e^{ix}(1-2\mu)$, then $C(z,\mu)$ is a polynomial with respect to $\mu$.  Notice that $\mu$ is real when $\epsilon$ is real, and when $x=\pi$ and $y=0$ we find $\mu=\epsilon$.  Using $\mu$ instead of $\epsilon$ in section 3 we find that paired eigenvalues are of the form $\lambda = \lambda_0e^{\pm i\sqrt{\mu}}+O(\mu)$.  However, because $\mu=O(\epsilon)$, this result is the same as the original: $\lambda = \lambda_0e^{\pm i\sqrt{\epsilon}}+O(\epsilon)$.  Otherwise, since $\sqrt{\mu}=O(\sqrt{\epsilon})$, $\mu=O(\epsilon)$, and so on, all of the other results in this paper remain the same.

\vspace{5mm}

{\bf Multiple Copies of $G$}

Assume there are $N$ total edges connected to the hub vertex, as before, but rather than 1 edge connected to $G$ there are $M$ edges connected to $M$ copies of $G$.  These various copies of $G$ can be connected to each other, but (for this generalization) in such a way that in the automorphism graph there is only one remaining edge on the Left and Right sides.

The reflection and transmission coefficients are $R_L = r+(N-M-1)t$, $R_R=r+(M-1)t = -1+2\frac{M}{N}$, and $T = t\sqrt{M}\sqrt{N-M}$.  Plugging in the standard solutions, $r=-1+\frac{2}{N}$ and $t=\frac{2}{N}$, yields $R_L = 1-2\frac{M}{N}$, $R_R=-1+2\frac{M}{N}$, and $T = 2\sqrt{\frac{M}{N}-\left(\frac{M}{N}\right)^2}$, where again $R_L$ and $R_R$ are the Left and Right side reflection coefficients respectively.  Clearly, by using $\epsilon=\frac{M}{N}$ instead of $\epsilon=\frac{1}{N}$ all of the behavior of a single copy of $G$ is recovered (see section 2).  Using the general solution for $r$ and $t$ doesn't provide quite such a clean result.  For example, $T = \frac{-2\cos{(x-y)}\sqrt{\frac{M}{N}-\left(\frac{M}{N}\right)^2}}{\sqrt{1 - 4\sin^2{(x-y)} ( \frac{1}{N} -  \frac{1}{N}^2)}}e^{iy}$.  But notice that to first order, like the standard solution, this is proportional to $\sqrt{\frac{M}{N}}$.

The important result to take away from this generalization is that we can expect searches to take a time of $O\left(\sqrt{\frac{N}{M}}\right)$.  $\frac{M}{N}$ is the "effective degree" of the hub vertex.

\vspace{5mm}

{\bf Multiple Subgraphs}

If there are multiple subgraphs that share a common eigenvalue but have different forms, then it can be show that they will behave collectively as though they were a single subgraph.  The probability of a particular subgraph being the result of a search increases with increasing $c$, as one might expect.

\vspace{5mm}

{\bf General Highly-Symmetric Graphs}

The star graph gives us a way of looking at the behavior of a hub vertex in depth.  For highly symmetric graphs with bounded diameter (the maximum distance between any pair of vertices is bound), we will find at least one hub vertex.  Using the techniques of this paper it should be fairly straight forward to generalize to automorphism graphs with multiple hubs.  For example, in the investigation of the behavior of finite-depth tree graphs.

\pagebreak

\subsection*{Acknowledgements}

I'd like to thank Professors Edgar Feldman, Janos Bergou, and Sylvain Cappell for their support and input.  But in particular I'd like to thank Mark Hillery, who helped with every step of the process of writing and publishing this paper, pointed out mistakes, and answered every question no matter how off-topic.  Professor Hillery is patience and generosity personified and this paper wouldn't exist without his help and the foundation provided in previous papers by Hillery, Feldman, and Bergou.

\pagebreak

\pagebreak

\setcounter{section}{9}

\section*{Appendices}\

\subsection{In-Depth Example: the Bolo Subgraph}

The bolo graph (which resembles a bolo tie) has a bound state, and 4 Right side active eigenvectors.  Using techniques from the paper we can quickly decide on the best eigenvalue and Left side eigenvector to use as an approximate initial state to ensure the quickest search.  For the purposes of this example, $N=10^6$.

\begin{figure}[h!]
\centering
\includegraphics[keepaspectratio=true, scale = 0.5]{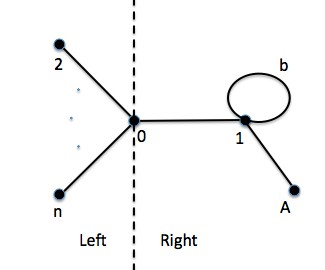}
\caption{The Right side of this graph is the "Bolo Graph".}
\end{figure}

We can get all the information we need from ${\bf U}_0$, so we can ignore the Left side entirely.  Define the basis vectors as:

$\begin{array}{ll}
|\Psi_1\rangle = |0,1\rangle \\
|\Psi_2\rangle = |A,1\rangle \\
|\Psi_3\rangle = |b\rangle \\
|\Psi_4\rangle = |1,A\rangle \\
|\Psi_5\rangle = |1,0\rangle \\
\end{array}$

And in this basis define the effect of ${\bf U}_0$ as:

${\bf U}_0=\left(\begin{array}{ccccc}
0&0&0&0&-1\\
0&0&0&-1&0\\
\frac{2}{3}&\frac{2}{3}&-\frac{1}{3}&0&0\\
\frac{2}{3}&-\frac{1}{3}&\frac{2}{3}&0&0\\
-\frac{1}{3}&\frac{2}{3}&\frac{2}{3}&0&0
\end{array}\right)$

This is an equal scattering in three directions at vertex 1.  The difference is that a signal returns from the $b$ arm in 1 time step, and from the $A$ arm in 2.

\vspace{5mm}

Step 1) Find the eigenvalues and eigenvectors of the Right side.

The characteristic polynomial is $C_0(z) = z^5+\frac{1}{3}z^4-\frac{2}{3}z^3+\frac{2}{3}z^2-\frac{1}{3}z-1$, and the five eigenvalues are then found to be: $\lambda = \left\{-1, -1, 1, \frac{1}{3}\left(1 + i2 \sqrt{2}\right), \frac{1}{3}\left(1 - i2 \sqrt{2}\right)\right\}$.

Already we know that there must be at least one bound eigenstate with eigenvalue -1, since the active eigenvector for each eigenvalue is unique and -1 is a degenerate eigenvalue.  The eigenvectors, in the same order, are:

$|bound\rangle = \frac{1}{\sqrt{3}}\left[|A,1\rangle - |b\rangle + |1,A\rangle\right]$

$\left|\mathfrak{r}^{(-1)}\right\rangle = \sqrt{\frac{3}{8}}\left[|0,1\rangle - \frac{1}{3}|A,1\rangle - \frac{2}{3}|b\rangle - \frac{1}{3}|1,A\rangle + |1,0\rangle\right]$

$\left|\mathfrak{r}^{(1)}\right\rangle = \frac{1}{2}\left[ - |0,1\rangle + |A,1\rangle - |1,A\rangle + |1,0\rangle\right]$

$\left|\mathfrak{r}^{\left(\frac{1}{3} + i \frac{2\sqrt{2}}{3}\right)}\right\rangle = \frac{\sqrt{3}}{4}\left[\left( -\frac{1}{3} + \frac{2\sqrt{2}}{3} i\right)|0,1\rangle + \left( -\frac{1}{3} + \frac{2\sqrt{2}}{3} i\right) |A,1\rangle + \left( \frac{2}{3} + \frac{2\sqrt{2}}{3} i\right) |b\rangle + |1,A\rangle + |1,0\rangle\right]$

$\left|\mathfrak{r}^{\left(\frac{1}{3} - i \frac{2\sqrt{2}}{3}\right)}\right\rangle = \frac{\sqrt{3}}{4}\left[\left( -\frac{1}{3} - \frac{2\sqrt{2}}{3} i\right)|0,1\rangle + \left( -\frac{1}{3} - \frac{2\sqrt{2}}{3} i\right) |A,1\rangle + \left( \frac{2}{3} - \frac{2\sqrt{2}}{3} i\right) |b\rangle + |1,A\rangle + |1,0\rangle\right]$

All of the last four eigenvectors are active eigenvectors.  This is immediately obvious because $\langle0,1|\mathfrak{r}^{(j)}\rangle\ne0$ for each of them, so they are hub adjacent.  The $\lambda_0=-1$ active eigenvector can be found by first finding the unique bound eigenvector.  What remains in the $-1$ eigenspace must be the active eigenvector.

The active eigenspace is 6-dimensional and is spanned by these four Right side and the two Left side active eigenvectors.  Paired, or otherwise dependent on $\epsilon$, eigenvectors are always expressible as superpositions of these six active eigenvectors.  Even if the Left side were replaced with something more interesting, these four vectors would provide all the information necessary to analyze how the Right side interacts.

\vspace{5mm}

Step 2) Select a target eigenvalue.

We already have enough information to see which of these states, or rather which of these eigenvalues, is the best target for a search.  The state that ``leans'' toward the hub is always the best choice.  There are two reasons: first, because the optimal number of iterations for a quadratic search is is given by $m = \lfloor\frac{\pi}{2c}\sqrt{N}\rfloor$, and since $c = |\langle\ell_0|{\bf U}_1|\mathfrak{r}_0\rangle| = \sqrt{2}|\langle1,0|\mathfrak{r}_0\rangle|$, the state that overlaps $|1,0\rangle$ most will yield the shortest search time.  And second, a large value of $c^{(j)}$ means that the Right side active eigenvector is concentrated closer to the hub, which means that states on the edge between $0$ and $1$ are more likely to be measured.

$\begin{array}{ll}
\sqrt{\frac{3}{4}} = c^{(-1)} = \sqrt{2} \left| \left\langle1,0 |\mathfrak{r}^{(-1)}\right\rangle \right| \\[2mm]
\sqrt{\frac{1}{2}} = c^{(1)} = \sqrt{2} \left| \left\langle1,0 |\mathfrak{r}^{(1)}\right\rangle \right| \\[2mm]
\sqrt{\frac{3}{8}} = c^{\left(\frac{1}{3} + i \frac{2\sqrt{2}}{3}\right)} = \sqrt{2} \left| \left\langle1,0 \Big|\mathfrak{r}^{\left(\frac{1}{3} + i \frac{2\sqrt{2}}{3}\right)}\right\rangle \right| \\[4mm]
\sqrt{\frac{3}{8}} = c^{\left(\frac{1}{3} - i \frac{2\sqrt{2}}{3}\right)} = \sqrt{2} \left| \left\langle1,0 \Big|\mathfrak{r}^{\left(\frac{1}{3} - i \frac{2\sqrt{2}}{3}\right)}\right\rangle \right|
\end{array}$

Clearly, $\lambda_0=-1$ is the best choice, since $max\{c^{(j)}\} = c^{(-1)} = \sqrt{\frac{3}{4}}$.  We know that there is always some $c^{(j)}\ge\sqrt{\frac{2}{d}} = \sqrt{\frac{2}{4}} = \frac{1}{\sqrt{2}}$, so a value of $\sqrt{\frac{3}{4}}$ is not surprising.

\vspace{5mm}

Step 3) Tune the Left side eigenvalues.

The Left side $\lambda_0=-1$ active eigenvector is $\left|\ell^{(-1)}\right\rangle = \frac{1}{\sqrt{2}}\left(|out\rangle - |in\rangle\right)$, and this eigenstate is only possible when $e^{i\phi} = \lambda_0^2 = (-1)^2 = 1$.  So, by setting $\phi =0$, the eigenvalues on the Left become $\pm1$.  This does mean that, since both sides now share both 1 and -1 as eigenvalues, paired states can exist for both eigenspaces.  However, by initializing with the -1 eigenstate, the +1 state is unimportant.  We want to match the -1 state because it is the fastest (largest value of $c^{(j)}$).
 
\vspace{5mm}

Step 4) Initialize the system with the state $|\Psi\rangle = \frac{1}{\sqrt{2\cdot 10^6}}\sum_{j=1}^{10^6} \left(|0,j\rangle - |j,0\rangle\right)$.

To within an error of $O(0.1\%) = O\left(\frac{1}{\sqrt{10^6}}\right)$, this is equal to the Left side $\lambda_0=-1$ eigenstate, $\left|\ell^{(-1)}\right\rangle = \frac{1}{\sqrt{2(10^6-1)}}\sum_{j=2}^{10^6} \left(|0,j\rangle - |j,0\rangle\right)$.  Since we assume that we don't know to which edge the Bolo graph is attached, we can't start {\it entirely} in the -1 Left active eigenstate.  That said, $\langle\ell^{(-1)}|\Psi\rangle = \frac{2(10^6-1)}{2\cdot 10^3\sqrt{(10^6-1)}} = 0.9999995$, so these two states are essentially equal.

\vspace{5mm}

Step 5) Iterate the time step operator, ${\bf U}$.

Use ${\bf U}$ to step time forward $m = 1813 = \left\lfloor\frac{\pi}{\sqrt{3}}\sqrt{10^6}\right\rfloor = \left\lfloor\frac{\pi}{2 c^{(-1)}}\sqrt{10^6}\right\rfloor$ times.  This process will cause the state to rotate from $\left|\ell^{(-1)}\right\rangle$ to $\left|\mathfrak{r}^{(-1)}\right\rangle$, to within an additional error of $O(0.1\%)$ produced by the rounding of the floor function, and the fact that the paired eigenvetors, $|V^{\pm}\rangle$, are only very closely approximated by combinations of $\left|\ell^{(-1)}\right\rangle$ and $\left|\mathfrak{r}^{(-1)}\right\rangle$.

\vspace{5mm}

Step 6) Measure the system.  The probability of the particle being detected on the edge between vertices $0$ and $1$ is $p = \left|\left\langle 0,1 \Big|\mathfrak{r}^{(-1)}\right\rangle\right|^2 + \left|\left\langle 1,0 \Big|\mathfrak{r}^{(-1)}\right\rangle\right|^2 = \left|\sqrt{\frac{3}{8}}\right|^2 + \left|\sqrt{\frac{3}{8}}\right|^2 = \frac{3}{4}$.  This "$\frac{3}{4}$" is good news, since it means that the algorithm won't need to be repeated extensively.

\vspace{5mm}

Finally, the Left and Right eigenvalues need not be exact.  According to theorem 5.5, the search will still work more than half of the time when the difference between the Left and Right eigenvalues, $\delta$, satisfies $\delta < c^{(-1)}\sqrt{\frac{2}{N}} = \sqrt{\frac{3}{4}}\sqrt{\frac{2}{10^6}} \approx 0.001$.  So, for $N=10^6$, the complex phase of the Left eigenvalue should be in the range $[\pi-0.001,\pi+0.001]$.

\vspace{5mm}

Notice that, aside from finding the value of $m$ and estimating errors, $N$ (or $\epsilon$) was never considered at all.  Indeed the the Left and Right sides are handled separately from beginning to end.

\subsection{Proofs from section 3 (Algebraic Functions and the Behavior of Zeros)}

\begin{thma} {\bf 3.1}
If $F(z)$ is globally analytic in an annulus around $0$, and is $m$-valued, then $F(z)$ can be expressed as a Puiseux series (a Laurent series with certain rational powers) of the form $F(z)=\sum_{n=-\infty}^\infty A_n z^\frac{n}{m}$.  Moreover, the m different branches of $F$, $F^{(j)}$, can be separated by an arbitrary branch cut through the annulus and expressed as $F^{(j)}(z) = \sum_{n=-\infty}^\infty A_n \omega^{jn} z^\frac{n}{m}$, where $\omega$ is a primitive $m$th root of unity, $\omega = e^{i\frac{2\pi}{m}}$.
\end{thma}

{\it Proof} Consider the annulus in the $z$-plane defined by $D_z=\{z:0<r<|z|<R\}$, and the mapping $z=\zeta^m$.  Define $G(\zeta) \equiv F(\zeta^m)$ on the annulus $D_\zeta=\{\zeta:0<r^{\frac{1}{m}}<|\zeta|<R^{\frac{1}{m}}\}$.  $G$ inherits its analyticity from $F$, since $\frac{d}{d\zeta}G(\zeta) = \frac{d}{d\zeta}F(\zeta^m) = \frac{dz}{d\zeta}\frac{d}{dz}F(z)$.
In addition, $G$ is single valued.  Defining any one of the branches of $F$ to be the principle branch, $f_0$, and taking the value of $G(x)=F^{(0)}(x^m)$, $x\in \mathbb{R}$, we can then define $G(z)$ to be the terminal value of any analytic continuation of $G$ from $x$.  While $F(z)$ has $m$ branches, $G(\zeta)$ has $m$ corresponding $\frac{2\pi}{m}$ wedges.

Continuation along a path, $C_\zeta$, that starts at $x\in\mathbb{R}^+$ and traverses once around the circle $|\zeta| = x$ corresponds to traversing the circle $C_z$, defined by $|z|=x^m$, in $D_z$, $m$ times.  But since $F(z)$ is $m$ valued, traversing $|z|=x^m$ $m$ times will return $F$ to the principle branch.  Thus, $G(xe^{2\pi i})=G(x)$, and more generally $G(\zeta e^{2\pi i})=G(\zeta)$, for $\zeta\in D_\zeta$.

Since $G(\zeta)$ is analytic and single-valued in the annulus $D_\zeta$, it admits a Laurent series: $G(\zeta)=\sum_{n=-\infty}^\infty A_n \zeta^n$.  Therefore, the globally analytic $m$-valued function $F$ can be written as $F(z)=G(z^\frac{1}{m})=\sum_{n=-\infty}^\infty A_n z^\frac{n}{m}$.

Notice that if the initial branch, $F^{(0)}(z)=\sum_{n=-\infty}^\infty A_n z^\frac{n}{m}$, is analytically continued once around $C_z$ a $2\pi$ phase is added to $z$ and we find the function taking values on the next branch cut (so a very natural ordering of the branch cuts is here defined by subsequent loops around $z=0$).  We find that $F^{(1)}(z)=\sum_{n=-\infty}^\infty A_n (e^{2\pi i})^\frac{n}{m}z^\frac{n}{m} = \sum_{n=-\infty}^\infty A_n \omega^n z^\frac{n}{m}$, where $\omega =e^{i\frac{2\pi}{m}}$.  $j$ loops around $z=0$ yields $F^{(j)}(z)= \sum_{n=-\infty}^\infty A_n \omega^{j n} z^\frac{n}{m}$.

$\square$

\vspace{5mm}

\begin{thma} {\bf 3.2}
There exists an open disk $D$, containing 0, and $d$ analytic functions, $f^{(1)}, \cdots, f^{(d)}$, such that:

$\begin{array}{ll}
(i) & P(f^{(k)}(\epsilon), \epsilon)=0, \epsilon\in D\\
(ii) & f^{(k)}(0) = \lambda^{(k)} \\
(iii) & P(\lambda, \epsilon)=0, \epsilon\in D \Rightarrow \lambda=f^{(k)}(\epsilon), for\; some\; k
\end{array}$
\end{thma}

{\it Proof} Note that $f^{(k)}$ are indexed functions, and not necessarily branches of the same globally analytic function.  While it is true that when $P(z,\epsilon)$ is not simultaneously reducible in both $z$ and $\epsilon$ the zeros, as functions or $\epsilon$, are all branches of the same globally analytic function, it isn't necessary to know that here.  What is important is that some of these 

{\it Proof} Since the zeros of $P(z,0)$ are distinct, $\exists \delta$ such that $|z -\lambda^{(k)}|\le\delta$ contains no zeros other than $\lambda^{(k)}$.  Define $C_k$ to be the loop defined by $|z - \lambda^{(k)}|=\delta$.  By the argument principle $\frac{1}{2\pi i}\oint_{C_k} \frac{\partial_z P(z, 0)}{P(z,0)}\, dz=1$.  Here we are assuming that the zeros of $P(z,0)$ are all of degree 1, and there is only one zero inside of $C_k$.  The case of higher degree zeros is dealt with in the next theorem.

The zeros of $P(z,\epsilon)$ are continuous functions of the coefficients of $P(z,\epsilon)$ (other than $a_d$ at $a_d=0$), and the coefficients of $P(z,\epsilon)$ are continuous functions of $\epsilon$.  By definition $\exists \sigma>0$ such that when $\epsilon<\sigma$, $|f^{(k)}(\epsilon)-\lambda^{(k)}|<\delta$.  In other words, the zero will stay within $C_k$, so for small values of $\epsilon$, $\frac{1}{2\pi i}\oint_{C_k}\frac{\partial_z P(z, \epsilon)}{P(z,\epsilon)}\, dz = 1$.  Notice that we've only used the definition of $f_k(\epsilon)$ as the zero of $P(z,\epsilon)$ corresponding to $\lambda^{(k)}$, independent of its analytic properties.

This integral can be used to pick out the value of the zero, $f^{(k)}(\epsilon)$, by multiplying the argument of the integral by $z$.  By the residue calculus:

$f^{(k)}(\epsilon) = \frac{1}{2\pi i}\oint_{C_k} z\frac{\partial_z P(z, \epsilon)}{P(z,\epsilon)}\, dz$

The important thing to notice here is that, since $P(z,\epsilon)$ is a polynomial of $\epsilon$ near $\epsilon=0$, $f_k(\epsilon)$ is analytic.

Repeating this process for each of the $d$ simple zeros of $P(z,0)$ yields the $d$ analytic functions $f^{(1)}(\epsilon), \cdots, f^{(d)}(\epsilon)$ that are the zeros of $P(z,\epsilon)$.

$\square$

\vspace{5mm}

\begin{thma} {\bf 3.3}
If $P(z,\epsilon)$ is an irreducible polynomial in $z$ and $\epsilon$, then all of the double roots of $P$ are isolated in the $\epsilon$-plane.  That is, if for some value $\epsilon_0$, $P(z,\epsilon_0)$ has a double root, then there exists $\delta>0$ such that when $0<|\epsilon-\epsilon_0|<\delta$, $P(z,\epsilon)$ does not have a double root in $z$.
\end{thma}

The proof of this requires the introduction of a new object: the discriminant.

\vspace{5mm}

{\bf Definition} The "discriminant of $P$", $\mathcal{D}[P]$, is a function of the coefficients of a polynomial $P$, and $\mathcal{D}[P]\equiv a_d^{2d-2}\prod_{j>k}(\lambda^{(j)}-\lambda^{(k)})^2$, where $a_d$ is the leading coefficient of $P$, and $\{\lambda^{(k)}\}$ are the zeros of $P$.

Clearly the discriminant is zero if and only if $P$ has a repeated root, and this is the property that makes it so appealing.

{\it Proof} A known fact about the discriminant is that it is the resultant of $P$ and $P_z$, which is in some sense like being the $gcd(P,P_z)$.  One of the methods of finding the discriminant is very much like Euclid's algorithm for finding the $gcd$ of two numbers, but rather than finding combinations of the two numbers that produce the lowest non-zero value, the discriminant involves finding the combination of polynomials that produces the lowest degree non-zero polynomial.  For example, if $P(z)=Az^2+Bz+C$, then $P_z(z)=2Az+B$, and

$\begin{array}{ll}
2P-zP_z = Bz+2C \\
\Rightarrow 2A(2P-zP_z) -BP_z = 4AC - B^2
\end{array}$

Which is the well-known discriminant for quadratic equations.  So, $\mathcal{D}[P] = B^2-4AC = (-4A)P+(2Az+B)P_z$.  

\vspace{5mm}

The things to keep in mind here are:

$\begin{array}{ll}
i) & \textrm{$\mathcal{D}=0$ if and only if $P(z)$ has a repeated zero.} \\
ii) & \textrm{$\mathcal{D}$ is \underline{not} a function of $z$.} \\
iii) & \textrm{$\mathcal{D}$ is a polynomial in the coefficients of $P$.} \\
iv) & \textrm{If the coefficients of $P$ are polynomials of $\epsilon$, then so is $\mathcal{D}$.} \\
\end{array}$

\vspace{5mm}

{\it Proof} $\mathcal{D}(\epsilon)$ inherits its analyticity from $P(z,\epsilon)$.  Because $\mathcal{D}$ is analytic, if it has a non-isolated zero, then it must be identically zero.  But if the discriminant of $P$ is always zero, then $P$ always has a double root, and must therefore be reducible into factors.  Because the original assumption was that $P(z,\epsilon)$ is irreducible, $\mathcal{D}$ cannot be identically zero, and therefore any zeros of $\mathcal{D}(\epsilon)$ are isolated.  This means that changing $\epsilon$ splits a multiple zero into simple zeros in general, and if $P(z,0)$ has a repeated root, then within a small punctured disk about $\epsilon=0$, $P(z,\epsilon)$ has only simple roots.

$\square$

\vspace{5mm}

\begin{thma} {\bf 3.4}
In the neighborhood of a zero of $P(z,\epsilon)$ of multiplicity $s>1$, the zeros take the form $f^{(j)}(\epsilon) = \sum_{n=-\infty}^\infty A_n\omega^{jn}\epsilon^\frac{n}{H}$, where $H<s$.  Specifically, the zeros are branches of one or more $H_i$-valued global analytic functions, with the given Puiseux series expansion, such that $\sum H_i=s$.
\end{thma}

{\it Proof} Without loss of generality, assume that the zero of multiplicity $s$ is found at $\epsilon=0$, so $f^{(1)}(0)=f^{(2)}(0)=\cdots=f^{(s)}(0)$.  In the last theorem it was shown that within a small disk excluding $\epsilon=0$ these $s$ functions are different.  For some $\epsilon_0$ within this punctured disk we can apply theorem 3.2 almost verbatim to show that $f^{(1)}(\epsilon), \cdots, f^{(s)}(\epsilon)$ are analytic functions.

However, they may not necessarily be single-valued.  If they are analytically continued in a loop around $\epsilon=0$ they may come back with a different value.  However, by definition $P(f^{(k)}(\epsilon), \epsilon)=0$, and there are only $d$ possible such functions.  As a result, if $f^{(k)}(\epsilon)$ does not return to it's original value it must return as one of the other functions.
Therefore, looping around $\epsilon=0$ permutes $f^{(1)}, \cdots, f^{(s)}$.  If $H$ loops brings $f_k$ back to its original value, then $f^{(k)}$ is a branch of an $H$-valued global analytic function.  Clearly, $H<s$ since there are only $s$ available functions to permute.  The $s$ different functions can be grouped according to which of the global analytic functions they're branches of, $\{f^{(1)}, f^{(2)},\cdots, f^{(H_1)}\}\{f^{(H_1+1)},\cdots, f^{(H_1+H_2)}\}\cdots$, where $\sum H_i = s$еее

Since these functions are branches of a global analytic function in an annulus, by theorem 3.4 these functions must be of the form $f^{(j)}(\epsilon) = \sum_{n=-\infty}^\infty A_n\omega^{jn}\epsilon^\frac{n}{H}$.

$\square$

\vspace{5mm}

\subsection*{Proof That the Characteristic Polynomial is Affine in $\epsilon$}

For a hub vertex with reflection coefficients $r$ and $t$, we know that unitarity implies $|r|^2+(N-1)|t|^2=1$ and $2Re(\overline{r}t) + (N-2)|t|^2=0$.  The most commonly used solution, and the one assumed throughout this paper, is $r=-1+2\epsilon$, $t=2\epsilon$.  The generalized solution is handled in section 7.

In the automorphism graph we found that the reflection coefficients are $R_L=r+\left(N-2\right)t = 1-2\epsilon$ and $R_R = r = -1+2\epsilon$, for the Left and Right sides respectively, and the transmission coefficients between the two sides is $T= t\sqrt{N-1} = 2\sqrt{\epsilon-\epsilon^2}$.

\begin{thma} {\bf 3.5}
Assume that ${\bf U}$ is a time step matrix as described so far.  That is, there is a Left and Right side and these are connected only through a hub vertex with $N$ edges where the reflection and transmission coefficients are $r=-1+\frac{2}{N}$ and $t=\frac{2}{N}$.  Then $C(z, \epsilon) = \left|{\bf U} - z {\bf I}\right|$ is an affine  polynomial of $\epsilon=\frac{1}{N}$, and can be written $C(z,\epsilon) = C_0(z) + \epsilon f(z)$.
\end{thma}

{\it Proof} First, some cumbersome notation.  Define $|{\bf M}|$ to be the determinant of the matrix ${\bf M}$, $|{\bf M}|_{(i,j)}$ to be the determinant with the $(i,j)$ term replaced with a zero, and $|{\bf M}|_{<i,j>}$ to be the determinant of ${\bf M}$ with the $i$th column and $j$th row removed.  Note that if ${\bf M}$ is an $n\times n$ matrix, then $|{\bf M}|_{(i,j)}$ is also $n\times n$, and $|{\bf M}|_{<i,j>}$ is $(n-1)\times (n-1)$.

In what follows we'll make use of the fact that we can remove an element from the determinant of ${\bf M}$, but must include a determinant of a minor matrix.  For example,

$\left|\begin{array}{ccc}
a&b&c\\
d&e&f\\
g&h&i
\end{array}\right|=a\left|\begin{array}{ccc}
e&f\\
h&i
\end{array}\right|-d\left|\begin{array}{ccc}
b&c\\
h&i
\end{array}\right|+g\left|\begin{array}{ccc}
b&c\\
e&f
\end{array}\right|=\left|\begin{array}{ccc}
a&b&c\\
d&e&f\\
0&h&i
\end{array}\right|+g\left|\begin{array}{ccc}
b&c\\
e&f
\end{array}\right|$

This is can be more succinctly written, $|{\bf M}| = |{\bf M}|_{(1,3)} + g|{\bf M}|_{<1,3>}$.  Notice that $|{\bf M}|_{(1,3)}$ is $3\times 3$, while $|{\bf M}|_{<1,3>}$ is $2\times 2$.

\vspace{5mm}

If we list all of the Left side states first, then

${\bf U} \equiv \left(\begin{array}{c|c}
{\bf U}_{L}& T_{ij} \\ \hline
T_{kl} & {\bf U}_{R}
\end{array}\right)$

where ${\bf U}_{L}$ and ${\bf U}_{R}$ are the restrictions of ${\bf U}$ to the Left and Right sides, and $T_{ij}$ and $T_{kl}$ are all zero except for a single $T=2\sqrt{\epsilon-\epsilon^2}$ at the indicated coordinate.  Note that if the $T$'s are at the coordinates $(i,j)$ and $(k,l)$, then the reflection coefficient for the left side, $R_L$, can be found at $(k,j)$ and similarly $R_R$ can be found at $(i,l)$.  It is now straightforward to see,

$C(z,\epsilon) = \left|\begin{array}{c|c}
{\bf U}_{L} - z{\bf I} & T_{ij} \\ \hline
T_{kl} & {\bf U}_{R} - z{\bf I}
\end{array}\right|$

$= \left|\begin{array}{c|c}
{\bf U}_{L} - z{\bf I} & 0 \\ \hline
T_{kl} & {\bf U}_{R} - z{\bf I}
\end{array}\right| + (-1)^{i+j}T\left|\begin{array}{c|c}
{\bf U}_{L} - z{\bf I} & 0 \\ \hline
T_{kl} & {\bf U}_{R} - z{\bf I}
\end{array}\right|_{<i,j>}$

$= \left|\begin{array}{c|c}
{\bf U}_{L} - z{\bf I} & 0 \\ \hline
0 & {\bf U}_{R} - z{\bf I}
\end{array}\right| + (-1)^{k+l}T\left|\begin{array}{c|c}
{\bf U}_{L} - z{\bf I} & 0 \\ \hline
0 & {\bf U}_{R} - z{\bf I}
\end{array}\right|_{<k,l>} + (-1)^{i+j}T\left|\begin{array}{c|c}
{\bf U}_{L} - z{\bf I} & 0 \\ \hline
T_{kl} & {\bf U}_{R} - z{\bf I}
\end{array}\right|_{<i,j>}$

There's a slight abuse of notation in this step.  Since the third term has had the $j$th row removed, the coordinate of $T_{kl}$ is not $(k,l)$, but is instead $(k,l-1)$.  It follows that,

$= \left|\begin{array}{c|c}
{\bf U}_{L} - z{\bf I} & 0 \\ \hline
0 & {\bf U}_{R} - z{\bf I}
\end{array}\right| 
+ (-1)^{k+l}T\left|\begin{array}{c|c}
{\bf U}_{L} - z{\bf I} & 0 \\ \hline
0 & {\bf U}_{R} - z{\bf I}
\end{array}\right|_{<k,l>} 
+ (-1)^{i+j}T\left|\begin{array}{c|c}
{\bf U}_{L} - z{\bf I} & 0 \\ \hline
0 & {\bf U}_{R} - z{\bf I}
\end{array}\right|_{<i,j>}
+ (-1)^{i+j+k+(l-1)}T^2\left|\begin{array}{c|c}
{\bf U}_{L} - z{\bf I} & 0 \\ \hline
0 & {\bf U}_{R} - z{\bf I}
\end{array}\right|_{<i,j>,<k,l>}$

The second and third terms of this last line are both equal to zero.  In the second term the $l$th column is removed, but since ${\bf U}_{R} - z{\bf I}$ is a square matrix, removing a row leaves the columns linearly dependent.  Thus the matrix is degenerate, and the determinant is zero.  A similar argument holds for the third term.  Both of the non-zero blocks of the fourth term are square matrices, and are therefore not necessarily zero.
We now have,

\begin{equation}
C(z, \epsilon) = \left|\begin{array}{c|c}
{\bf U}_{L} - z{\bf I} & 0 \\ \hline
0 & {\bf U}_{R} - z{\bf I}
\end{array}\right| 
- (-1)^{i+j+k+l}T^2\left|\begin{array}{c|c}
{\bf U}_{L} - z{\bf I} & 0 \\ \hline
0 & {\bf U}_{R} - z{\bf I}
\end{array}\right|_{<i,j>,<k,l>}
\end{equation}

This is enough to show that $C(z, \epsilon)$ is a polynomial in both $z$ and $\epsilon$.  However, this can be taken a step further.  Repeating the same trick we find that

$\left|\begin{array}{c|c}
{\bf U}_{L} - z{\bf I} & 0 \\ \hline
0 & {\bf U}_{R} - z{\bf I}
\end{array}\right|$

$= \left|\begin{array}{c|c}
{\bf U}_{L} - z{\bf I} & 0 \\ \hline
0 & {\bf U}_{R} - z{\bf I}
\end{array}\right|_{(i,l)} + (-1)^{i+l}R_R\left|\begin{array}{c|c}
{\bf U}_{L} - z{\bf I} & 0 \\ \hline
0 & {\bf U}_{R} - z{\bf I}
\end{array}\right|_{<i,l>}$

$= \left\{\begin{array}{ll}
\left|\begin{array}{c|c}
{\bf U}_{L} - z{\bf I} & 0 \\ \hline
0 & {\bf U}_{R} - z{\bf I}
\end{array}\right|_{(i,l)} \\+ (-1)^{i+l}R_R\left(\left|\begin{array}{c|c}
{\bf U}_{L} - z{\bf I} & 0 \\ \hline
0 & {\bf U}_{R} - z{\bf I}
\end{array}\right|_{<i,l>,(k,j)} + (-1)^{j+k}R_L\left|\begin{array}{c|c}
{\bf U}_{L} - z{\bf I} & 0 \\ \hline
0 & {\bf U}_{R} - z{\bf I}
\end{array}\right|_{<i,l>,<k,j>}\right)\end{array}\right.$

$= \left\{\begin{array}{ll}
\left|\begin{array}{c|c}
{\bf U}_{L} - z{\bf I} & 0 \\ \hline
0 & {\bf U}_{R} - z{\bf I}
\end{array}\right|_{(i,l),(k,j)} + (-1)^{j+k}R_L\left|\begin{array}{c|c}
{\bf U}_{L} - z{\bf I} & 0 \\ \hline
0 & {\bf U}_{R} - z{\bf I}
\end{array}\right|_{(i,l),<k,j>} \\
+ (-1)^{i+l}R_R\left|\begin{array}{c|c}
{\bf U}_{L} - z{\bf I} & 0 \\ \hline
0 & {\bf U}_{R} - z{\bf I}
\end{array}\right|_{<i,l>,(k,j)} + (-1)^{i+j+k+l}R_RR_L\left|\begin{array}{c|c}
{\bf U}_{L} - z{\bf I} & 0 \\ \hline
0 & {\bf U}_{R} - z{\bf I}
\end{array}\right|_{<i,l>,<k,j>}\end{array}\right.$

By definition we know that $\left|\begin{array}{c|c}
{\bf U}_{L} - z{\bf I} & 0 \\ \hline
0 & {\bf U}_{R} - z{\bf I}
\end{array}\right|_{<i,l>,<k,j>} = \left|\begin{array}{c|c}
{\bf U}_{L} - z{\bf I} & 0 \\ \hline
0 & {\bf U}_{R} - z{\bf I}
\end{array}\right|_{<i,j>,<k,l>}$ since these matrices are missing the same rows and columns.  We can now write,

$C(z,\epsilon)= \left\{\begin{array}{ll}
\left|\begin{array}{c|c}
{\bf U}_{L} - z{\bf I} & 0 \\ \hline
0 & {\bf U}_{R} - z{\bf I}
\end{array}\right|_{(i,l),(k,j)} \\
+ (-1)^{j+k}R_L\left|\begin{array}{c|c}
{\bf U}_{L} - z{\bf I} & 0 \\ \hline
0 & {\bf U}_{R} - z{\bf I}
\end{array}\right|_{(i,l),<k,j>} + (-1)^{i+l}R_R\left|\begin{array}{c|c}
{\bf U}_{L} - z{\bf I} & 0 \\ \hline
0 & {\bf U}_{R} - z{\bf I}
\end{array}\right|_{<i,l>,(k,j)} \\
+ (-1)^{i+j+k+l}\left(R_RR_L-T^2\right)\left|\begin{array}{c|c}
{\bf U}_{L} - z{\bf I} & 0 \\ \hline
0 & {\bf U}_{R} - z{\bf I}
\end{array}\right|_{<i,l>,<k,j>}\end{array}\right.$

And finally, using the fact that $R_R = -1+2\epsilon$, $R_L = 1-2\epsilon$, and $T=2\sqrt{\epsilon-\epsilon^2}$,

\begin{equation}
C(z,\epsilon)= \left\{\begin{array}{ll}
\left|\begin{array}{c|c}
{\bf U}_{L} - z{\bf I} & 0 \\ \hline
0 & {\bf U}_{R} - z{\bf I}
\end{array}\right|_{(i,l),(k,j)} \\
+ (1-2\epsilon)\left((-1)^{j+k}\left|\begin{array}{c|c}
{\bf U}_{L} - z{\bf I} & 0 \\ \hline
0 & {\bf U}_{R} - z{\bf I}
\end{array}\right|_{(i,l),<k,j>} - (-1)^{i+l}\left|\begin{array}{c|c}
{\bf U}_{L} - z{\bf I} & 0 \\ \hline
0 & {\bf U}_{R} - z{\bf I}
\end{array}\right|_{<i,l>,(k,j)}\right) \\
- (-1)^{i+j+k+l}\left|\begin{array}{c|c}
{\bf U}_{L} - z{\bf I} & 0 \\ \hline
0 & {\bf U}_{R} - z{\bf I}
\end{array}\right|_{<i,l>,<k,j>}\end{array}\right.
\end{equation}

It's worth noting that $R_RR_L-T^2=-1$ is not a coincidence dependent on how the solutions to the unitarity condition are chosen, but is in fact the unitarity condition itself.  Thus, if we had used a solution different from $r=-1+2\epsilon$, $t=2\epsilon$,  we would find that $R_RR_L-T^2$ is always a constant.

The long manipulation in this proof is essentially just a careful removal of every $\epsilon$-dependent element of ${\bf U}$, so the explicit $\epsilon$ in the above equation is in fact the {\it only} remaining $\epsilon$.  Clearly, the characteristic polynomial is a polynomial in $z$ and an affine polynomial in $\epsilon$.  With $C_0(z)\equiv \left|{\bf U}_0 - z{\bf I}\right|$ we can now write

\begin{equation}
C(z,\epsilon) \equiv \left|{\bf U}-z{\bf I}\right| = C_0(z) + \epsilon f(z)
\end{equation}

$\square$

\subsection{Proofs from section 4 (Pairing)}

\begin{thma} {\bf 4.2}
$||{\bf P}^{(k)}(\epsilon) - {\bf P}^{(k)}(0)||=O(\sqrt{\epsilon})$ and $|V^{(k)}(\epsilon)\rangle=|V^{(k)}(0)\rangle+O(\sqrt{\epsilon})$.
\end{thma}

{\it Proof} The "resolvent" is a matrix defined as ${\bf R}(\zeta,\epsilon) = ({\bf U} - \zeta{\bf I})^{-1}$.

When $\zeta_0$ is not an eigenvalue of ${\bf U}$,

$\begin{array}{ll}
{\bf U} - \zeta{\bf I} \\
= {\bf U}_0 - \zeta_0{\bf I} - (\zeta-\zeta_0){\bf I} + ({\bf U}-{\bf U}_0) \\
= {\bf U}_0 - \zeta_0{\bf I} - (\zeta{\bf I} - \zeta_0{\bf I} + {\bf U}-{\bf U}_0){\bf R}(\zeta_0,0)({\bf U}_0 - \zeta_0{\bf I}) \\
= [{\bf I} - (\zeta{\bf I} - \zeta_0{\bf I} + {\bf U} - {\bf U}_0){\bf R}(\zeta_0,0)]({\bf U}_0 - \zeta_0{\bf I}) \\
\Rightarrow {\bf R}(\zeta,\epsilon) = {\bf R}(\zeta_0,0) [{\bf I} - ((\zeta - \zeta_0){\bf I} + {\bf U} - {\bf U}_0){\bf R}(\zeta_0,0)]^{-1} \\
\end{array}$

Since ${\bf U}_0$ is a known unitary matrix, for which $\zeta_0$ is not an eigenvalue, we know that ${\bf R}(\zeta_0,0)$ is well-defined.  The entries of ${\bf U}$ can be expanded as power series in $\sqrt{\epsilon}$, and since the entries of ${\bf U}$ are continuous functions of $\epsilon$, $||{\bf U} - {\bf U}_0||$ can be made arbitrarily small.  Therefore, for small values of $\epsilon$ and $(\zeta-\zeta_0)$ we find that $||(\zeta - \zeta_0){\bf I} + ({\bf U} -{\bf U}_0)||<||{\bf R}(\zeta_0,0)||^{-1}$, and therefore ${\bf R}(\zeta,\epsilon)$ can be written as a double power series in $\zeta$ and $\sqrt{\epsilon}$.

Interesting things happen when $\zeta=\lambda^{(k)}$.  The projection operator onto the $\lambda^{(k)}$-eigenspace can be expressed as ${\bf P}^{(k)} = -\frac{1}{2\pi i}\oint {\bf R}(\zeta,\epsilon)d\zeta$, where the integral is taken over a curve that loops once around $\lambda^{(k)}$, and no other eigenvalues.  This can be shown by applying ${\bf P}^{(k)}$ to the eigenvector $|V^{(j)}\rangle$.

Since $|V^{(j)}\rangle = {\bf R}(\zeta,\epsilon)({\bf U} - \zeta{\bf I})|V^{(j)}\rangle = (\lambda^{(j)} - \zeta){\bf R}(\zeta,\epsilon)|V^{(j)}\rangle$ we know that ${\bf R}(\zeta,\epsilon)|V^{(j)}\rangle = \frac{1}{\lambda^{(j)} - \zeta}|V^{(j)}\rangle$.

It follows that,

$\begin{array}{ll}
{\bf P}^{(k)}|V^{(j)}\rangle \\
= -\frac{1}{2\pi i}\oint {\bf R}(\zeta,\epsilon)d\zeta|V^{(j)}\rangle \\
= -\frac{1}{2\pi i}\oint {\bf R}(\zeta,\epsilon)|V^{(j)}\rangle d\zeta \\
= -\frac{1}{2\pi i}\oint  \frac{1}{\lambda^{(j)} - \zeta}|V^{(j)}\rangle d\zeta \\
= \frac{1}{2\pi i}\oint  \frac{1}{\zeta - \lambda^{(j)}}d\zeta |V^{(j)}\rangle \\
= \delta_{jk} |V^{(j)}\rangle \\
\end{array}$

The last step follows from the residue theorem, and the assumption that the integral is over a path that encloses only $\lambda^{(k)}$.  $[{\bf P}^{(k)}]^2={\bf P}^{(k)}$, so ${\bf P}^{(k)}$ is a projection, and ${\bf P}^{(j)}{\bf P}^{(k)} = \delta_{jk}{\bf P}^{(k)}$, by the orthogonality of eigenvectors with different eigenvalues.  Notice that this isn't a projection onto a particular eigenvector with eigenvalue $\lambda^{(k)}$, but is a projection onto the eigenspace for $\lambda^{(k)}$.

The important thing is that since ${\bf R}(\zeta,\epsilon)$ is expressible as a power series in $\sqrt{\epsilon}$, then so is ${\bf P}^{(k)}$.

Away from $\epsilon=0$ the eigenvalues are distinct, and thus the eigenprojections are 1-dimensional and can be written ${\bf P}^{(k)} = |V^{(k)}\rangle\langle V^{(k)}|$.  So if the entries of ${\bf P}^{(k)}$ take the form of $\sum_{n=0}^\infty c_n(\sqrt{\epsilon})^n$, then so do the entries in the corresponding eigenvector.

$\square$

\vspace{5mm}

\begin{thma} {\bf 4.4}
Define ${\bf U}|V^{(j)}(\epsilon)\rangle = \lambda^{(j)}(\epsilon)|V^{(j)}(\epsilon)\rangle$ for all $j$, $\mathcal{S} = span\{|V^{(1)}(0)\rangle, \cdots,|V^{(j)}(0)\rangle\}$, and ${\bf P}_{\mathcal{S}}$ as the projection operator onto $\mathcal{S}$.

If $|u\rangle \in \mathcal{S}$, then $\forall m$

$i) \quad{\bf P}_{\mathcal{S}^\perp}{\bf U}_0^m|u\rangle = 0$

$ii) \quad{\bf P}_{\mathcal{S}^\perp}{\bf U}^m|u\rangle = O(\sqrt{\epsilon})$

That is, if $|u\rangle\in\mathcal{S}$, then ${\bf U}_0^m|u\rangle$ is also in $\mathcal{S}$, and ${\bf U}^m|u\rangle$ is almost entirely in $\mathcal{S}$.
\end{thma}

{\it Proof} $i)$

$\begin{array}{ll}
|u\rangle\in\mathcal{S} \\
\Rightarrow |u\rangle = \alpha_1|V^{(1)}(0)\rangle + \cdots + \alpha_j|V^{(j)}(0)\rangle \\
\Rightarrow {\bf U}_0^m|u\rangle = \alpha_1\left[\lambda^{(1)}\right]^m(0)|V^{(1)}(0)\rangle + \cdots + \alpha_j\left[\lambda^{(j)}\right]^m(0)|V^{(j)}(0)\rangle \\
\Rightarrow {\bf U}_0^m|u\rangle \in \mathcal{S} \\
\end{array}$

$ii)$

Notice first that ${\bf U}_0$ is unitary, so $||{\bf U}_0||=1$.  Since it has entries of $O(\sqrt{\epsilon})$, $||{\bf U} - {\bf U}_0||=O(\sqrt{\epsilon})$ and therefore $||{\bf U}||\le||{\bf U}_0||+O(\sqrt{\epsilon})=1+O(\sqrt{\epsilon})$.  When on the "unitary preserving path", $||{\bf U}||=1$.

$\begin{array}{ll}
|u\rangle\in\mathcal{S} \\
\Rightarrow |u\rangle = \alpha_1|V^{(1)}(0)\rangle + \cdots + \alpha_j|V^{(j)}(0)\rangle \\
= |u\rangle = \alpha_1|V^{(1)}(\epsilon)\rangle + \cdots + \alpha_j|V^{(j)}(\epsilon)\rangle +O(\sqrt{\epsilon}) & \textrm{By thm. 4.3} \\
\Rightarrow {\bf U}^m|u\rangle = \alpha_1[\lambda^{(1)}(\epsilon)]^m|V^{(1)}(\epsilon)\rangle + \cdots + \alpha_j[\lambda^{(j)}(\epsilon)]^m|V^{(j)}(\epsilon)\rangle + {\bf U}^m O(\sqrt{\epsilon}) \\
= \alpha_1[\lambda^{(1)}(\epsilon)]^m|V^{(1)}(\epsilon)\rangle + \cdots + \alpha_j[\lambda^{(j)}(\epsilon)]^m|V^{(j)}(\epsilon)\rangle + O(\sqrt{\epsilon}) & \textrm{By unitrarity} \\
= \alpha_1[\lambda^{(1)}(\epsilon)]^m|V^{(1)}(0)\rangle + \cdots + \alpha_j[\lambda^{(j)}(\epsilon)]^m|V^{(j)}(0)\rangle + O(\sqrt{\epsilon}) & \textrm{By thm. 4.3} \\
\end{array}$

Since $\alpha_1[\lambda^{(1)}(\epsilon)]^m|V^{(1)}(0)\rangle + \cdots + \alpha_j[\lambda^{(j)}(\epsilon)]^m|V^{(j)}(0)\rangle\in\mathcal{S}$, ${\bf P}_{\mathcal{S}^\perp}{\bf U}^m|u\rangle = O(\sqrt{\epsilon})$.

$\square$

\begin{thma} {\bf 4.8}

Assume that the Right side $\lambda_0$-eigenspace of ${\bf U}_0$ is $D$ dimensional.

1) If the $\lambda_0$-eigenspace of ${\bf U}_0$ is bound in $G$, then the $\lambda_0$-eigenspace of ${\bf U}$ is $D$ dimensional and all of the associated eigenvectors are constant.  This is case i) of the Three Case Theorem.

2) If the $\lambda_0$-eigenspace of ${\bf U}_0$ is in contact with the hub vertex, then the Right sided $\lambda_0$-eigenspace of ${\bf U}$ is $D$-1 dimensional and the $D$-1 associated eigenvectors are constant and bound in $G$.  This leaves one eigenvector which is non-constant in $\epsilon$, and is in contact with the hub vertex.  This is either case ii or case iii of the Three Case Theorem.

\end{thma}

{\it Proof} The first result is trivial.  If an eigenvector is bound in $G$, then varying $\epsilon$ (which only affects reflection and transmission across the hub vertex) can't have any impact on it.  So, for eigenvectors bound in $G$, ${\bf U}|V\rangle = {\bf U}_0|V\rangle = \lambda_0|V\rangle$.

\vspace{5mm}

For the second result we assume that the $\lambda_0$-eigenspace is in contact with the hub vertex, and we'll use the set up described in section 2.  Define $|V\rangle = \alpha|in\rangle + \beta|out\rangle + \gamma|1,0\rangle + \delta|0,1\rangle + |G\rangle$, where $|G\rangle = {\bf P}_G|V\rangle$ and ${\bf U}|V\rangle = \lambda_0|V\rangle$.  So $|V\rangle$ is in contact with the hub vertex, is an eigenvector of ${\bf U}$, and has a constant eigenvalue.  It may seem too restrictive to assume that the Left side has a particular form, but we'll find that it makes no difference.

The rough idea of the proof is to show that if $|V_0\rangle$ is a hub-adjacent $\lambda_0$ eigenvector of ${\bf U}_0$, then $|V\rangle$ cannot be a $\lambda_0$ eigenvector of ${\bf U}$.  Instead, the eigenvalue must be a non-constant function of $\epsilon$.  Since a $\lambda_0$ eigenvector is lost when we change from the $\epsilon=0$ case to the $\epsilon\ne0$ case, we can say that if the $\lambda_0$-eigenspace of ${\bf U}_0$ is $D$ dimensional, then the $\lambda_0$-eigenspace of ${\bf U}$ is $D$-1 dimensional.

\vspace{5mm}

Assume that there is no Left side $\lambda_0$-eigenspace.  By the Three Case theorem and theorem 4.7, the only possibilities are that all of the eigenvectors in the $\lambda_0$ family are constant, or there is one unique non-constant eigenvector.  Since the $\lambda_0$ space of ${\bf U}_0$ is entirely Right sided, $|V_0\rangle$ takes the form $|V_0\rangle = \gamma_0|1,0\rangle + \delta_0|0,1\rangle + |G_0\rangle$.  It follows that,

${\bf U}_0|V_0\rangle = \lambda_0|V_0\rangle$

$\Rightarrow -\gamma_0|0,1\rangle + {\bf U}_0\left[\delta_0 |0,1\rangle + |G_0\rangle\right] = \gamma_0\lambda_0|1,0\rangle + \delta_0\lambda_0|0,1\rangle + \lambda_0|G_0\rangle$

$\Rightarrow \left\{\begin{array}{ll}
\gamma_0 = -\delta_0 \lambda_0 \\
{\bf U}_0\left[\delta_0 |0,1\rangle + |G_0\rangle\right] = \gamma_0\lambda_0|0,1\rangle + \lambda_0|G_0\rangle
\end{array}\right.$

Keep both of these innocent looking results in mind for a moment.

Now extending to the $\epsilon\ne0$ case,

${\bf U}|V\rangle=$

$\alpha\left[(1-2\epsilon)|out\rangle + 2\sqrt{\epsilon-\epsilon^2}|0,1\rangle\right] + \beta e^{i\phi}|in\rangle + \gamma\left[(-1+2\epsilon)|0,1\rangle + 2\sqrt{\epsilon-\epsilon^2}|out\rangle \right] + {\bf U}\left[\delta |0,1\rangle + |G\rangle\right]$

${\bf U}|V\rangle=\lambda_0|V\rangle\Rightarrow\left\{\begin{array}{ll}
|in\rangle: & \beta e^{i\phi} = \alpha\lambda_0 \\
|out\rangle: & \alpha(1-2\epsilon) + \gamma2\sqrt{\epsilon-\epsilon^2} = \beta\lambda_0 \\
|0,1\rangle: & \alpha2\sqrt{\epsilon-\epsilon^2} + \gamma(-1+2\epsilon) = \delta\lambda_0 \\
& {\bf U}\left[\delta |0,1\rangle + |G\rangle\right] = \gamma\lambda_0|1,0\rangle + \lambda_0|G\rangle
\end{array}\right.$

\vspace{5mm}

This is difficult to solve directly, but fortunately the last of these four relations can be used to eliminate a variable.  Notice that ${\bf U}\left[\delta |0,1\rangle + |G\rangle\right] = {\bf U}_0\left[\delta |0,1\rangle + |G\rangle\right]$ since both $|0,1\rangle$ and $|G\rangle$ are unaffected by the hub vertex.  Beginning with the last relation from both the $\epsilon=0$ and $\epsilon\ne0$ cases,

$\begin{array}{ll}
\left\{\begin{array}{ll}
{\bf U}_0\left[\delta_0 |0,1\rangle + |G_0\rangle\right] = \gamma_0\lambda_0|0,1\rangle + \lambda_0|G_0\rangle \\
{\bf U}\left[\delta |0,1\rangle + |G\rangle\right] = \gamma\lambda_0|1,0\rangle + \lambda_0|G\rangle
\end{array}\right.\\[2mm]
\Rightarrow\left\{\begin{array}{ll}
{\bf U}_0\left[\delta_0 |0,1\rangle + |G_0\rangle\right] = \gamma_0\lambda_0|0,1\rangle + \lambda_0|G_0\rangle \\
{\bf U}_0\left[\delta |0,1\rangle + |G\rangle\right] = \gamma\lambda_0|1,0\rangle + \lambda_0|G\rangle
\end{array}\right.\\[2mm]
\Rightarrow \left[\bar{\delta_0} \langle0,1| + \langle G_0|\right]{\bf U}_0^\dagger{\bf U}_0\left[\delta |0,1\rangle + |G\rangle\right] = \left(\bar{\gamma_0}\bar{\lambda_0}\langle1,0| + \bar{\lambda_0}\langle G_0|\right) \left(\gamma\lambda_0|0,1\rangle + \lambda_0|G\rangle\right) \\[2mm]
\Rightarrow \bar{\delta_0} \delta \langle0,1|0,1\rangle + \langle G_0|G\rangle = \bar{\gamma_0} \gamma\bar{\lambda_0}\lambda_0\langle1,0|1,0\rangle+ \bar{\lambda_0} \lambda_0\langle G_0|G\rangle\\[2mm]
\Rightarrow \bar{\delta_0} \delta + \langle G_0|G\rangle = \bar{\gamma_0}\gamma + \langle G_0|G\rangle\\[2mm]
\Rightarrow \bar{\delta_0} \delta = \bar{\gamma_0}\gamma \\[2mm]
\Rightarrow \bar{\delta_0} \delta = \left(-\bar{\delta_0} \bar{\lambda_0}\right)\gamma \\[2mm]
\Rightarrow \delta = - \bar{\lambda_0}\gamma \\[2mm]
\Rightarrow -\delta\lambda_0 = \gamma \\[2mm]
\end{array}$

\vspace{5mm}

So we now have four straightforward equations and four variables:

$\left\{\begin{array}{ll}
\beta e^{i\phi} = \alpha\lambda_0 \\[2mm]
\alpha(1-2\epsilon) + \gamma2\sqrt{\epsilon-\epsilon^2} = \beta\lambda_0 \\[2mm]
\alpha2\sqrt{\epsilon-\epsilon^2} + \gamma(-1+2\epsilon) = \delta\lambda_0 \\[2mm]
\gamma = -\delta\lambda_0
\end{array}\right.\\[2mm]$

$\Rightarrow\left\{\begin{array}{ll}
\beta = \alpha\lambda_0e^{-i\phi} \\[2mm]
\alpha(1-2\epsilon) + (-\delta\lambda_0)2\sqrt{\epsilon-\epsilon^2} = \beta\lambda_0 \\[2mm]
\alpha2\sqrt{\epsilon-\epsilon^2} + (-\delta\lambda_0)(-1+2\epsilon) = \delta\lambda_0
\end{array}\right.\\[2mm]$

$\Rightarrow\left\{\begin{array}{ll}
\alpha(1-2\epsilon) - \delta\lambda_02\sqrt{\epsilon-\epsilon^2} = (\alpha\lambda_0e^{-i\phi})\lambda_0 \\[2mm]
\alpha2\sqrt{\epsilon-\epsilon^2} = \delta\lambda_02\epsilon
\end{array}\right.\\[2mm]$

$\Rightarrow\left\{\begin{array}{ll}
\alpha(1 - \lambda_0^2e^{-i\phi} - 2\epsilon) = \delta\lambda_02\sqrt{\epsilon-\epsilon^2} \\[2mm]
\alpha2\sqrt{\epsilon-\epsilon^2} = \delta\lambda_02\epsilon
\end{array}\right.\\[2mm]$

$\Rightarrow\left\{\begin{array}{ll}
\alpha(1 - \lambda_0^2e^{-i\phi} - 2\epsilon) = \delta\lambda_02\sqrt{\epsilon-\epsilon^2} \\[2mm]
\delta = \alpha\frac{\sqrt{\epsilon-\epsilon^2}}{\lambda_0\epsilon}
\end{array}\right.\\[2mm]$

$\Rightarrow \alpha(1 - \lambda_0^2e^{-i\phi} - 2\epsilon) = \left(\alpha\frac{\sqrt{\epsilon-\epsilon^2}}{\lambda_0\epsilon}\right)\lambda_02\sqrt{\epsilon-\epsilon^2}$

$\Rightarrow \alpha(1 - \lambda_0^2e^{-i\phi} - 2\epsilon) = \alpha(2-2\epsilon)$

$\Rightarrow \alpha(\lambda_0^2 + e^{i\phi}) = 0$

Unless $\phi$ was specifically chosen so that $e^{i\phi} + \lambda_0^{2} = 0$ it follows that $\alpha=\beta=\gamma=\delta=0$, which contradicts the statement that $|V_0\rangle$ is in contact with the hub vertex.  What has just been shown is that if $|V_0\rangle$ is a $\lambda_0$ eigenvector of ${\bf U}_0$ in contact with the hub vertex, then $|V\rangle$ either has a different eigenvalue or $e^{i\phi} + \lambda_0^{2} = 0$.

First assume that $\lambda_0^2 + e^{i\phi} \ne 0$ and $\lambda_0^2 - e^{i\phi} \ne 0$ (there is no Left side $\lambda_0$-eigenspace).

If the $\lambda_0$ eigenspace of ${\bf U}_0$ is in contact with the hub vertex, then by moving from the $\epsilon=0$ case to the $\epsilon\ne0$ case at least one eigenvector is lost.  Since $\lambda_0^2 - e^{i\phi} \ne 0$, by thm. 4.7 there can be no pairing.  The Three Case Theorem then implies that there is at most only one non-constant eigenvalue in the $\lambda_0$ family.  So, only one eigenvector is lost and if the Right side $\lambda_0$-eigenspace of ${\bf U}_0$ is $D$ dimensional, then the Right side $\lambda_0$-eigenspace of ${\bf U}$ must be $D$-1 dimensional.

$\square$

\subsection*{The Fundamental Pairing Theorem}

\begin{thma} {\bf 4.10}
The $\lambda_0$-eigenspace is in contact with both the Left and Right sides of the hub vertex if and only if there exists paired vectors $|V^\pm\rangle$ with eigenvalues of the form $\lambda_0e^{\pm ic\sqrt{\epsilon}}+O(\epsilon)$.
\end{thma}

{\it Proof} Both sides of the $\lambda_0$-eigenspace have an active eigenvector, because both sides are in contact with the hub (thm. 4.8).  Define $|\ell_0\rangle$ and $|\mathfrak{r}_0\rangle$ to be the Left and Right side active eigenvectors.

Assume that $|P\rangle$ and $|Q\rangle$ are eigenvectors with eigenvalues in the $\lambda_0$ family, and that both are in contact with the hub vertex.  They may be paired, or one or both may be constant.  $|P\rangle$ and $|Q\rangle$ are in the active subspace, and $|P_0\rangle$ and $|Q_0\rangle$ are in the $\lambda_0$-eigenspace of ${\bf U}_0$, so it follows that $span\{|P_0\rangle, |Q_0\rangle\}=span\{|\ell_0\rangle,|\mathfrak{r}_0\rangle\}$.  Define the eigenvalues of $|P\rangle$ and $|Q\rangle$ to be $\lambda_0e^{ip(\epsilon)}$ and $\lambda_0e^{iq(\epsilon)}$, where $p(0)=q(0)=0$.

Because the spans are equal and two dimensional there exists a transformation between the orthonormal bases, $\{|\ell_0\rangle,|\mathfrak{r}_0\rangle\}$ and $\{|P_0\rangle,|Q_0\rangle\}$, which we can write

$\left(\begin{array}{cc} |\ell_0\rangle \\ |\mathfrak{r}_0\rangle \end{array}\right) = \left(\begin{array}{cc} e^{i\gamma}cos(\omega) & e^{i\delta}sin(\omega) \\ e^{i\gamma+i\eta}sin(\omega) & -e^{i\delta+i\eta}cos(\omega) \end{array}\right) \left(\begin{array}{cc} |P_0\rangle \\ |Q_0\rangle \end{array}\right)$.

By carefully choosing the relative phases between all four vectors, we can assume without loss of generality that $\left(\begin{array}{cc} |\ell_0\rangle \\ |\mathfrak{r}_0\rangle \end{array}\right) = \left(\begin{array}{cc} cos(\omega) & sin(\omega) \\ -sin(\omega) & cos(\omega) \end{array}\right) \left(\begin{array}{cc} |P_0\rangle \\ |Q_0\rangle \end{array}\right)$, with $0\le\omega\le\frac{\pi}{2}$, and use this to define $|\ell\rangle$ and $|\mathfrak{r}\rangle$ as: $\left(\begin{array}{cc} |\ell\rangle \\ |\mathfrak{r}\rangle \end{array}\right) = \left(\begin{array}{cc} cos(\omega) & sin(\omega) \\ -sin(\omega) & cos(\omega) \end{array}\right) \left(\begin{array}{cc} |P\rangle \\ |Q\rangle \end{array}\right)$.  

Notice that while $|\ell_0\rangle$ and $|\mathfrak{r}_0\rangle$ are eigenvectors of ${\bf U}_0$, $|\ell\rangle$ and $|\mathfrak{r}\rangle$ are \underline{not} eigenvectors of ${\bf U}$, they're merely defined in terms of the eigenvectors $|P\rangle$ and $|Q\rangle$.  Since eigenvectors can be expressed as power series in $\sqrt{\epsilon}$, we can write $|\ell\rangle = |\ell_0\rangle + \sqrt{\epsilon}|\ell_1\rangle+O(\epsilon)$, and similarly for $|\mathfrak{r}\rangle$.

In the $\{|P\rangle,|Q\rangle\}$ basis ${\bf U} = \lambda_0\left(\begin{array}{cc} e^{ip(\epsilon)} & 0 \\ 0 & e^{iq(\epsilon)} \end{array}\right)$

and a similarity transform allows us to write this in the $\{|\ell\rangle,|\mathfrak{r}\rangle\}$ basis,

${\bf U} = \left(\begin{array}{cc} cos(\omega) & sin(\omega) \\ -sin(\omega) & cos(\omega) \end{array}\right) \left(\begin{array}{cc} \lambda_0 e^{ip(\epsilon)} & 0 \\ 0 & \lambda_0 e^{iq(\epsilon)} \end{array}\right) \left(\begin{array}{cc} cos(\omega) & -sin(\omega) \\ sin(\omega) & cos(\omega) \end{array}\right) \\[2mm]
= \lambda_0 \left(\begin{array}{cc} cos(\omega) & sin(\omega) \\ -sin(\omega) & cos(\omega) \end{array}\right) \left(\begin{array}{cc} cos(\omega) e^{ip(\epsilon)} & -sin(\omega) e^{ip(\epsilon)} \\ sin(\omega) e^{iq(\epsilon)} & cos(\omega) e^{iq(\epsilon)} \end{array}\right) \\[2mm]
= \lambda_0 \left(\begin{array}{cc} cos^2(\omega) e^{ip(\epsilon)} + sin^2(\omega) e^{iq(\epsilon)} & sin(\omega)cos(\omega)\left( -e^{ip(\epsilon)} + e^{iq(\epsilon)} \right) \\ sin(\omega)cos(\omega)\left( -e^{ip(\epsilon)} + e^{iq(\epsilon)} \right) & cos^2(\omega) e^{iq(\epsilon)} + sin^2(\omega) e^{ip(\epsilon)} \end{array}\right) \\[2mm]$

Using the trig identities $cos^2(\omega) = \frac{1 + cos(2\omega)}{2}$, $sin^2(\omega) = \frac{1 - cos(2\omega)}{2}$, and $sin(\omega)cos(\omega) = \frac{sin(2\omega)}{2}$ we find

${\bf U} = \frac{\lambda_0}{2} \left(\begin{array}{cc} e^{ip(\epsilon)} + e^{iq(\epsilon)} + cos(2\omega) \left(e^{ip(\epsilon)} - e^{iq(\epsilon)}\right) & sin(2\omega) \left( -e^{ip(\epsilon)} + e^{iq(\epsilon)} \right) \\ sin(2\omega) \left( -e^{ip(\epsilon)} + e^{iq(\epsilon)} \right) & e^{ip(\epsilon)} + e^{iq(\epsilon)} - cos(2\omega) \left(e^{ip(\epsilon)} - e^{iq(\epsilon)}\right) \end{array}\right) \\[2mm]$

This is enough to determine the value of $\omega$.

Notice that $2|{\bf U}_{12}|=|sin(2\omega)||e^{ip(\epsilon)} - e^{iq(\epsilon)}|$ and $|{\bf U}_{11}-{\bf U}_{22}|=|cos(2\omega)||e^{ip(\epsilon)} - e^{iq(\epsilon)}|$.  It follows that $\frac{4|{\bf U}_{12}|^2}{4|{\bf U}_{12}|^2+|{\bf U}_{11}-{\bf U}_{22}|^2} = \frac{sin^2(2\omega)|e^{ip(\epsilon)} - e^{iq(\epsilon)}|^2}{sin^2(2\omega)|e^{ip(\epsilon)} - e^{iq(\epsilon)}|^2+cos^2(2\omega)|e^{ip(\epsilon)} - e^{iq(\epsilon)}|^2} = \frac{sin^2(2\omega)}{sin^2(2\omega)+cos^2(2\omega)} = sin^2(2\omega)$.

Or more simply, $sin^2(2\omega) = \left(1+\left|\frac{{\bf U}_{11} - {\bf U}_{22}}{2{\bf U}_{12}}\right|^2\right)^{-1}$.  We can calculate $|{\bf U}_{11}-{\bf U}_{22}|$ and $|{\bf U}_{12}|$ directly:

$\begin{array}{ll}
& |{\bf U}_{11}-{\bf U}_{22}|\\[2mm]
= & |\langle\ell|{\bf U}|\ell\rangle - \langle\mathfrak{r}|{\bf U}|\mathfrak{r}\rangle| \\[2mm]
= & |\langle\ell_0|{\bf U}_0|\ell_0\rangle - \langle \mathfrak{r}_0|{\bf U}_0|\mathfrak{r}_0\rangle \\[2mm]
&+\sqrt{\epsilon}\left(\langle\ell_1|{\bf U}_0|\ell_0\rangle + \langle\ell_0|{\bf U}_1|\ell_0\rangle + \langle\ell_0|{\bf U}_0|\ell_1\rangle - \langle \mathfrak{r}_1|{\bf U}_0|\mathfrak{r}_0\rangle - \langle \mathfrak{r}_0|{\bf U}_1|\mathfrak{r}_0\rangle - \langle \mathfrak{r}_0|{\bf U}_0|\mathfrak{r}_1\rangle\right)|+O(\epsilon) \\[2mm]
= & |\lambda_0\langle\ell_0|\ell_0\rangle - \lambda_0\langle \mathfrak{r}_0|\mathfrak{r}_0\rangle \\[2mm]
&+\sqrt{\epsilon}\left(\lambda_0\langle\ell_1|\ell_0\rangle + \langle\ell_0|{\bf U}_1|\ell_0\rangle + \lambda_0\langle\ell_0|\ell_1\rangle - \lambda_0\langle \mathfrak{r}_1|\mathfrak{r}_0\rangle - \langle \mathfrak{r}_0|{\bf U}_1|\mathfrak{r}_0\rangle - \lambda_0\langle \mathfrak{r}_0|\mathfrak{r}_1\rangle\right)|+O(\epsilon) \\[2mm]
= & |\lambda_0 - \lambda_0 +\sqrt{\epsilon}\left(\lambda_0[(\langle\ell_1|\ell_0\rangle + \langle\ell_0|\ell_1\rangle) - (\langle \mathfrak{r}_1|\mathfrak{r}_0\rangle + \langle \mathfrak{r}_0|\mathfrak{r}_1\rangle)] + \langle\ell_0|{\bf U}_1|\ell_0\rangle - \langle \mathfrak{r}_0|{\bf U}_1|\mathfrak{r}_0\rangle\right)|+O(\epsilon) \\[2mm]
= & |0+\sqrt{\epsilon}\left(\lambda_0[0-0] + \langle\ell_0|{\bf U}_1|\ell_0\rangle - \langle \mathfrak{r}_0|{\bf U}_1|\mathfrak{r}_0\rangle\right)|+O(\epsilon) \\[2mm]
= & \sqrt{\epsilon}|\langle\ell_0|{\bf U}_1|\ell_0\rangle - \langle \mathfrak{r}_0|{\bf U}_1|\mathfrak{r}_0\rangle|+O(\epsilon) \\[2mm]
= & \sqrt{\epsilon}|0-0|+O(\epsilon) \\[2mm]
= & O(\epsilon) \\
\end{array}$

In the last step here we used the fact that $|\ell_0\rangle$ and $|\mathfrak{r}_0\rangle$ are one-sided, and since ${\bf U}_1$ is only involved in transmitting between the Left and Right sides, $\langle\ell_0|{\bf U}_1|\ell_0\rangle = \langle \mathfrak{r}_0|{\bf U}_1|\mathfrak{r}_0\rangle = 0$.

\vspace{5mm}

$\begin{array}{ll}
& |{\bf U}_{12}| \\[2mm]
= & |\langle\mathfrak{r}|{\bf U}|\ell\rangle| \\[2mm]
= & |\langle\mathfrak{r}_0|{\bf U}_0|\ell_0\rangle + \sqrt{\epsilon}\left(\langle\mathfrak{r}_1|{\bf U}_0|\ell_0\rangle + \langle\mathfrak{r}_0|{\bf U}_1|\ell_0\rangle + \langle\mathfrak{r}_0|{\bf U}_0|\ell_1\rangle\right)| + O(\epsilon) \\[2mm]
= & |\lambda_0\langle\mathfrak{r}_0|\ell_0\rangle + \sqrt{\epsilon}\left(\lambda_0\langle\mathfrak{r}_1|\ell_0\rangle + \lambda_0\langle\mathfrak{r}_0|\ell_1\rangle + \langle\mathfrak{r}_0|{\bf U}_1|\ell_0\rangle\right)| + O(\epsilon) \\[2mm]
= & \sqrt{\epsilon}|\langle\mathfrak{r}_0|{\bf U}_1|\ell_0\rangle| + O(\epsilon) \\
\end{array}$

By construction, $\langle P|Q\rangle=0 \Rightarrow \langle\mathfrak{r}|\ell\rangle = 0$, and in the last step this fact is used twice.  First $\langle\mathfrak{r}_0|\ell_0\rangle=0$ and second, by theorem 4.6, $\langle\mathfrak{r}_0|\ell_1\rangle+\langle\mathfrak{r}_1|\ell_0\rangle=0$.  By assumption, $|\ell_0\rangle$ and $|\mathfrak{r}_0\rangle$ are adjacent to the hub vertex, $\langle\mathfrak{r}_0|{\bf U}_1|\ell_0\rangle\ne0$.

So we now have that $|{\bf U}_{12}| = O(\sqrt{\epsilon})$ and $|{\bf U}_{11}-{\bf U}_{22}|=O(\epsilon)$.  It follows that $sin^2(2\omega) = \left(1+\left|\frac{{\bf U}_{11} - {\bf U}_{22}}{2{\bf U}_{12}}\right|^2\right)^{-1} = \left(1+O(\epsilon)\right)^{-1}\Rightarrow sin^2(2\omega) =1$, since $\omega$ was originally defined independently of $\epsilon$ (e.g., $cos(\omega) = \langle \ell_0|V^+_0 \rangle$).  So, $sin(2\omega)=\pm1$, and since $0\le\omega\le\frac{\pi}{2}$, we find that $\omega=\frac{\pi}{4}$.  Now ${\bf U}$ takes a much simpler form in the $\{|\ell\rangle, |\mathfrak{r}\rangle\}$ basis:

${\bf U} = \frac{\lambda_0}{2}\left(\begin{array}{cc} (e^{ip(\epsilon)} +e^{iq(\epsilon)}) & (e^{ip(\epsilon)} - e^{iq(\epsilon)}) \\ (e^{ip(\epsilon)} - e^{iq(\epsilon)}) & (e^{ip(\epsilon)} + e^{iq(\epsilon)}) \end{array}\right)+O(\epsilon)$

Since $|e^{ip(\epsilon)} - e^{iq(\epsilon)}| = 2|{\bf U}_{12}| = O(\sqrt{\epsilon})$ it follows that $p(\epsilon)-q(\epsilon)=O(\sqrt{\epsilon})$.  So, $p(\epsilon)=O(\sqrt{\epsilon})$ or $q(\epsilon)=O(\sqrt{\epsilon})$.  Without loss of generality, assume that $p(\epsilon) = c\sqrt{\epsilon}+O(\epsilon)$.

In addition,

$\begin{array}{ll}
{\bf U}_{11} = \langle\ell|{\bf U}|\ell\rangle \\
= \langle\ell_0|{\bf U}_0|\ell_0\rangle + \sqrt{\epsilon}\left(\langle\ell_1|{\bf U}_0|\ell_0\rangle+\langle\ell_0|{\bf U}_1|\ell_0\rangle+\langle\ell_0|{\bf U}_0|\ell_1\rangle\right) +O(\epsilon) \\
= \lambda_0\langle\ell_0|\ell_0\rangle + \sqrt{\epsilon}\left(\lambda_0\langle\ell_1|\ell_0\rangle+0+\lambda_0\langle\ell_0|\ell_1\rangle\right) +O(\epsilon) \\
= \lambda_0 + \lambda_0\sqrt{\epsilon}\left(\langle\ell_1|\ell_0\rangle+\langle\ell_0|\ell_1\rangle\right) +O(\epsilon) \\
= \lambda_0 + O(\epsilon)
\end{array}$

Since $\frac{\lambda_0}{2}(e^{ip(\epsilon)} +e^{iq(\epsilon)}) = \lambda_0 + O(\epsilon)$, $p(\epsilon)+q(\epsilon)$ has no $\sqrt{\epsilon}$ term, and therefore $q(\epsilon) = -c\sqrt{\epsilon} +O(\epsilon)$.

Clearly, this situation is case iii in the three-case theorem, and since there are only two non-constant eigenvalues, and all eigenvalues that vary by $O(\sqrt{\epsilon})$ are paired with another eigenvalue, these two eigenvalues are paired to each other (and not merely coincidentally related).  So by definition $|P\rangle$ and $|Q\rangle$ are paired eigenvectors.  In fact, they are $|V^\pm\rangle$.

The reverse implication, that the $\lambda_0$-eigenspace of ${\bf U}_0$ is adjacent to both sides of the hub vertex if there exists paired eigenvectors, is a direct result of theorem 4.7.

$\square$

\subsection{Proofs from section 5 (Tolerances)}

With some foresight, define:

$\left\{\begin{array}{ll}
|V^+(\delta,\epsilon)\rangle = cos\left(\omega\right)|\ell_0\rangle + sin\left(\omega\right)|\mathfrak{r}_0\rangle + O(\delta, \sqrt{\epsilon})\\
|V^-(\delta,\epsilon)\rangle = -sin\left(\omega\right)|\ell_0\rangle + cos\left(\omega\right)|\mathfrak{r}_0\rangle + O(\delta, \sqrt{\epsilon})
\end{array}\right.$

\vspace{5mm}

\begin{thma} {\bf 5.4}
The angle between the paired eigenvectors and the active eigenvectors, $\omega$, is to lowest order a function of $\frac{\delta^2}{4c^2 \epsilon}$.
\end{thma}

{\it Proof} The (arbitrary) phase of each of the eigenvectors can be carefully chosen so that each of these amplitudes are real, and so that $0\le\omega\le\frac{\pi}{2}$.  

\vspace{5mm}

Define $\left\{\begin{array}{ll}
|\ell\rangle = cos\left(\omega\right)|V^+(\delta,\epsilon)\rangle - sin\left(\omega\right)|V^-(\delta,\epsilon)\rangle \\
|\mathfrak{r}\rangle = sin\left(\omega\right)|V^+(\delta,\epsilon)\rangle + cos\left(\omega\right)|V^-(\delta, \epsilon)\rangle
\end{array}\right.$
and
$\left\{\begin{array}{ll}
|V^+(\delta,\epsilon)\rangle = cos\left(\omega\right)|\ell\rangle + sin\left(\omega\right)|\mathfrak{r}\rangle \\
|V^-(\delta,\epsilon)\rangle = -sin\left(\omega\right)|\ell\rangle + cos\left(\omega\right)|\mathfrak{r}\rangle
\end{array}\right.$

$|\ell\rangle$ and $|\mathfrak{r}\rangle$ are projections of the corresponding active eigenvectors onto the space spanned by the paired eigenvectors.  As such, $Span\{|\ell\rangle, |\mathfrak{r}\rangle\} = Span\{|V^+\rangle, |V^-\rangle\}$ is an invariant subspace of ${\bf U}$.

\vspace{5mm}

${\bf U}\left(\begin{array}{cc}
|\ell\rangle \\
 |\mathfrak{r}\rangle \\
\end{array}\right)$

$= \left(\begin{array}{cc}
cos\left(\omega\right) & sin\left(\omega\right) \\
-sin\left(\omega\right) & cos\left(\omega\right) \\
\end{array}\right)
\left(\begin{array}{cc}
\lambda^{+} & 0 \\
0 & \lambda^{-} \\
\end{array}\right)
\left(\begin{array}{cc}
cos\left(\omega\right) & -sin\left(\omega\right) \\
sin\left(\omega\right) & cos\left(\omega\right) \\
\end{array}\right)
\left(\begin{array}{cc}
|\ell\rangle \\
 |\mathfrak{r}\rangle \\
\end{array}\right)$

$= \sqrt{\lambda_{\ell}\lambda_r} \left(\begin{array}{cc}
cos\left(\omega\right) & sin\left(\omega\right) \\
-sin\left(\omega\right) & cos\left(\omega\right) \\
\end{array}\right)
\left(\begin{array}{cc}
e^{ic\sqrt{\epsilon-\epsilon_0}} & 0 \\
0 & e^{-ic\sqrt{\epsilon-\epsilon_0}} \\
\end{array}\right)
\left(\begin{array}{cc}
cos\left(\omega\right) & -sin\left(\omega\right) \\
sin\left(\omega\right) & cos\left(\omega\right) \\
\end{array}\right)
\left(\begin{array}{cc}
|\ell\rangle \\
 |\mathfrak{r}\rangle \\
\end{array}\right) + O(\delta^2, \delta\sqrt{\epsilon-\epsilon_0}, \epsilon-\epsilon_0)$

$= \sqrt{\lambda_{\ell}\lambda_r} \left(\begin{array}{cc}
cos\left(\omega\right) & sin\left(\omega\right) \\
-sin\left(\omega\right) & cos\left(\omega\right) \\
\end{array}\right)
\left(\begin{array}{cc}
cos\left(\omega\right)e^{ic\sqrt{\epsilon-\epsilon_0}} & -sin\left(\omega\right)e^{ic\sqrt{\epsilon-\epsilon_0}} \\
sin\left(\omega\right)e^{-ic\sqrt{\epsilon-\epsilon_0}} & cos\left(\omega\right)e^{-ic\sqrt{\epsilon-\epsilon_0}} \\
\end{array}\right)
\left(\begin{array}{cc}
|\ell\rangle \\
 |\mathfrak{r}\rangle \\
\end{array}\right) + O(\delta^2, \delta\sqrt{\epsilon-\epsilon_0}, \epsilon-\epsilon_0)$

$= \sqrt{\lambda_{\ell}\lambda_r} 
\left(\begin{array}{cc}
cos^2\left(\omega\right)e^{ic\sqrt{\epsilon-\epsilon_0}} + sin^2\left(\omega\right)e^{-ic\sqrt{\epsilon-\epsilon_0}} & -sin\left(\omega\right)cos\left(\omega\right)\left(e^{ic\sqrt{\epsilon-\epsilon_0}} - e^{-ic\sqrt{\epsilon-\epsilon_0}} \right) \\
-sin\left(\omega\right)cos\left(\omega\right)\left(e^{ic\sqrt{\epsilon-\epsilon_0}} - e^{-ic\sqrt{\epsilon-\epsilon_0}} \right) & cos^2\left(\omega\right)e^{-ic\sqrt{\epsilon-\epsilon_0}} + sin^2\left(\omega\right)e^{ic\sqrt{\epsilon-\epsilon_0}} \\
\end{array}\right)
\left(\begin{array}{cc}
|\ell\rangle \\
 |\mathfrak{r}\rangle \\
\end{array}\right) + O(\delta^2, \delta\sqrt{\epsilon-\epsilon_0}, \epsilon-\epsilon_0)$

$= \sqrt{\lambda_{\ell}\lambda_r}
\left(\begin{array}{cc}
cos(c\sqrt{\epsilon-\epsilon_0}) + i\,cos(2\omega)sin(c\sqrt{\epsilon-\epsilon_0}) & -i\,sin(2\omega)sin\left(c\sqrt{\epsilon-\epsilon_0}\right) \\
-i\,sin(2\omega)sin\left(c\sqrt{\epsilon-\epsilon_0} \right) & cos(c\sqrt{\epsilon-\epsilon_0}) - i\,cos(2\omega)sin(c\sqrt{\epsilon-\epsilon_0}) \\
\end{array}\right)
\left(\begin{array}{cc}
|\ell\rangle \\
 |\mathfrak{r}\rangle \\
\end{array}\right) + O(\delta^2, \delta\sqrt{\epsilon-\epsilon_0}, \epsilon-\epsilon_0)$

\vspace{5mm}

Using the same technique used in the proof of the Fundamental Pairing theorem we can find an expression for $\omega$:

$\frac{4|{\bf U}_{lr}|^2}{|{\bf U}_{ll} - {\bf U}_{rr}|^2 + 4|{\bf U}_{lr}|^2}$

$=\frac{4sin^2(2\omega)sin^2(c\sqrt{\epsilon-\epsilon_0})}{4cos^2(2\omega)sin^2(c\sqrt{\epsilon-\epsilon_0}) + 4sin^2(2\omega)sin^2(c\sqrt{\epsilon-\epsilon_0})}$

$=\frac{sin^2(2\omega)sin^2(c\sqrt{\epsilon-\epsilon_0})}{sin^2(c\sqrt{\epsilon-\epsilon_0})}$

$=sin^2(2\omega)$

\vspace{5mm}

$\begin{array}{ll}
|{\bf U}_{ll} - {\bf U}_{rr}|^2 \\[2mm]
= |\langle \ell|{\bf U}|\ell\rangle - \langle \mathfrak{r}|{\bf U}|\mathfrak{r}\rangle|^2 \\[2mm]
= |\langle \ell_0|{\bf U}_0|\ell_0\rangle - \langle \mathfrak{r}_0|{\bf U}_0|\mathfrak{r}_0\rangle
+ \langle \ell_1|{\bf U}_0|\ell_0\rangle - \langle \mathfrak{r}_1|{\bf U}_0|\mathfrak{r}_0\rangle
+ \langle \ell_0|{\bf U}_0|\ell_1\rangle - \langle \mathfrak{r}_0|{\bf U}_0|\mathfrak{r}_1\rangle
+ \langle \ell_0|{\bf U}_1|\ell_0\rangle - \langle \mathfrak{r}_0|{\bf U}_1|\mathfrak{r}_0\rangle
+ O(\epsilon)|^2 \\[2mm]
= |\lambda_l - \lambda_r
+ \lambda_l\langle \ell_1|\ell_0\rangle - \lambda_r\langle \mathfrak{r}_1|\mathfrak{r}_0\rangle
+ \lambda_l\langle \ell_0|\ell_1\rangle - \lambda_r\langle \mathfrak{r}_0|\mathfrak{r}_1\rangle
+ 0 - 0
+ O(\epsilon)|^2 \\[2mm]
= |\lambda_l - \lambda_r + \lambda_l\left(\langle \ell_1|\ell_0\rangle + \langle \ell_0|\ell_1\rangle\right) - \lambda_r\left(\langle \mathfrak{r}_1|\mathfrak{r}_0\rangle + \langle \mathfrak{r}_0|\mathfrak{r}_1\rangle  \right) + O(\epsilon)|^2 \\[2mm]
= |\lambda_l - \lambda_r + O(\epsilon)|^2 \\[2mm]
= |\lambda_0e^{i\frac{\delta}{2}} - \lambda_0e^{-i\frac{\delta}{2}} + O(\epsilon)|^2 \\[2mm]
= |2i\lambda_0sin\left(\frac{\delta}{2}\right) + O(\epsilon)|^2 \\[2mm]
= 4sin^2\left(\frac{\delta}{2}\right) +O(\delta\epsilon,\epsilon^2) \\[2mm]
\end{array}$

\vspace{5mm}

$\begin{array}{ll}
4|{\bf U}_{lr}|^2 \\[2mm]
= |\langle \ell|{\bf U}|\mathfrak{r}\rangle|^2 \\[2mm]
= |\langle \ell_0|{\bf U}_0|\mathfrak{r}_0\rangle + \langle \ell_1|{\bf U}_0|\mathfrak{r}_0\rangle + \langle \ell_0|{\bf U}_0|\mathfrak{r}_1\rangle + \langle \ell_0|{\bf U}_1|\mathfrak{r}_0\rangle|^2 \\[2mm]
= |0 + \lambda_r\langle \ell_1|\mathfrak{r}_0\rangle + \lambda_l\langle \ell_0|\mathfrak{r}_1\rangle + \langle \ell_0|{\bf U}_1|\mathfrak{r}_0\rangle|^2 \\[2mm]
= |\lambda_0\left(e^{i\frac{\delta}{2}}\langle \ell_1|\mathfrak{r}_0\rangle + e^{-i\frac{\delta}{2}}\langle \ell_0|\mathfrak{r}_1\rangle\right) + \langle \ell_0|{\bf U}_1|\mathfrak{r}_0\rangle|^2 \\[2mm]
= |i\lambda_0sin\left(\frac{\delta}{2}\right)\left(\langle \ell_1|\mathfrak{r}_0\rangle - \langle \ell_0|\mathfrak{r}_1\rangle\right) + \langle \ell_0|{\bf U}_1|\mathfrak{r}_0\rangle|^2 \\[2mm]
= |i\lambda_0sin\left(\frac{\delta}{2}\right)\left(\langle \ell_1|\mathfrak{r}_0\rangle - \langle \ell_0|\mathfrak{r}_1\rangle\right)|^2 + 2Re\left[i\lambda_0sin\left(\frac{\delta}{2}\right)\left(\langle \ell_1|\mathfrak{r}_0\rangle - \langle \ell_0|\mathfrak{r}_1\rangle\right)\langle \mathfrak{r}_0|{\bf U}_1|\ell_0\rangle \right] + |\langle \ell_0|{\bf U}_1|\mathfrak{r}_0\rangle|^2 \\[2mm]
= sin^2\left(\frac{\delta}{2}\right)\left|\langle \ell_1|\mathfrak{r}_0\rangle - \langle \ell_0|\mathfrak{r}_1\rangle\right|^2 + 2Re\left[i\lambda_0sin\left(\frac{\delta}{2}\right)\left(\langle \ell_1|\mathfrak{r}_0\rangle - \langle \ell_0|\mathfrak{r}_1\rangle\right)\langle \mathfrak{r}_0|{\bf U}_1|\ell_0\rangle \right] + c^2\epsilon \\[2mm]
= c^2\epsilon + O\left(\delta^2\epsilon, \delta\epsilon \right) \\[2mm]
= c^2\epsilon + O\left( \delta\epsilon \right) \\[2mm]
\end{array}$

\vspace{5mm}

Plugging these in to the formula for $sin^2\left(2\omega\right)$,

$\begin{array}{ll}
sin^2(2\omega) = \frac{4|{\bf U}_{lr}|^2}{|{\bf U}_{ll} - {\bf U}_{rr}|^2 + 4|{\bf U}_{lr}|^2} \\[2mm]
= \left(1 + \frac{|{\bf U}_{ll} - {\bf U}_{rr}|^2}{4|{\bf U}_{lr}|^2}\right)^{-1} \\[2mm]
= \left(1 + \frac{4sin^2\left(\frac{\delta}{2}\right) +O(\delta\epsilon,\epsilon^2)}{4c^2\epsilon + O\left( \delta\epsilon \right)}\right)^{-1} \\[2mm]
= \left(1 + sin^2\left(\frac{\delta}{2}\right)\frac{1}{c^2\epsilon + O\left( \delta\epsilon \right)} + \frac{O(\delta\epsilon,\epsilon^2)}{c^2\epsilon + O\left( \delta\epsilon \right)}\right)^{-1}\\[2mm]
= \left(1 + \frac{sin^2\left(\frac{\delta}{2}\right)}{c^2\epsilon}\frac{1}{1 + O\left( \delta \right)} + \frac{O(\delta,\epsilon)}{1 + O\left( \delta \right)}\right)^{-1} \\[2mm]
= \left(1 + \frac{sin^2\left(\frac{\delta}{2}\right)}{c^2\epsilon} + O\left( \frac{\delta^3}{\epsilon} \right) + O(\delta,\epsilon)\right)^{-1}\\[2mm]
= \left(1 + \frac{\delta^2}{4c^2\epsilon} + O\left( \frac{\delta^3}{\epsilon}, \delta, \epsilon\right)\right)^{-1} \\[2mm]
\end{array}$

Since we have made no statement about how $\epsilon$ and $\delta$ are related, this cannot be further simplified.  However, assumeing that both variables are small, we can say that the largest term is $1 + \frac{\delta^2}{4c^2\epsilon}$.

$\square$

\vspace{5mm}

\begin{thma} {\bf 5.5}
There is a better than 50\% chance of a successful search of the $N$ edges of the hub vertex using the states $|\ell_0\rangle$ and $|\mathfrak{r}_0\rangle$ after $m=\left\lfloor \frac{\pi}{2c}\sqrt{N} \right\rfloor$ iterations of the time step operator, whenever

\begin{equation}
\delta < c\sqrt{\frac{2}{N}}
\end{equation}

where $\delta$ is the difference in phase between the Left and Right eigenvalues, and $c=\left|\langle\mathfrak{r_0|{\bf U}}|\ell_0\rangle\right|$.
\end{thma}

{\it Proof} First, we find an expression for $P(m,t)$,

$\begin{array}{ll}
P(m,t) \\[2mm]
= |\langle\mathfrak{r}_0|{\bf U}^m|\ell_0\rangle|^2 \\[2mm]
= \left|\left(sin\left(\omega\right)\langle V^+(\delta, \epsilon)| + cos\left(\omega\right)\langle V^-(\delta, \epsilon)|\right){\bf U}^m\left(cos\left(\omega\right)|V^+(\delta, \epsilon)\rangle - sin\left(\omega\right)|V^-(\delta, \epsilon)\rangle\right)\right|^2 + O\left(\sqrt{\epsilon}, \delta\right) \\[2mm]
= \left|sin\left(\omega\right)cos\left(\omega\right)\left(\lambda^+\right)^m - sin\left(\omega\right)cos\left(\omega\right)\left(\lambda^-\right)^m\right|^2 + O\left(\sqrt{\epsilon}, \delta\right) \\[2mm]
= \left|sin\left(\omega\right)cos\left(\omega\right)e^{imc\sqrt{\epsilon-\epsilon_0}} - sin\left(\omega\right)cos\left(\omega\right)e^{-imc\sqrt{\epsilon-\epsilon_0}}\right|^2 + O\left(\sqrt{\epsilon}, \delta\right) \\[2mm]
= \left|sin\left(\omega\right)cos\left(\omega\right)\right|^2\left|e^{imc\sqrt{\epsilon-\epsilon_0}} - e^{-imc\sqrt{\epsilon-\epsilon_0}}\right|^2 + O\left(\sqrt{\epsilon}, \delta\right) \\[2mm]

= \left|sin\left(\omega\right)cos\left(\omega\right)\right|^2\left|e^{imc\sqrt{(1+t)\epsilon}} - e^{-imc\sqrt{(1+t)\epsilon}}\right|^2 + O\left(\sqrt{\epsilon}, \delta\right) \\[2mm]
= \left|2sin\left(\omega\right)cos\left(\omega\right)\right|^2\left|sin\left(mc\sqrt{(1+t)\epsilon}\right)\right|^2 + O\left(\sqrt{\epsilon}, \delta\right) \\[2mm]
= sin^2\left(2\omega\right) sin^2\left(mc\sqrt{(1+t)\epsilon}\right) + O\left(\sqrt{\epsilon}, \delta\right) \\[2mm]
= \frac{1}{1+t} sin^2\left(mc\sqrt{(1+t)\epsilon}\right) + O\left(\sqrt{\epsilon}, \delta\right) \\[2mm]
\end{array}$

Notice that this isn't a function of $\epsilon$, it's a function of $(1+t)\epsilon$.  This raises issues, because we may chose the wrong value of $m$.  $m=\left\lfloor \frac{\pi}{2c\sqrt{\epsilon}} \right\rfloor$is the value that would be chosen if the graph was assumed to be "correctly tuned", with $\delta=0$.  Knowing only that the "error" between the eigenvalues is small, this value of $m$ is the natural choice.

$m = \left\lfloor \frac{\pi}{2c\sqrt{(1+t)\epsilon}} \right\rfloor$ is the value of $m$ that should be chosen if $\delta$ is known, and is being compensated for.  That is, if the exact difference between the eigenvalues is known, then the number of iterations can be adjusted to give a slightly better chance of success.

Taking into account the difference between the eigenvalues, 

$\begin{array}{ll}
P\left(\left\lfloor \frac{\pi}{2c\sqrt{(1+t)\epsilon}} \right\rfloor\right) \\[2mm]
= \frac{1}{1+t} sin^2\left(\left[\frac{\pi}{2c\sqrt{(1+t)\epsilon}}+O(1)\right]c\sqrt{(1+t)\epsilon}\right) + O\left(\sqrt{\epsilon} \right) \\[2mm]
= \frac{1}{1+t} sin^2\left(\frac{\pi}{2}+O\left(\sqrt{(1+t)\epsilon}\right)\right) + O\left(\sqrt{\epsilon}\right) \\[2mm]
= \frac{1}{1+t} cos^2\left(O\left(\sqrt{(1+t)\epsilon}\right)\right) + O\left(\sqrt{\epsilon}\right) \\[2mm]
= \frac{1}{1+t} + O\left((1+t)\epsilon, \sqrt{\epsilon}\right) \\[2mm]
= \frac{1}{1+t} + O\left(\sqrt{\epsilon}\right) \\[2mm]
\end{array}$

\vspace{5mm}

And not taking into account the difference $\delta$, but instead assuming that $\delta=0$,

$\begin{array}{ll}
P\left(\left\lfloor \frac{\pi}{2c\sqrt{\epsilon}} \right\rfloor\right) \\[2mm]
=  \frac{1}{1+t} sin^2\left(\left[\frac{\pi}{2c\sqrt{\epsilon}}+O(1)\right]c\sqrt{(1+t)\epsilon}\right) + O\left(\sqrt{\epsilon} \right) \\[2mm]
= \frac{1}{1+t} sin^2\left(\frac{\pi}{2}\sqrt{1+t}+O\left(\sqrt{(1+t)\epsilon}\right)\right) + O\left(\sqrt{\epsilon}\right) \\[2mm]
= \frac{1}{1+t} sin^2\left(\frac{\pi}{2}\sqrt{1+t}\right) + O\left(\sqrt{(1+t)\epsilon}, \sqrt{\epsilon}\right) \\[2mm]
= \frac{1}{1+t} sin^2\left(\frac{\pi}{2}\sqrt{1+t}\right) + O\left(\sqrt{\epsilon}\right) \\[2mm]
\end{array}$

\vspace{5mm}

$\frac{1}{2} < \frac{1}{1+t} sin^2\left(\frac{\pi}{2}\sqrt{1+t}\right) \le \frac{1}{1+t}$ over the interval $0\le t \le \frac{1}{2}$.  This condition can be rewritten,

$\begin{array}{ll}
t \le \frac{1}{2} \\[2mm]
\Rightarrow \frac{\delta^2}{4c^2\epsilon} \le \frac{1}{2} \\[2mm]
\Rightarrow \frac{\delta^2N}{4c^2} \le \frac{1}{2} \\[2mm]
\Rightarrow \delta^2 \le \frac{2c^2}{N} \\[2mm]
\Rightarrow \delta \le c \sqrt{\frac{2}{N}} \\[2mm]
\end{array}$

This means that $P\left(\left\lfloor \frac{\pi}{2c}\sqrt{N} \right\rfloor\right) > \frac{1}{2}$ whenever $\delta < c \sqrt{\frac{2}{N}}$.

$\square$

\end{document}